\documentclass[10pt,letterpaper]{article}
\usepackage[numbers]{natbib}
\usepackage[utf8]{inputenc}
\usepackage[english]{babel}
\usepackage{amsmath}
\usepackage{amsfonts}
\usepackage{amssymb}
\usepackage{graphicx}
\usepackage[left=2cm,right=2cm,top=2cm,bottom=2cm]{geometry}
\usepackage{multicol}
\usepackage{multirow}
\usepackage{color}
\usepackage{natbib}
\usepackage{graphicx}
\usepackage{graphics}
\usepackage{hyperref} 
\usepackage{float}

\begin{document}

\title{\Large \bf A Dissipative Photochemical Origin of Life:\\ The UVC Abiogenisis of Adenine}
\author{Karo Michaelian}

\maketitle

\begin{center}
{\small{}Department of Nuclear Physics and Application of Radiation, Instituto de F\'{i}sica, Universidad Nacional Aut\'onoma de M\'exico, Circuito Interior de la Investigaci\'{o}n Cient\'{i}fica, Cuidad Universitaria, M\'{e}xico D.F., Mexico, C.P. 04510.}

\end{center}

\date{}
\begin{center}
\textbf{\footnotesize{}karo@fisica.unam.mx}
\par\end{center}{\footnotesize \par}

\abstract{I describe the non-equilibrium thermodynamics and the photochemical mechanisms which may have been involved in the dissipative structuring, proliferation and evolution of the fundamental molecules at the origin of life from simpler and more common precursor molecules under the impressed UVC photon flux of the Archean. Dissipative structuring of the fundamental molecules is evidenced by their strong and broad wavelength absorption bands and rapid radiationless dexcitation in this wavelength region. Proliferation arises from the auto- and cross-catalytic nature of the intermediate products. Evolution towards states of concentration profiles of generally increasing photon disspative efficacy arises since the system has numerous stationary states, due to the non-linearity of the photochemical and chemical reactions with diffusion, which can be reached by amplification of a molecular concentration fluctuation near a bifurcation. An example is given of photochemical dissipative abiogenisis of adenine from the precursors HCN and H$_2$O within a fatty acid vesicle on a hot ocean surface, driven far from equilibrium by the impressed UVC light. The kinetic equations are resolved under different environmental conditions and the results analyzed within the framework of Classical Irreversible Thermodynamic theory.}

\ \\
{\bf keywords}: origin of life; disspative structuring; prebiotic chemistry; abiogenisis;  adenine; non-equilibrium thermodynamics\\

\section{Introduction}
There exists many proposals for the abiogenisis of the fundamental molecules of life (those found in all three domains of life) supported by a large body of empirical data for the exogenous delivery (comets, meteorites, and space dust) \cite{Anders1989,ChybaEtAl1990} or endogenous synthesis (atmospheric, ocean surface, warm ponds, hydrothermal vents) \cite{ChybaSagan1992, Brack1998, Sutherland2017}. Free energy sources proposed for affecting synthesis include; meteoric shock impact, electric discharge, temperature gradient, pH gradient, particle radiations, gamma rays, UV light, organocatalysis, micro forces, etc. \cite{MillerEtAl1973,OroEtAl1990,CleavesMiller2007}. A robust explanation of the origin of life, however, requires a clear understanding of not only how biologically important molecules spontaneously emerged, but also how they proliferated and evolved together into ever more complex dissipative structures, eventually leading to the global dissipative processes known as the {\em biosphere}, incorporating both biotic and abiotic components.

Contemporary proposals for the origin of life consider ``replication first'' scenarios supposing fortuitous synthesis of a self-replicating or autocatalytic molecular system undergoing incipient Darwinian-type evolution based on selection of molecular stability, fidelity, or chemical sequestration. ``Metabolism first'' scenarios assume an energy metabolizing system arising first which could have facilitated later heritable replication, also under Darwinian-like selection. No detailed theory based on physical principle has yet been proposed, however, and certainly no indulging empirical chemical reaction set has yet been found. 

Schrödinger's 1944 book ``What is Life'' \cite{Schrodinger1944} inspired consideration of life process from universal physical principles contained within classical thermodynamic, kinetic, and quantum theories (see reference \cite{Moberg2019} for a recent review). Even before Schrödingers book, however, a sophisticated non-equilibrium thermodynamic theory was being developed by Théophile de Donder and Lars Onsager.  Later, Ilya Prigogine, Paul Glansdorff, Grégoire Nicolis, Agnessa Babloyantz, and others from this ``Brussels school'' completed the program, now known as Classical Irreversible Thermodynamics (CIT). This theory has proven useful for understanding living systems and their dynamics, including the origin of life \cite{Glansdorff1977,Michaelian2009, Michaelian2011,MichaelianSimeonov2015, Michaelian2016, Michaelian2017, Michaelian2018, MichaelianSantillan2019, MichaelianRodriguez2019, MejiaMichaelian2020}, the cell \cite{PrigogineEtAl1972}, cell differentiation \cite{BabloyantzHiernaux1975}, ecosystems \cite{Michaelian2005, KleidonEtAl2010}, the biosphere \cite{PrigogineNicolis1971, Prigogine2EtAl1972, Kleidon2009,Michaelian2012, Michaelian2012a} and even the synthesis of organic molecules detected in space \cite{MichaelianSimeonov2017}. 

In this paper, I employ CIT theory to analyze the abiogenisis of adenine from HCN in a water solvent environment under the imposed Archean UVC photon flux. Our model consists of a set of non-linear photochemical and chemical reactions with diffusion occurring within a fatty acid vesicle, driven far from equilibrium by the impressed UVC light. The system has multiple stationary states consisting of different intermediate product concentration profiles through which it evolves in response to external fluctuation. selection being related to the dissipative efficacy of the profile.

There are only two classes of structures in nature; {\em equilibrium structures} and {\em non-equilibrium} structures. Equilibrium structures arise naturally and their synthesis from arbitrary distributions of material can be described through the minimization of an internal thermodynamic potential, for example, a crystal structure, or a lipid vesicle, arising from the minimization of the Gibbs potential at constant temperature and pressure. The second class is that of non-equilibrium structures (or processes) known as {\it dissipative structures} which also arise naturally, but through the optimization of the dissipation of an externally imposed generalized thermodynamic potential \cite{Prigogine1967}, for example the ``spontaneous'' emergence of convection cells arising to increase the thermal dissipation at a critical value of an externally imposed temperature gradient, or the water cycle which arises to dissipate the externally incident solar photon spectrum.

Life, although incorporating equilibrium structures, is fundamentally a non-equilibrium process and therefore its origin, proliferation and evolution are wholly determined by the dissipation of one or more thermodynamic potentials from its environment. Boltzmann recognized this almost 125 years ago \cite{Boltzmann1886} and suggested that life dissipates the solar photon potential. Present day life dissipates other thermodynamic potentials accessible on Earth's surface, for example, chemical potentials available in organic or inorganic molecules or available in concentration, temperature or pH gradients at hydrothermal vents. However, all ecosystems in which such organisms are embedded ultimately depend on the dissipation of the solar photon spectrum. 

At the origin of life around 3.9 thousand million years ago, UV photons provided approximately three orders of magnitude more free energy for dissipation as compared to that available from volcanic activity including hydrothermal vents, electric discharge, or meteoritic impact, combined \cite{MillerEtAl1973,OroEtAl1990,CleavesMiller2007}, irrespective of a more radioactive Archean Earth. This UV solar photon flux was continually available at Earth's surface during the Archean and thus could have provided the dissipative potential for not only molecular synthesis of the fundamental molecules, but also for molecular proliferation and the evolution towards structures of increasingly greater photon dissipation.

\begin{figure}[H]
\centering
\includegraphics[width=16cm]{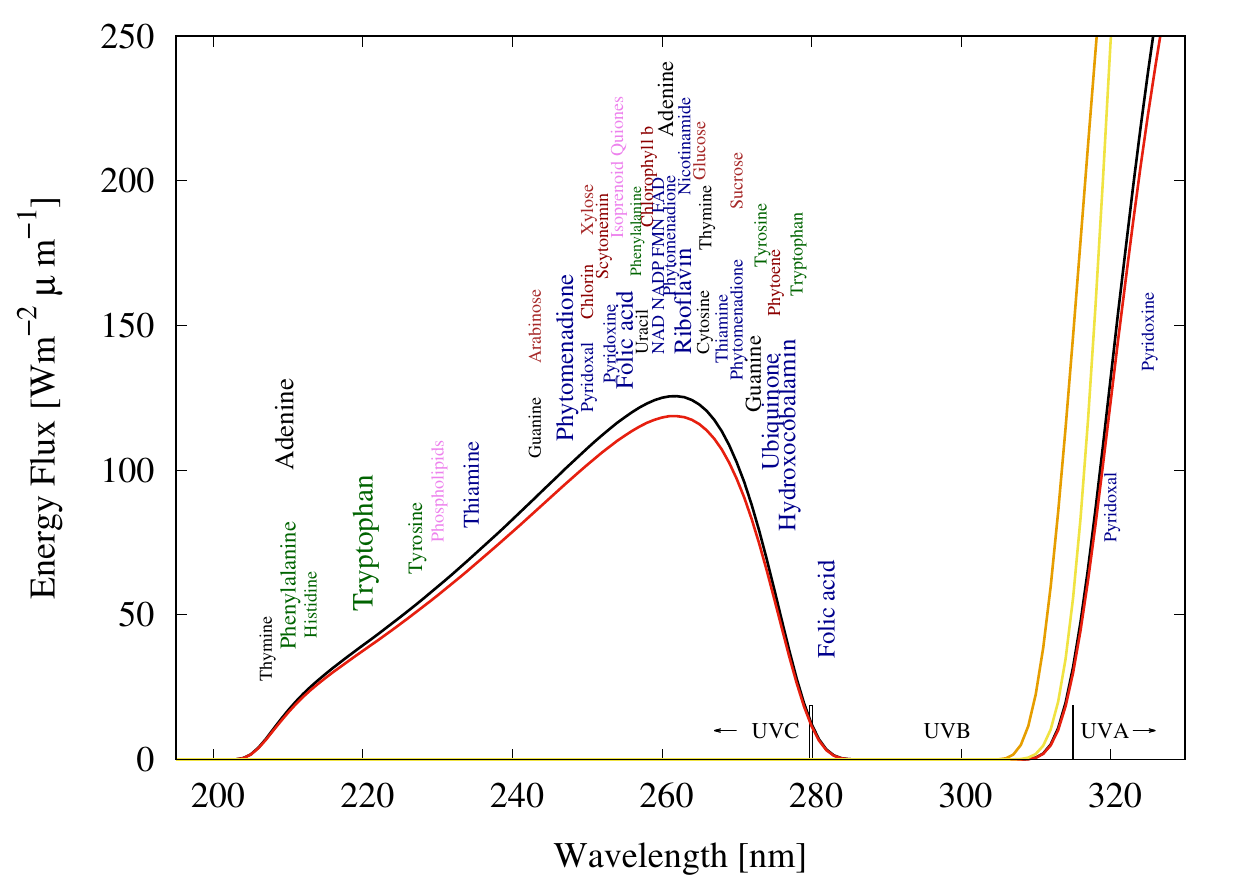}
\caption{The spectrum of light available in the UV region at Earth's surface before the origin of life at approximately 3.9 Ga and until at least 2.9 Ga (curves black and red respectively) during the Archean. CO$_2$ and probably some H$_2$S were responsible for absorption at wavelengths shorter than $\sim$ 205 nm and atmospheric aldehydes (comon photochemical products of CO$_2$ and water) absorbed between about 285 and 310 nm \cite{Sagan1973}, approximately corresponding to the UVB region. Around 2.2 Ga (yellow curve), UVC light at Earth's surface had been extinguished by oxygen and ozone resulting from organisms performing oxygenic photosynthesis. The green curve corresponds to the present surface spectrum. Energy fluxes are for the sun at the zenith. The names of the fundamental molecules of life are plotted at their wavelengths of maximum absorption; nucleic acids (black), amino acids (green), fatty acids (violet), sugars (brown), vitamins, co-enzymes and co-factors (blue), and pigments (red)  (the font size of the letter roughly corresponds to the relative size of their molar extinction coefficient). An intriguing indication of dissipative structuring is that the absorption wavelengths of these fundamental molecules coincide with the Archean UV surface spectrum. Adapted from Michaelian and Simeonov \cite{MichaelianSimeonov2015}.}
\label{fig:Pigments}
\end{figure}

We have identified the long wavelength part of the UVC region ($\sim$ 210-285 nm), plus the long wavelength part of the UVB and short wavelength part of the UVA regions ($\sim$ 310-360 nm), as the thermodynamic potential which probably drove the dissipative structuring, proliferation, and evolution relevant to the origin of life \cite{Michaelian2009, Michaelian2011}. This light prevailed at Earth's surface from the Hadean, before the probable origin of life near the beginning of the Archean ($\sim3.9$ Ga), and for at least 1000 million years \cite{BerknerMarshall1964,Sagan1973,CnossenEtAl2007} (figure \ref{fig:Pigments}) until the formation of an ozone layer when natural oxygen sinks (for example, free hydrogen and Fe$^{+2}$) became overwhelmed by organisms performing oxygenic photosynthesis. A strong argument for the relevance of this particular region of the solar spectrum, corresponding to the Archean atmospheric transparency at the origin of life, is that longer wavelength photons do not contain sufficient free energy to directly break covalent bonds of carbon based molecules, while shorter wavelengths contain enough energy to destroy these molecules through successive ionization or fragmentation.

Empirical evidence supports our conjecture of the dissipative structuring of the fundamental molecules of life under these wavelengths; first, the maximum in the strong absorption spectrum of many of these molecules coincides with the predicted window in the Archean atmosphere (Fig. \ref{fig:Pigments}) \cite{MichaelianSimeonov2015}. Secondly, many of the fundamental molecules of life are endowed with {\em peaked conical intersections} (section \ref{sec:ConicalInt}) giving them a broad absorption band and high quantum yield for rapid (picosecond) dissipation of the photon-induced electronic excitation energy into vibrational energy of molecular atomic coordinates, and finally into the surrounding water solvent \cite{SchuurmanStolow2018}. Perhaps the most convincing evidence of all, however, is that many photochemical routes to the synthesis of nucleic acids \cite{FerrisOrgel1966}, amino acids \cite{SaganKhare1971}, fatty acids \cite{MichaelianRodriguez2019}, sugars \cite{Ruiz-BermejoEtAl2013}, and other pigments \cite{MichaelianSimeonov2015} from common precursor molecules have been identified at these wavelengths and the rate of photon dissipation within the Archean window generally increases after each incremental step on route to synthesis, a behavior strongly suggestive of dissipative structuring \cite{Michaelian2017,MichaelianRodriguez2019} (section \ref{sec:Thermodynamics}).

In contradistinction to the generally held view that UV wavelengths were detrimental to early life and thereby induced extreme selection pressure for mechanisms or behavioral traits that protected life from, or made life tolerable under, these photons \cite{Sagan1973,CleavesMiller1998,Cockell2000,MulkidjanianEtAl2003}, we argue that these wavelengths were not only fundamental to the photochemical synthesis of life's first molecules (as suggested with increasing sophistication by Oparin \cite{Oparin1924}, Haldane \cite{Haldane1929}, Urey \cite{Urey1952}, Sagan \cite{Sagan1957} and Mulkidjanian \cite{MulkidjanianEtAl2003} and supported experimentally by Baly \cite{BalyEtAl1927}, Miller \cite{Miller1953}, Oro and Kimball \cite{OroKimball1962}, Ponnamperuma et al. \cite{PonnamperumaEtAl1963,PonnamperumaEtAl1963b,PonnamperumaEtAl1963c}, Ferris and Orgel \cite{FerrisOrgel1966}, and Sagan and Khare \cite{SaganKhare1971} as well as others) but that this UV light was fundamental to the origin and early existence of the entire thermodynamic dissipative process known as ``life'', comprising of synthesis, proliferation, and evolution (section \ref{sec:Thermodynamics}) leading to concomitant  increases in biosphere photon dissipation over time.

Thus, rather than requiring refuge or protection from this UV light, it is argued here that UV-induced molecular transformations providing innovations which allowed early molecular life to maximize UV exposure; e.g. buoyancy at the ocean surface, larger molecular antennas for capturing this light, increases in the width of the wavelength absorption band, and peaked conical intersections providing extraordinarily low antenna dead-times, all would have been selected for \cite{MejiaMichaelian2018} through non-equilibrium thermodynamic principles to be presented in section \ref{sec:Thermodynamics}. In fact, UVC photons provide orders of magnitude more free energy and many more pathways for carbon covalent bond transformation of precursor molecules than do thermal reactions (section \ref{sec:Photochemistry}). Furthermore, there exists empirical evidence suggesting selection for traits optimizing UV exposure for particular amino acids complexed with their RNA or DNA cognate codons or anticodons, particularly for those amino acids displaying the strongest stereochemical affinity to these \cite{MejiaMichaelian2018}. This has led us to suggest \cite{MejiaMichaelian2018,MejiaMichaelian2020} that UVC photon dissipation was the basis of the initial specificity in the amino acid - nucleic acid association during an early stereochemical era \cite{YarusEtAl2009}. The origin of the translation is considered as one of the enduring mysteries of molecular biology \cite{VitasDobovisek2018}.

The perspective taken here, therefore, is that the origin of life was not a scenario of organic material organization driven by natural selection leading to ``better adapted'' organisms, or to greater chemically stability (e.g. UV resistant organisms), but rather a scenario of the dissipative structuring of material under the imposed UV solar photon potential leading to a structuring of material in space and time (biosynthetic pathways) in such a manner so as to provide a more efficient route to the dissipation of the externally imposed photon potential. Similar dissipative synthesis of an ever larger array of photochemical catalysts and cofactors, would allow ever more complex biosynthetic pathways to emerge through this {\em thermodynamic selection} to promote the synthesis of novel pigments for dissipating not only the fundamental UVC and other UV regions, but the entire short wavelength region of the solar photon spectrum \cite{MichaelianSimeonov2015,MichaelianSimeonov2017}, eventually reaching the red-edge ($\sim$ 700 nm), which is the approximate limit of biological photon dissipation on Earth today. 

The thermodynamic dissipation theory for the origin of life \cite{Michaelian2011,Michaelian2016} as summarized above, employed as the framework here, assigns an explicit thermodynamic function to life; {\em life is the dissipative structuring, proliferation, and evolution of molecular pigments and their complexes from common precursor carbon based molecules under the imposed short wavelength solar photon potential for performing the explicit thermodynamic function of dissipating this light into long wavelength infrared light (heat)}. The external photon potential supplied continuously by the environment, and its dissipation into heat by the spontaneously-assembled dissipative structures, are both integral components necessary for understanding life.

In section \ref{sec:Thermodynamics} I review the non-equilibrium thermodynamics needed to understand the abiogenisis of dissipative structuring, proliferation, and evolution of life and how this led to increases in solar photon dissipation, roughly corresponding to increases in global entropy production. In section \ref{sec:Photochemistry} I describe in detail the photochemistry and the particular transformation mechanisms which are employed in dissipative structuring of molecules. In section \ref{sec:Adenine} I give the explicit example of the photochemical dissipative structuring of adenine and show that if intermediate product molecules on route to the dissipative synthesis of adenine become catalysts for the chemical or photochemical reactions, then this leads to their proliferation, as well as to that of their final product molecule adenine. Selection for greater efficacy in photon dissipation on route to the product molecule is probabilistic and determined by fluctuations near instabilities and the widths of phase-space paths to conical intersections (section \ref{sec:ConicalInt}) leading to the intermediate molecules. This evolution in efficacy is guided by the {\em universal evolutionary criterion} of Glansdorff and Prigogine (section \ref{sec:Thermodynamics}) and this, along with auto- and cross-catalytic proliferation, provides a mechanism for evolution which may be termed {\em dissipative selection}, or more generally, {\em thermodynamic selection}. Dissipative structuring, dissipative proliferation, and dissipative selection, are the necessary and sufficient elements for a non-equilibrium thermodynamic framework from within which the origin and evolution of life can be explained in purely physical and chemical terms \cite{Michaelian2009,Michaelian2011,Michaelian2016}.

\section{Thermodynamic Foundations}
\label{sec:Thermodynamics}
\subsection{Classical Irreversible Thermodynamics}
 
Irreversible processes can be identified by the redistribution (flow) of conserved quantities (e.g. energy, momentum, angular momentum, charge, etc.) over an increasing number of microscopic degrees of freedom, often involving, at the macroscopic scale, spatial coordinate degrees of freedom. Corresponding to a given flow there exists a conjugate generalized thermodynamic force. For example, to the macroscopic flows of heat, matter, and charge, over coordinate space, there corresponds the conjugate forces of minus the gradient of the inverse of temperature, of the mass density (concentration gradient), and of the electric charge density (the electrostatic potential) respectively. 

Flows of the conserved quantities can occur not only over macroscopic coordinate degrees of freedom, but also over molecular degrees of freedom \cite{Prigogine1967}, such as over electronic or vibrational coordinates, spin coordinates, and reaction coordinates (ionizations, deprotonations, charge transfer, disassociations, isomerizations, tautomerizations, rotations around covalent bonds, sigmatrophic shifts, etc.), obeying statistical quantum mechanical rules. The corresponding conjugate forces to these flows of the conserved quantities involved in the molecular structuring processes of life are electromagnetic in nature, for example, the chemical and photochemical potentials. Since, for covalent, strongly bonded organic material, access to these molecular degrees of freedom usually requires the deposition of a large amount of the conserved quantity (e.g. energy) locally (e.g. on a particular region of a molecule), such flow, and any resulting dissipative structuring occurring at the origin of life (before the evolution of complex biosynthetic pathways) was necessarily associated with ultraviolet photon absorption.

The existence of any macroscopic flow, or equivalently any unbalanced generalized thermodynamic force, necessarily implies that the system is not in thermodynamic equilibrium. Under the assumption of {\em local thermodynamic equilibrium} (e.g., local Maxwell-Bolzmann distribution of particle velocities or excited vibrational states), Onsager, Prigogine, Glansdorff , Nicolis, and others developed the mathematical framework to treat out-of-equilibrium  phenomena known as “Classical Irreversible Thermodynamics” (CIT) \cite{Prigogine1967}. In this framework, the total internal (to the system) entropy production $P$ per unit volume $V$, $\sigma\equiv P/V = (d_iS/dt)/V$, of all irreversible processes occurring within the volume due to $n$ generalized thermodynamic forces $k=1,n$ is simply the sum of all forces $X_k =A_k/T$ (where $A_k$ are the affinities and $T$ is the temperature) multiplied by their conjugate flows $J_k$. This sum, by the local formulation of the second law of thermodynamics \cite{Prigogine1967}, in any macroscopic volume, is positive definite for irreversible processes and equal to zero for reversible processes (those occurring in thermodynamic equilibrium),
\begin{equation}
\sigma \equiv {P \over V}={d_iS/dt \over V} = \sum_{k=1,n} X_k J_k = \sum_{k=1,n}  {A_k\over T} J_k \ge 0.
\label{eq:entprod}
\end{equation}
The assumption of local equilibrium for the case studied here, of molecular photochemical dissipative structuring of the fundamental molecules, is valid if the absorbed energy of the incident photon becomes distributed with Boltzmann statistics over the nuclear vibrational degrees of freedom implicated in molecular transformations (hot ground or excited state reactions -- the Franck-Condon principle implies that the electronically excited molecule will most likely also be in a vibrationally excited state). Organic materials in the liquid or condensed phase are generally ``soft materials'' in the sense that their vibrational degrees of freedom in the electronic excited state couple significantly to their vibrational degrees of freedom in the electronic ground state (unlike in the case of inorganic material). This nonadiabatic coupling is mediated by conical intersections (section \ref{sec:ConicalInt}) which allow for ultra-fast equilibration of the photon energy over the vibrational degrees of freedom of the electronic ground state, often on femtosecond time scales \cite{SchuurmanStolow2018}, leaving small molecules for a short time (depending on the nature of their surroundings) with an effective vibrational temperature of 2000-4000 K. This time for vibrational equilibration is generally less than the time required for a typical chemical transformation and therefore the irreversible process of molecular dissipative photochemical structuring can be justifiably treated under the CIT framework in the non-linear regime. Indeed, Prigogine has shown that irrespective of the imposed affinities, chemical reactions in the electronic ground state can be treated successfully under CIT theory as long as the reactants retain a Maxwell-Boltzmann distribution of their velocities, which is the case for all but very exothermic (explosive) reactions \cite{Prigogine1967}.

The time change of the total entropy production $P$ for any out-of-equilibrium system can be split into two parts, one depending on the time change of the forces $X$, and the other on the time change of the flows $J$,
\begin{equation}
{dP \over dt}= {d_XP \over dt} + {d_JP \over dt},
\label{eq:EntProdChg}
\end{equation}
where, for a continuous system within a volume $V$,
\begin{equation}
{d_XP \over dt}  = \int{\sum_{k=1,n} J_k {dX_k \over dt} dV},   \hspace{1cm}  {d_JP \over dt} = \int{\sum_{k=1,n} X_k {dJ_k \over dt} dV}, \label{eq:dxp}
\end{equation}

For the case of constant external constraints over the system, for example when affinities ${\bf \mathcal{A}}=\{A_k$; $k=1,c\}$ are externally imposed and held constant, CIT theory indicates that the system will evolve towards a stationary state in which its thermodynamic state variables (for example, the internal energy $E$, entropy $S$, and entropy production $P=d_iS/dt$) become time invariant. For flows linearly related to their forces, it is easy to show that there is only one stationary state and that the entropy production in this stationary state takes on its minimal value with respect to variation of the free affinities ${\bf A}= \{A_k$; $k=c+1,n\}$ in the system \cite{Prigogine1967}. This principle of minimum dissipation for {\em linear} systems was first proposed by Lord Rayleigh in 1873 \cite{Rayleigh1873}.

However, if the flows are {\em non-linearly} related to the forces, then, depending on the number of degrees of freedom and how non-linear the system is, at a certain value of a variable of the system (e.g. overall affinity), labeled a {\em critical point}, the system becomes unstable and new, possibly many, different stationary states become available, each with a possibly different value of internal entropy production $P=d_iS/dt$. In this case, stationary states are only locally stable in some variables of the system, or even unstable in all variables. The non-linear dynamics is such that different stationary states, corresponding to different sets of flows ${\bf J}_\alpha$, ${\bf J}_\beta$, etc. conjugate to their sets of free affinities ${\bf A}_\alpha$, ${\bf A}_\beta$, etc. become available through current fluctuations $\delta{\bf J}_\alpha$ , $\delta {\bf J}_\beta$, etc., at the critical instability point (or bifurcation point) along a particular variable of the system, because, unlike in the equilibrium or in the linear non-equilibrium regimes, in the non-linear non-equilibrium regime these microscopic fluctuations $\delta{\bf J}_\alpha$ on their original flows ${\bf J}_\alpha$ can be amplified through feedback (e.g. autocatalysis) into new macroscopic flows ${\bf J}_\beta$ \cite{Prigogine1967}.

Since for such a non-linear system, under an externally imposed thermodynamic force, multiple stationary states are available, an interesting question arises concerning the stability of the system and how the system may evolve over time between different stationary states. Because the system harbors critical points at which microscopic fluctuations can be amplified into macroscopic flows leading the system towards a new stationary state, it cannot be expected that there exists a potential for the system whose optimization could predict its evolution. What could be hoped for, however, is an understanding of the stochastic and deterministic components governing relative probabilities for the different evolutionary trajectories over the stationary states.

Prigogine and co-workers have shown that, although in general no optimizable total differential (thermodynamic potential) exists for these non-linear systems, there does, however, exist a non-total differential, the time variation of the entropy production with respect to the time variation of the free forces $d_XP/dt$ (see equation \ref{eq:EntProdChg}), which always has a definite sign,
\begin{equation}
 {d_XP \over dt} \le 0.
 \label{eq:GPC}
 \end{equation}
This is the most general result so far obtained from CIT theory, valid in the whole domain of its applicability, independent of the nature of the relation between the flows and forces. It is known as the {\em universal evolution criterion}, or sometimes called the {\em Glansdorff-Prigogine criterion}. This criterion indicates that the free forces always arrange themselves within a system such that this arrangement contributes to a decrease in the entropy production. However, in general, there is no such constraint on the total entropy production of the system because this also includes a component due to the corresponding rearrangement of the flows (see Eqn. (\ref{eq:EntProdChg})) which has no definite sign. The total entropy production may either increase or decrease during evolution in the nonlinear regime, depending on the relative signs and sizes of the two terms in equation (\ref{eq:EntProdChg}). In the restricted regime of {\em linear} phenomenological relations (a linear relation between the flows and forces), it is easy to show \cite{Prigogine1967} that  $d_JP/dt = d_XP/dt$ and thus the universal evolutionary criterion, Eq. (\ref{eq:GPC}), correctly predicts the theorem of minimum entropy production alluded to above, $dP/dt \le 0$. The stability in the Lyapunov sense of the one and only stationary state in this linear regime is then guaranteed by the fact that the entropy production is a Lyapunov function (i.e. $P>0$ and $dP/dt \le 0)$.

In the {\em nonlinear} regime, however, bifurcations can be reached leading to multiple stationary states which, for the case studied here of the photochemical dissipative structuring of the fundamental molecules of life, corresponds to different concentration profiles of the distinct molecular configurations involved in the synthesis, each profile having a potentially different rate of dissipation of the applied photon potential. It is therefore pertinent to inquire if there indeed exists certain stochastic and deterministic components (as alluded to above), giving probabilities for the different paths of the evolution of the system over stationary states, which may be related to dissipation or entropy production. 

\subsection{Stability, Evolution, and Entropy Production in the Nonlinear Regime}
\label{sec:StabTheo}

As will be seen in section \ref{sec:Adenine}, our model system demonstrates evolution over different concentration profiles of the intermediate product molecules on route to adenine such that the photon dissipation, or the entropy production, of the entire system steadily increases, with large discontinuous increases at perturbations. The purpose of this subsection is to understand this evolutionary dynamics from within CIT theory.

Even though $d_XP$ or $d_X\sigma$ (per unit volume) is not a total differential, the Glansdorff-Prigogine criterion, $d_X\sigma /dt \le 0$, can still be used to determine the nature and local stability of each stationary state, not only in the linear regime as shown above, but also in the non-linear regime. To illustrate this, consider a set of chemical reactions with rates $v_k$ and affinities $A_k$, and allow for fluctuations of the affinities $\delta A_k$ and rates $\delta v_k$ around particular stationary state values  $v^0_k$ and $A^0_k$,
\begin{eqnarray}
v_k &=v^0_k+\delta v_k, \\
A_k &=A^0_k + \delta A_k,
\end{eqnarray}
where $k$ specifies the particular reaction. We can define what is called the {\em excess entropy production} per unit volume $\delta_X\sigma$ due to random fluctuation of the free forces $X_k$ (in this case the affinities over the temperature) about their stationary state values as (see Eq. (\ref{eq:dxp}));
\begin{equation}
\delta_X\sigma = {1\over T}\Sigma_k v_k \delta A_k.
\label{eq:eep}
\end{equation}
At the stationary state, we must have,
\begin{equation}
\delta_X\sigma = {1\over T}\Sigma_k v^0_k \delta A_k = 0,
\end{equation}
which implies that, for independent affinities, all the $v^0_k=0$. Therefore, we can write the excess entropy production, Eq. (\ref{eq:eep}), as (see also ref. \cite{PrigogineNicolis1971} p. 119)
\begin{equation}
\delta_X\sigma = {1\over T}\Sigma_k \delta v_k \delta A_k.
\end{equation}
If the stationary state is stable, then excess entropy production due to a random fluctuation must be positive definite, if it were not, then the natural evolution of the system defined by the general evolution criterion $d_X\sigma /dt \le 0$ (Eq. (\ref{eq:GPC}) will not bring the system back to the stationary state but rather will amplify the fluctuation. Stationary state stability therefore requires the excess entropy production be positive definite, 
\begin{equation}
\delta_X\sigma = {1\over T}\Sigma_k \delta v_k \delta A_k \ge 0.
\end{equation}
If processes become physically possible which give a negative contribution to the excess entropy production, then the system may become unstable and this can occur for autocatalytic and cross-catalytic reaction systems (demonstrated in reference \cite{GlansdorffPrigogine1971}, p. 81). 

The above, however, says nothing about the total entropy production which may either increase, decrease, or stay the same since there is a second contribution to the entropy production related to the changes in the flows which has no definite sign (Eq. (\ref{eq:EntProdChg})). In general, to determine the direction of evolution, a complete stability analysis must be performed around the stationary state, linear stability analysis does not suffice (see Appendix). For chemical reactions, or coupled chemical and photochemical reactions, exhibiting positive feedback, i.e. auto-catalysis or cross-catalysis (e.g. the chemical transformation of precursor or intermediate molecules providing a new route for photon dissipation), a concentration fluctuation could be amplified, taking the system to a state of greater dissipative efficacy and thereby moving the system even father from equilibrium. The impressed photochemical and chemical affinities are thus dissipated more rapidly in this new stationary state  The more probable stationary states present larger ``catchment basins'' in the generalized phase space of the system. In our particular case of the dissipative structuring of molecules under UV light, the size of the catchment basin is related to the sum of the widths of paths leading to the conical intersections which connect the potential energy surface of the electronically excited molecule to that of the ground state, which may be either the atomic coordinate transformed ground state of the molecule after the photochemical reaction or the ground state of the same molecule after internal conversion. Auto-, or cross-, catalysis increases the widths of these paths. Only statistical probabilities for evolution can be determined once the allowed reactions are delineated and weighted by the size of their catchment basin. Quantitatively, these will be specified by the empirically determined quantum efficiency for the particular photochemical reaction or internal conversion.

Since some recent works have considered a statistical mechanics approach employing fluctuation theorems with linear stability theory to describe evolution in living systems, in the Appendix I outline the relationship between this and the more complete Classical Irreversible Thermodynamic theory presented above. It was, in fact, shown in the early 1970s by Glansdorff, Prigogine, and Nicolis \cite{GlansdorffPrigogine1971, PrigogineNicolis1971} that, contrary to what has been stated in the recent literature, statistical fluctuation theorems employing linear stability theory are not sufficient to describe the evolution of non-linear dissipative systems.

To summarize, for isolated systems governed by equilibrium thermodynamics, the evolution of the system is determined only by the thermodynamics, the final state being independent of the initial conditions or kinetic factors. However, in out-of-equilibrium thermodynamics described above, kinetics plays a very important role in the evolution of the system. Kinetic factors, like auto- or cross-catalytic activity, can become more important than thermodynamic improbability \cite{GlansdorffPrigogine1971}. In the non-linear regime, there are multiple, locally stable, stationary states available and instability can arise after fluctuations which give a negative contribution to the excess entropy production. The system will evolve by amplifying the fluctuation according to the Glansdorff-Prigogine criterion, leading it from one stationary state to another, depending on the size of the fluctuation and the kinetic factors involved. For such auto- or cross-catalytic systems this results in an increase in dissipation or entropy production, taking the system father from equilibrium. A specific example of the dissipative structuring, proliferation and evolution leading to the UVC chromophore adenine, one of the fundamental molecules of life, will be given after first describing the photochemistry available to carbon based organic molecules undergoing photochemical dissipative structuring.

\section{Photochemistry}
\label{sec:Photochemistry}
\unskip
\subsection{Quantum Selection Rules}
 
Absorption by an organic molecule of a visible or UV photon of the required energy $E = h \nu$ leads to an electronic spin singlet or triplet excited state. The width of the allowed transition $\Delta E$ is determined by the natural line width dependent on the natural lifetime $\Delta t$ of the excited state, as given by the Heisenberg uncertainty relation $\Delta E \Delta t \ge \hbar$. In condensed material or at high pressure, further broadening occurs due to dexcitation through collisions with neighboring molecules, reducing further the lifetime. There is also a broadening due to the Doppler effect which increases with temperature. Contributing most to the broadening for the organic molecules, however, is the coupling of electronic degrees to the vibrational degrees of freedom of the molecule (vibronic or non-adiabatic coupling). 

Excitation to the triplet state is a spin forbidden transition but can occur due to spin-orbit coupling or interaction with a paramagnetic solvent molecule, for example oxygen in its spin-triplet ground state. Under laboratory conditions and for organic molecules, however, the singlet state is favored over the triplet state by $\sim 1000:1$. Moreover, since electronic excitations are affected by the electronic dipole transition which is first order in the coordinates $x$ (i.e. the dipole moment is an odd function $f(x)\ne f(-x)$), and since an additional quantum selection rule is that transitions must be symmetric, the symmetries of the wavefunctions of the molecule in the initial and final state must be different (e.g. even $\rightarrow$ odd) giving rise to the electronic angular momentum selection rule $\Delta l = \pm 1$. For example, a $1S \rightarrow 2S$ transition is forbidden while a $1S \rightarrow 2P$ transition is allowed.  

\subsection{Conical Intersections}
\label{sec:ConicalInt}
The Born-Oppenheimer approximation in molecular structure calculations assumes independence of the electronic and nuclear motions. However, such an approximation is obviously not valid for chemical reactions where nuclear reconfiguration is coupled to electronic redistribution and particularly not valid for photochemical reactions where the potential energy surface of an electronic excited state is reached. 

\begin{figure}[htbp]
\begin{center}
\includegraphics[width=10 cm]{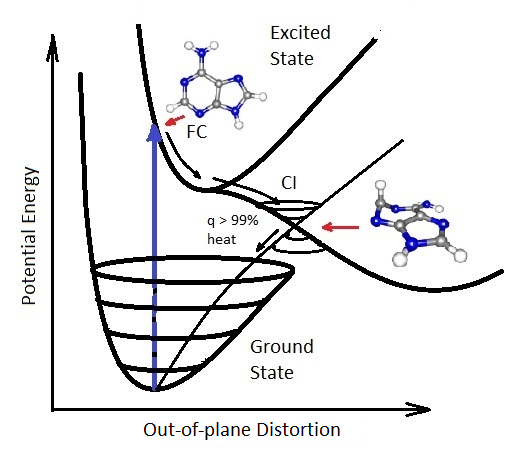}
\end{center}
\caption{Conical Intersection (CI) for adenine showing the degeneracy of the electronic excited state with the electronic ground state after a UVC photon absorption event (blue arrow) which induces a nuclear coordinate deformation from its original structure in the Franck-Condon (FC) region known as {\em pyrimidilization}. Conical intersections provide rapid (sub-picosecond) dissipation of the original electronic excitation energy into heat. The quantum efficiency (q) for this dissipative route is very large, making adenine photochemically stable but more importantly very efficient at photon dissipation. Another common form of coordinate transformation associated with conical intersections are proton transfers within the molecule or with the solvent.  Based on data  from Andrew Orr-Ewing \cite{Orr-Ewing}  Roberts et al. \cite{RobertsEtAl2014}, Kleinermanns et al. \cite{KleinermannsEtAl2013} and Barbatti et al. \cite{BarbattiEtAl2010}}
\label{fig:ConicalInt}
\end{figure}

Conical intersections are multi-dimensional seams in nuclear coordinate space where the adiabatic potential energy surface of the electronic excited state becomes degenerate with the potential energy surface of the electronic ground state of the same spin multiplicity, resulting from a normally barrier-less out of plane distortion of the nuclear coordinates (e.g. bond length stretching or rotation about a bond) \cite{SchuurmanStolow2018}. A common distortion of the nuclear coordinates for the excited state of the nucleobases is ring puckering as shown for adenine in figure \ref{fig:ConicalInt}. This multi-dimensional seam, defining the energy degeneracy, allows for rapid (sub-picosecond) radiationless dexcitation of the photon-induced electronic excited state, distributing the electronic energy over nuclear vibrational modes of the molecule as either a prelude to a photochemical transformation, or to the harmless dissipation of the energy into the solvent and leaving the molecule in its original ground state ready for another photon absorption event. 

The conical intersection seams thus define the photoisomerization or photoreaction products that can be reached after an electronic excitation event. Since conical intersections are located energetically down-hill from the Franck-Condon region, the direction and velocities of approach of the nuclear coordinates to a conical intersection are important in defining the outcome \cite{SchuurmanStolow2018}. (This is what I referred to in the Introduction as the ``catchment basin'' in a generalized phase space, and the reason that not all fluctuations are identical, and therefore both stochastic and deterministic components to evolution exist, see Appendix.) For example, it is known that for the molecule retinal in rhodopsin the photoexcited molecule reaches the conical intersection extremely fast (75 femtoseconds) implying that the conical intersection must be peaked (inverted cone-like on the excited state potential energy surface) and, overwhelmingly, only one reaction product is reached, which for the case of retinal, as well as for the fundamental molecules of life, is the original ground state configuration \cite{PolliEtAl2010}. A more extended seam with different minima can lead to different reaction products \cite{Serrano-PerezEtAl2013} such as those intermediates on route to the photochemical synthesis of adenine which will be described in section \ref{sec:Adenine}. The final product in the photochemical synthesis of the fundamental molecules of life must, however, always have a {\em peaked} conical intersection so that they become the {\em final} and {\em photo-stable} product of dissipative structuring in the relevant region of the solar spectrum.

It has been a recurrent theme in the literature that the rapid (sub-picosecond) dexcitation of the excited nucleobases due to their conical intersections had evolutionary utility in providing stability under the high flux of UV photons that penetrated the Archean atmosphere \cite{Sagan1973,MulkidjanianEtAl2003} since a peaked conical intersection reduces the lifetime of the excited state to the point where further chemical reactions are improbable. However, photo-stability is never complete, and photochemical reactions under UVC light do still occur for the fundamental molecules of life, particularly after excitation to the long lived triplet state, for example in the formation of cyclobutane pyrimidine dimers in RNA and DNA. An apparently more optimal and simpler solution for avoiding radiation damage with its concomitant degradation in biological function, therefore, would have been the synthesis of molecules transparent to, or reflective to, the offending UV light. From the thermodynamic perspective of dissipation presented here, however, a large antenna for maximum UVC photon absorption and a peaked conical intersection for its rapid dissipation into heat are, in fact, precisely the design goals of dissipative structuring.

\subsection{Excited State Reaction Mechanisms}
\label{sec:ReacMech}

The photochemistry of molecules in electronic excited states is much richer than the thermal chemistry of their ground state, because; 1) the absorbed photon energy allows very endothermic reactions to occur, 2) anti-bonding orbitals can be reached, allowing reactions to occur which are prohibited in the ground state, 3) triplet states can be reached from the electronic excited state, allowing intermediates that cannot be accessed in thermal reactions, 4) electronically exicted molecules are often converted into radicals, making them much more reactive. For example, a molecule in its excited state can be a much stronger oxidizer or reductor having a pK$_a$ value substantially different from that of its ground state (e.g. if the pK$_a$ value becomes more acidic, proton transfer to an acceptor solvent water OH$^-$ ion becomes much more probable). Singlet excited states have a particularly rich chemistry, while triplet states have a more restricted chemistry. This richness in photochemistry is, in itself, yet another strong argument in favor of the complex molecules of life arising out of photon-induced reactions occurring at the surface of the ocean rather than out of thermal reactions occurring at the bottom of the ocean.

Photochemical processes that arise after photon-induced excitation can be classified into dissasociations, rearrangement, additions and substitutions. Each process within these classes constitutes a particular mechanism for molecular transformation which could have been employed in the photochemical dissipative structuring of the fundamental molecules at the origin of life under the UVC photon potential. Indeed, these mechanisms still occur today in many important photochemical processes of life, albeit in the near UV or visible regions of the spectrum and through more complex biosynthetic pathways.

The photochemical transformations listed above generally have a strong dependence on wavelength due to the particular absorption characteristics of the inherent chromophores of the precursor molecules. However, it is not only the absorption coefficient of the chromophore which is important since within a given wavelength region there may be two or more such molecular transformational processes in competition, and therefore the particular conformation of the electronic ground state before excitation may be relevant. This conformation could depend on the temperature, viscosity, polarity, ionic strength and pH of the solvent, all of which are determinant in the yields of the final photoproducts.

Some of the molecular transformations mentioned above do not belong exclusively to the domain of photochemical reactions but can also occur through thermal reactions at high temperature, albeit with lower yield and less variety of product. Therefore, as well known, some of the fundamental molecules of life could have been produced through thermal mechanisms without recourse to the incident light, for example at ocean floor hydrothermal vents. However, as emphasized in the Introduction, the mere efficient synthesis of the fundamental molecules is not sufficient to bootstrap the irreversible dissipative process known as life. The continuous dissipation of an external thermodynamic potential is a necessary condition for the structuring, proliferation, and evolution of life, as it is for any sustained irreversible process.

\section{Example: The Dissipative Structuring of Adenine}
\label{sec:Adenine}

\subsection{The Model}

HCN is a common molecule found throughout the cosmos and its production during the Hadean and Archean on Earth was probably a result of the solar Lyman alpha line (121.6 nm) photo-lysing N$_2$ in the upper atmosphere which then attacks CH or CH$_2$ to form HCN  \cite{TrainerEtAl2012}, or the UV (145 nm) photolysis of CH$_4$ leading to a CH$^*$ radical which attacks N$_2$ \cite{TrainerEtAl2012}. HCN and its hydrolysis product formamide are now recognized as probable precursors of many of the fundamental molecules of life, including nucleic acids, amino acids, fatty acids \cite{Ruiz-BermejoEtAl2013}, and even simple sugars \cite{RitsonSutherland2012,DasEtAl2019}. As early as 1875 E. Pflüger suggested that life may have followed from ``cyanogen compounds'' \cite{Pfluger1875}. The ubiquity of different chemical and photochemical routes from HCN to the fundamental molecules discovered over the last 60 years has led to the suggestion of an``HCN World''  \cite{MinardMatthews2004,Matthews2004} occurring before the postulated``RNA World'' \cite{NeveuEtAl2013}.  

\begin{figure}[htbp]
\centering
\includegraphics[width=17 cm]{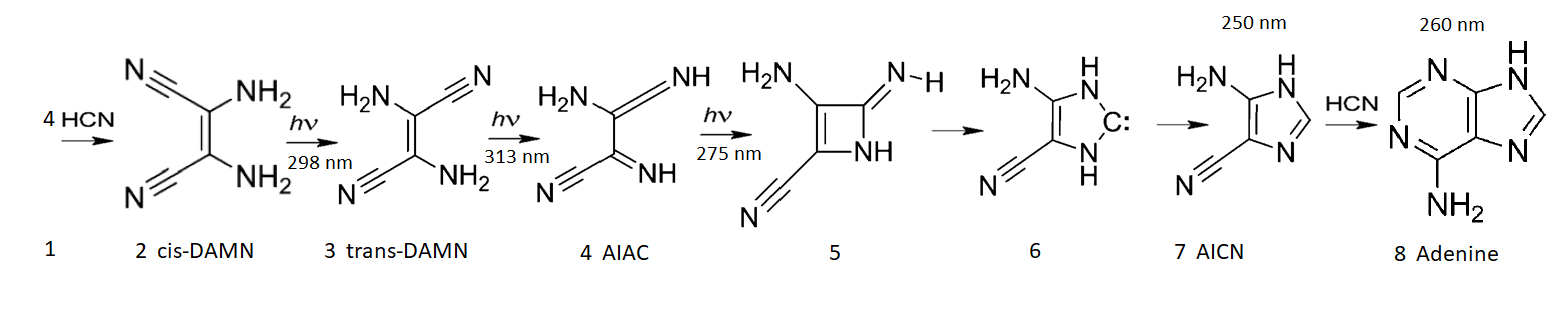} 
\caption{The photochemical synthesis of adenine from 5 molecules of hydrogen cyanide (HCN) in water, as discovered by Ferris and Orgel (1966) \cite{FerrisOrgel1966,BoulangerEtAl2013}. Four molecules of HCN are transformed into the smallest stable oligomer (tetramer) of HCN, known as  cis-2,3-diaminomaleonitrile (cis-DAMN) (2), which, under a constant UV-C photon flux isomerizes into trans-DAMN (3) (diaminofumaronitrile, DAFN) which may be further converted on absorbing two more UV-C photons into an imidazole intermediate, 4-amino-1H-imidazole-5-carbonitrile (AICN) (7).  Hot ground state thermal reactions with another HCN molecule or its hydrolysis product formamide (or ammonium formate) leads to the purine adenine (8). This is a microscopic dissipative structuring process which ends in adenine \cite{Michaelian2017}, a pigment with a large molar extinction coefficient at 260 nm and a peaked conical intersection which promotes the dissipation of photons at the wavelength of maximum intensity of the Archean solar UVC spectrum (figure \ref{fig:Pigments}). Adapted from Ferris and Orgel (1966)\cite{FerrisOrgel1966}.}
\label{fig:AdenineSyn}
\end{figure} 

The synthesis of adenine from HCN has been studied by numerous groups since the first experimental observations of the chemical reaction at high temperatures by Or\'o in 1960 \cite{Oro1960} and photochemically at moderate temperatures by Ferris and Orgel in 1966 \cite{FerrisOrgel1966, SanchezEtAl1967, SanchezEtAl1968, RoyEtAl2007, BoulangerEtAl2013}. Adenine is a pentamer of HCN and the overall reaction from 5 HCN to adenine is exothermic ($\Delta G = -53.7$ kcal mol$^{-1}$ \cite{RoyEtAl2007}) but presents a number of large kinetic barriers which can be overcome at high temperatures or at low temperatures if UV photons are absorbed. The reactions on route to adenine are in competition with hydrolysis and UV lysis, and these relative rates are dependent on concentrations, temperature, pH, metal ion- and product- catalysis, and the wavelength dependent intensity of the incident UV spectrum. The complexities involved in the photochemical reactions leading to adenine have been studied by Sanchez et al. \cite{SanchezEtAl1967, SanchezEtAl1968}. 

An apparent difficulty exists with respect to the synthesis of the purines from HCN in that, for dilute concentrations of HCN ($<0.01$ M), hydrolysis of HCN occurs at a rate greater than its polymerization, e.g. its tetramization (step 1 to 2, figure \ref{fig:AdenineSyn}), the first required step on route to adenine. Hydrolysis is proportional to the HCN concentration whereas tetramization is proportional to the square of the concentration \cite{SanchezEtAl1967}. Stribling and Miller \cite{StriblingMiller1986} estimated that atmospheric production of HCN and subsequent loss to hydrolysis and recycling through thermal vents, would have led to ocean concentrations, at neutral pH, of no greater than about $1.0\times 10^{-12}$ M at 100 $^\circ$C and $1.0\times 10^{-4}$ M at 0$^\circ$C for an ocean of 3 Km average depth. This led Sanchez, Miller, Ferris, Orgel \cite{SanchezEtAl1966,SanchezEtAl1967} to conclude that eutectic concentration of HCN (through freezing of the water solvent) would have been the only viable route to synthesis of the purines, and this is the primary reason why subsequent analyses favored a cold scenario for the origin of life \cite{MillerLazcano1995, BadaLazcano2002, MiyakawaEtAlb2002}, not withstanding the geochemical evidence to the contrary, and even though this severely reduces all reaction rates and inhibits diffusion. 

However, it is now known that the top $\sim 100\ \mu$m of the ocean surface (known as the microlayer) is a unique environment with organic material densities as large as 10$^4$ times that of bulk water below. This is due to lowering of the free energy of fatty acids and other amphipathic molecules at the air/water interface, as well as  Eddy currents and air bubbles from raindrops bringing organic material to the surface \cite{Hardy1982,GrammatikaZimmerman2001}. Furthermore, it has been shown that even though HCN is very soluble in water (and even in non-polar solvents), it tends to  concentrate at a water surface and is observed to align itself through a dipole-dipole interaction in such a manner so as to facilitate polymerization. Molecular dynamic simulations of HCN in water have shown that it can form patches of significantly higher density in both the lateral and vertical dimensions at the surface, due to this strong inter-molecular dipole-dipole interaction \cite{FabianEtAl2014}. 

\begin{figure}[H]
\centering
\includegraphics[width=8 cm]{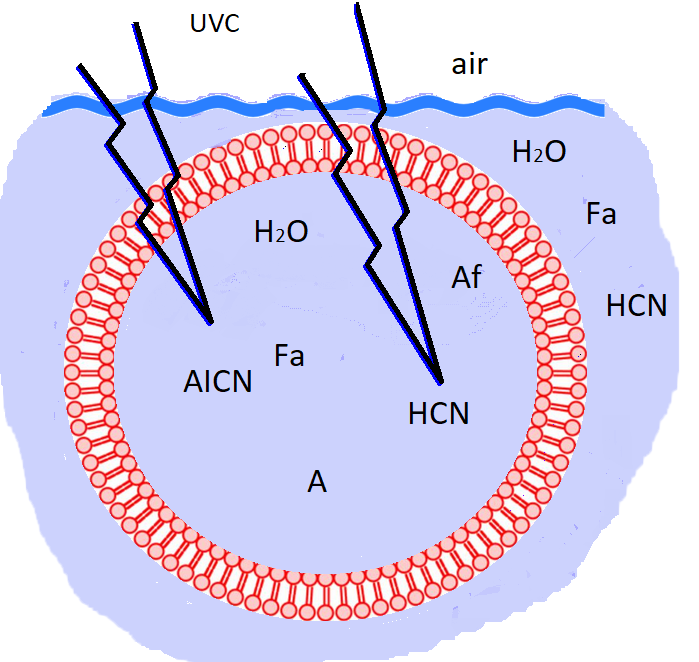} 
\caption{Fatty acid vesicle of $\sim 100$ $\mu$m diameter floating at the ocean surface microlayer, transparent to UVC light and permeable to H$_2$O, HCN and formimidic acid (Fa) but impermeable to the photochemical reaction products (e.g. ammonium formate (Af), AICN, adenine (A)) which are larger in size and have larger dipole moments (Table \ref{tab:Abbrev}).}
\label{fig:Vesicle}
\end{figure}

Rather than invoking eutectic concentration to increase the solute HCN concentration to values sufficient for significant adenine production, here we assume instead the existence at the surface of fatty acid vesicles of $\sim 100\ \mu$m diameter which would allow the incident UVC light, as well as small molecules such as HCN and H$_2$O,  to enter or leave relatively unimpeded by permeating its bi-layer wall (figure \ref{fig:Vesicle}), while trapping within the vesicle the photochemical reaction products due to their larger sizes and larger dipole moments (table \ref{tab:Abbrev}). This would allow these molecules, as well as the heat from their UV photon dissipation, to accumulate within the vesicle. 

The existence of amphipathic hydrocarbon chains, which through Gibb's free energy minimization spontaneously form lipid vesicles at the ocean surface, is a common assumption in origin of life scenarios \cite{Oro1995,WaldeEtAl1994,Deamer2017} and their abiotic production during the Archean could be attributed to heat activated Fischer-Tropsch polymerization of smaller hydrocarbon chains such as ethylene at very high temperatures at deep ocean hydrothermal vents,or to dissipative structuring under UVC photons of HCN and CO$_2$ saturated water at moderate temperatures on the ocean surface \cite{MichaelianRodriguez2019}. In order to maintain vesicle integrity at the hot surface temperatures considered here of $\sim 80 ^\circ$C these fatty acids would necessarily have been long ($\sim 18$ C atoms) and cross linked through UVC light which improves stability at high temperatures and over a wider range of pH values \cite{FanEtAl2014,MichaelianRodriguez2019}. There is, in fact, a predominance of 16 and 18 carbon atom fatty acids in the whole available Precambrian fossil record \cite{HanCalvin1969,VanHoevenEtAl1969}.

In the following subsection I present a simplified out-of-equilibrium kinetic model for our 5HCN $\rightarrow$ adenine photochemical reaction system occurring within a fatty acid vesicle floating within the surface microlayer of a hot ($\sim 80^\circ$C \cite{KarhuEpstein1986, Knauth1992, KnauthLowe2003}) Archean ocean under the UV surface spectrum of figure \ref{fig:Pigments}. I assume that the system is under a diurnal 8 hr flux of radiation followed by an 8 hour period of darkness during which thermal reactions occur but not photochemical reactions. The system is assumed to be perturbed by the existence of sparse patches of relatively high concentration ($0.1$ M) of HCN and formimidic acid (Fa) (a photon-induced tautomer of its hydrolysis product formamide (F)) into which our vesicle is assumed to drift for a short period (120 seconds) only once during 30 Archean days.

The kinetic equations for the above model of chemical and photochemical reactions are resolved numerically, and the stationary state solutions obtained. For such a perturbed non-linear reaction-diffusion system it will be shown in section \ref{sec:Results} that various stationary state solutions exist with the highest concentration of adenine at the center of the vesicle, which could facilitate a subsequent UVC polymerization of nucleobases into oligos, admitting the possibility of UVC-assisted synthesis of ribose from similar precursor molecules \cite{DasEtAl2019} and a temperature \cite{Schoffstall1976} or formamide-catalyzed \cite{CostanzoEtAl2007} phosphorylation (these latter reactions will not be considered here). Stationary state coupling of reactions to diffusion, leading to particular regions of high concentration of the products, was shown to occur for purely thermal reactions with different activator and inhibitor diffusion rates by Turing \cite{Turing1952} and studied more generally as dissipative structures under the framework of CIT theory by Glansdorff and Prigogine  \cite{GlansdorffPrigogine1971}.

\subsection{The Kinetic Equations}

Nomenclature, chemical formula, and abbreviations used throughout the text, for the concentrations of the participating chemical species of the photochemical reactions leading to adenine shown in figure \ref{fig:AdenineSyn}, along with their photon extinction coefficients and size and dipole moments, related to vesicle permeability, are given in table \ref{tab:Abbrev}.

\begin{table}[H]
\caption{Nomenclature, chemical formula, abbreviation in the text and in kinetic equations, position in figure \ref{fig:AdenineSyn}, wavelength of maximum absorption $\lambda_{max}$ (within the spectrum of figure \ref{fig:Pigments}), molar extinction coefficient at that wavelength $\epsilon_{max}$, electric dipole moment $\mu$, and the topological polar surface area (TPSA), of the molecules involved in the photochemical synthesis of adenine. Values marked with ``*'' are estimates obtained by comparing to similar molecules since no data have been found in the literature.}
\centering
\resizebox{17cm}{!}{
\begin{tabular}{|l|l|l|l|l|l|l|l|l|}
\hline
Name & chemical &abbrev. &abbrev. in&Fig. \ref{fig:AdenineSyn}& $\lambda_{max}$& $\epsilon_{max} $ & $\mu$ & TPSA\\
 & formula &in text & kinetics       & & nm & M$^{-1}$ cm$^{-1}$  &[D] &[\AA$^2$]\\
\hline
hydrogen cyanide & HCN  & HCN & H & 1 & & & 2.98 & 23.8\\
formamide &  H$_2$N-CHO & formamide  & F &  & 220 & 60 \cite{PetersenEtAl2008, BaschEtAl1968} & 4.27 \cite{LeljAdamo1995} & 43.1 \\
formimidic acid  & H(OH)C=NH & formimidic acid (trans)  & Fa  & & 220 &60 & 1.14 \cite{LeljAdamo1995} & 43.1 *\\
ammonium formate & NH$_4$HCO$_2$  & ammonium formate & Af & &  & & +/-, 2.0 *& 41.1 \\
diaminomaleonitrile & C$_4$H$_4$N$_4$ &  cis-DAMN (DAMN) & C & 2  &298 &14000 \cite{KochRodehorst1974} & 6.80 \cite{GuptaTandon2012} & 99.6\\
diaminofumaronitrile & C$_4$H$_4$N$_4$  & trans-DAMN (DAFN) & T & 3 &313 &8500 \cite{KochRodehorst1974} &  1.49 \cite{GuptaTandon2012} & 99.6 \\
2-amino-3-iminoacrylimidoyl cyanide & C$_4$H$_4$N$_4$ & AIAC & J & 4  & 275& 9000 \cite{BoulangerEtAl2013,SanchezEtAl1967}& 1.49 & 99.6 *\\
4-aminoimidazole-5-carbonitrile & C$_4$H$_4$N$_4$ & AICN & I & 7 &250 &10700 \cite{KochRodehorst1974} & 3.67 & 78.5 \\
4-aminoimidazole-5-carboxamide& C$_4$H$_6$N$_4$O & AICA & L &  &266 \cite{FerrisEtAl1978} &10700 *  & 3.67 * & 97.8 \\
5-(N'-formamidinyl)-1H-imidazole-4-carbonitrileamidine & C$_5$H$_5$N$_5$ & amidine &  Am &  & 250 &10700 \cite{GlaserEtAl2007} & 6.83 *  & 80.5 *\\
adenine & C$_5$H$_5$N$_5$ & adenine &  A & 8 & 260 &15040 \cite{CavaluzziBorer2004} & 6.83 \cite{FranzGianturco2014} & 80.5\\
hypoxanthine & C$_5$H$_4$N$_4$O & hypoxanthine & Hy &  &250 &12500 \cite{StimsonReuter1943}& 3.16 & 70.1 \\
\hline
\end{tabular}}
\label{tab:Abbrev}
\end{table}

Under non-coherent light sources, photochemical reactions can be treated using elementary kinetics equations of the balance type in the product and reactant concentrations. From a detailed analysis of the experiments and calculations performed in the literature, the chemical and photochemical reactions listed in table \ref{tab:Reactions} will occur in the photochemical dissipative structuring of adenine from HCN and are described in detail below.

\begin{table}[H]
\caption{Reactions involved in the photochemical synthesis of adenine (see figure \ref{fig:AdenineSyn}). Temperature $T$ is in K and all kinetic parameters were obtained at pH 7.0, see description and derivations after table.}
\centering
\resizebox{17cm}{!}{
\begin{tabular}{|l|l|l|}
\hline
\# & reaction & reaction constants \\
\hline
   &      &   \\
1 & H $\stackrel{k_1}{\rightharpoonup}$ F & $k_1=\exp(-14039.0/T + 24.732)$; s$^{-1}$; hydrolysis of HCN \cite{SanchezEtAl1967,MiyakawaEtAla2002,KuaThrush2016}\\
2 & $\gamma_{220} +$ F $\rightarrow$ Fa &  $q_2=0.05$ \cite{PetersenEtAl2008,BaschEtAl1968} \cite{MaierEndres2000, DuvernayEtAl2005, BarksEtAl2010}\\
3 & $\gamma_{220} +$ Fa $\rightarrow$ H + H$_2$O & $q_3=0.03$  \cite{BarksEtAl2010,DuvernayEtAl2005,GingellEtAl1997}\\
4 & F $\stackrel{k_4}{\rightharpoonup}$ Af  & $k_4=\exp(-13587.0/T + 23.735)$; s$^{-1}$; hydrolysis of formamide \cite{KuaThrush2016, BarksEtAl2010} \\
5 & 4H $\stackrel{k_5}{\rightharpoonup}$ C  & $k_5 =1/(\exp (-\Delta E/RT)+1) \cdot\exp(-10822.37/T + 19.049)$; M$^{-1}$ s$^{-1}$; $\Delta E =0.61$ kcal mol$^{-1}$ \cite{SanchezEtAl1967} \\
6 & 4H $\stackrel{k_6}{\rightharpoonup}$ T & $k_6=1/(\exp (+\Delta E/RT)+1) \cdot \exp(-10822.37/T + 19.049)$;  M$^{-1}$ s$^{-1}$; tetramization\cite{SanchezEtAl1967}\\
7 & 4H + T  $\stackrel{k_7}{\rightharpoonup}$ C+T & $k_7 =(1.0/(1.0\cdot0.01))\exp(-(10822.37-728.45)/T + 19.049)$; M$^{-2}$ s$^{-1}$ \cite{SanchezEtAl1967}\\
8 & 4H + T  $\stackrel{k_8}{\rightharpoonup}$ 2T & $k_8 = k_7$; M$^{-2}$ s$^{-1}$ \cite{SanchezEtAl1967} \\
9a & $\gamma_{298}$ + C   $\rightarrow$ T & $q_9=0.045$ \cite{KochRodehorst1974}\\
9b & $\gamma_{313}$ + T $\rightarrow$ C & $q_{9r}=0.020$ \cite{KochRodehorst1974,BoulangerEtAl2013,SanchezEtAl1967}\\
10 & $\gamma_{313}$ + T $\rightarrow$ J & $q_{10}=0.006$ \cite{KochRodehorst1974,BoulangerEtAl2013,SanchezEtAl1967}\\
11 & $\gamma_{275}$ + J $\rightarrow$ I & $q_{11}=0.583$; T $\rightarrow I$; $q_{10}\times q_{11}=0.0034$ \cite{BoulangerEtAl2013,SanchezEtAl1967}\\
12 & I ${\stackrel{k_{12}}{\rightharpoonup}}$ L & $k_{12}=\exp(-E_a/RT + 12.974)$; s$^{-1}$; $E_a=19.93$ kcal mol$^{-1}$; hydrolysis of AICN \cite{SanchezEtAl1968}\\
13 & I:F + Af ${\stackrel{k_{13}}{\rightharpoonup}}$ A + F  & $k_{13}=\exp(- E_a/RT + 12.973)$; M$^{-1}$ s$^{-1}$; $E_a=6.68$ kcal mol$^{-1}$  \cite{YonemitsuEtAl1974,ZubayMui2001}\\
14 & I:F + Fa ${\stackrel{k_{14}}{\rightharpoonup}}$ Am + Fa +H$_2$O & $k_{14}=\exp(- E_a/RT + 12.613)$; M$^{-1}$ s$^{-1}$; $E_a=19.90$ kcal mol$^{-1}$  \cite{WangEtAl2013}\\
15 & $\gamma_{250}$ + Am $\rightarrow$ A & $q_{15}=0.060$ \cite{GlaserEtAl2007}\\
16 & A $\stackrel{k_{16}}{\rightharpoonup}$ Hy & $k_{16} = 10^{(-5902/T + 8.15)}$; s$^{-1}$; valid for pH within 5 to 8; hydrolysis of adenine \cite{LevyMiller1998,WangHu2016} \\
\hline
17 & $\gamma_{298}$ + C   $\rightarrow$ C & $q_{17}=0.955$\\
18 & $\gamma_{313}$ + T $\rightarrow$ T & $q_{18}=0.972$\\
19 & $\gamma_{275}$ + J $\rightarrow$ J & $q_{19}=0.417$\\
20 & $\gamma_{250}$ + Am $\rightarrow$ Am & $q_{20}=0.940$\\
21 & $\gamma_{250}$ + I $\rightarrow$ I & $q_{21}=1.000$\\
22 & $\gamma_{266}$ + L $\rightarrow$ L & $q_{22}=1.000$\\
23 & $\gamma_{260}$ + A $\rightarrow$ A & $q_{23}=1.000$\\
24 & $\gamma_{250}$ + Hy $\rightarrow$ Hy & $q_{24}=1.000$\\
\hline
\end{tabular}}
\label{tab:Reactions}
\end{table}

The following is a detailed description of each reaction given in table \ref{tab:Reactions}   by reaction number:
\begin{enumerate}
\item Hydrogen cyanide HCN (H) hydrolyses to formamide H$_2$NCOH (F) with a half-life dependent on temperature and pH \cite{SanchezEtAl1967}. The temperature dependent rate equation used here was determined by Kua and Thrush \cite{KuaThrush2016} at pH 7.0 from the experimental data of Miyakawa et al. \cite{MiyakawaEtAla2002}.

\item A photon-induced tautomerization coverts formamide (F) into formimidic acid (Fa). Basch et al. \cite{BaschEtAl1968} have measured the electronic excitation spectrum of formamide (F) and find a peak in absorption at 55,000 cm$^{-1}$ (182 nm) with a molar extinction of 11,000 M$^{-1}$ cm$^{-1}$. However, a shoulder exists on the main absorption peak which extends down to 40,000 cm$^{-1}$ (250 nm). Duvernay et al. \cite{DuvernayEtAl2005} suggest that this shoulder arises from the resonant excitation of the forbidden $n\rightarrow \pi^*$ transition located at 219 nm (130 kcal mol$^{-1}$) and not from the main $\pi\rightarrow \pi^*$ transition located at 182 nm. Maier and Endres \cite{MaierEndres2000} have determined that irradiation of formamide (F) at 248 nm rapidlly converts it into basically two tautomeric isomers of formimidic acid (Fa), H(OH)C=NH, which are both about 3.6 kcal mol$^{-1}$ in energy above formamide and separated from it by a transition barrier of height of $E_a=45.4$ kcal mol$^{-1}$ (gas phase). Similarly, Duvernay et al. \cite{DuvernayEtAl2005} have shown that under UVC light of 240 nm, formamide (F) tautomerizes into formimidic acid (Fa) and their calculation gives a similar transition state barrier height of 47.8 kcal mol$^{-1}$. Wang et al. calculate a transition state barrier of 49.8 kcal mol$^{-1}$ \cite{WangEtAl1991} but show that this is reduced to to $22.6$ kcal mol$^{-1}$ in the presence of only a single water molecule. This energy needed to overcome this barrier is in the infrared (1265 nm) but Cataldo et al. have shown that there is no evidence of thermal excitation until about 220 $^\circ$C \cite{CataldoEtAl2009}. Our model, therefore, assumes that the F $\rightarrow$ Fa tautomerization requires the absorption of a photon and we take the wavelength region for tautomerization due to the $n\rightarrow \pi^*$ transition of $220\pm$ 10 nm and assign an average molar extinction coefficient to that region of 60 M$^{-1}$ cm$^{-1}$ as measured by Basch et al. \cite{BaschEtAl1968} and also by Petersen et al. \cite{PetersenEtAl2008}.

\item Duvernay et al. \cite{DuvernayEtAl2005} have shown that formimidic acid (Fa) can, in turn, be photo-lysed into HCN (H), or HNC, plus H$_2$O, (dehydration) with maximal efficiency at about $198$ nm \cite{GingellEtAl1997}. However, the absorption spectrum of formimidic acid also has a shoulder extending to about 250 nm due to the same $n\rightarrow \pi^*$ excitation as in formamide. For example, Duvernay et al. observe a small amount of dehydration of formimidic acid at 240 nm. Given that our surface solar spectrum during the Archean (figure \ref{fig:Pigments}) is extinguished below about 205 nm, here we likewise assume an absorption wavelength for photo-lysing of $220\pm20$ nm and a similar average molar extinction coefficient as for the tautomerization of fomamide (F) of 60 M$^{-1}$ cm$^{-1}$ which is in accordance with the findings of Gingell et al. \cite{GingellEtAl1997}. Combining photo-reactions \#2 and \#3, we thus recuperate some of the HCN lost to thermal hydrolysis as described by reaction \#1 (see figure \ref{fig:FormamidicAcid}). Barks et al. \cite{BarksEtAl2010} have shown that if neat formamide is heated (130 $^\circ$C), thereby exciting vibrational states, a photon-induced excitation at even longer wavelengths (254 nm) also leads to the disintegration of formamide into HCN and H$_2$O, and they believe that this is the route to the production of the purines, adenine, guanine, and hypoxanthine, that they detect. Their yields are increased when including the inorganic catalysts sodium pyrophosphate and calcium carbonate, indicating that heating and inorganic catalysts can improve the photochemical reaction steps \#2 and \#3. Formamide also disintegrates thermally into HCN and H$_2$O, without requiring the absorption of a photon, but only at temperatures greater than about 220 $^\circ$C \cite{CataldoEtAl2009} because of high barriers \cite{KuaThrush2016}.

\begin{figure}[h]
\centering
\includegraphics[width=10 cm]{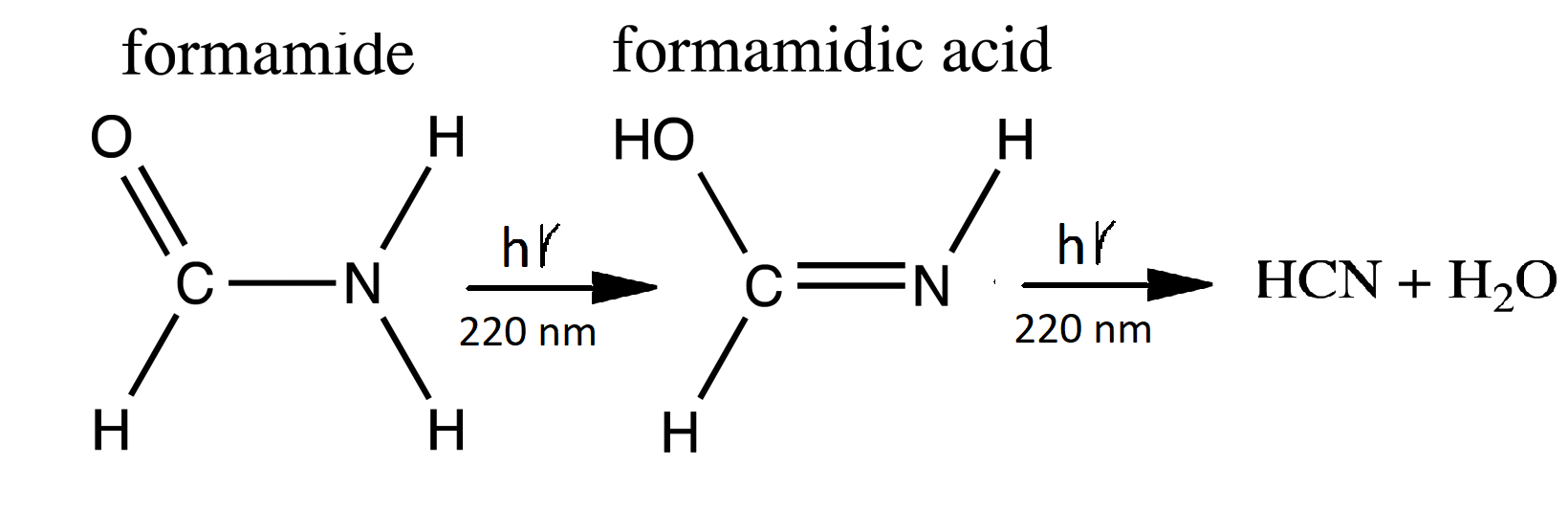} 
\caption{The production of formamidic acid (Fa) from formamide (F) (photoreaction \#2) and its subquent decay into HCN (H) and water (photoreaction \#3).}
\label{fig:FormamidicAcid}
\end{figure}

\item Formamide (F) hydrolyzes to ammonium formate (Af) at a rate of 1\% in 24 hr at $100^\circ$C \cite{BarksEtAl2010}. The temperature dependent rate equation given in table \ref{tab:Reactions} was determined by Kua and Thrush \cite{KuaThrush2016} at pH 7.0 from the experimental data of Miyakawa et al. \cite{MiyakawaEtAla2002}. Ammonium and formate from this salt become useful for the thermal reaction leading to the final addition of an HCN (H) to  AICN (I) catalyzed by formamide (F) to give adenine (A) (reaction \#13) \cite{YonemitsuEtAl1974,ZubayMui2001}. 

\item HCN (H) thermally polymerizes into (HCN)$_x$ with its most stable tetramer ($x=4$) known as cis-DAMN (C) being the preferred polymer from which more complex polymers can be synthesized \cite{CataldoEtAl2009}. The tetramization of 4HCN is an exothermic thermal reaction and occurs most rapidly at a solvent pH at its pKa value, which decreases with increasing temperature (pKa $=8.5$ at 60 $^\circ$C and $7.9$ at 100$^\circ$C) \cite{SanchezEtAl1967}. The tetramization of HCN into DAMN is not elementary but involves successive polymerization of HCN with H$^+$ and CN$^-$ ions \cite{SanchezEtAl1967} so is second order in the concentration of HCN. The temperature dependence of the rate of conversion of HCN to DAMN has been measured by Sanchez et al. \cite{SanchezEtAl1967}.  We assume transition state theory and an Arrhenius equation of form,
\begin{equation}
k_5=exp(- E_a/RT +\ln A).
\end{equation}
From the conversion rates for a 1 M solution of HCN with 0.01 M tetramer catalyst as given in table 5 of Sanchez et al. \cite{SanchezEtAl1967}, a straight line can be fitted to the graph of $\ln(1/T)$ vs $\ln(k_5)$ giving values of $\ln A = 19.049$ and $E_a/R = 9964.3$, or $E_a=19.8$ kcal mol$^{-1}$. However, this would be the rate equation for tetramerization of HCN at its pKa value which would be about 8.2 at 80  $^\circ$C \cite{SanchezEtAl1967}. To obtain the rate equation at the lower pH value assumed here of 7.0 we note that from Fig. 14 of Sanchez et al. \cite{SanchezEtAl1967} the half-lives for tetramization of HCN are the same for pH 7.0 at 80 $^\circ$C as they are for pH 8.2 at 64 $^\circ$C  (1 day). Setting the rate constants the same for these two conditions leads to a value of $E_a/R = 10853.01$ for pH 7.0. Finally, since HCN can polymerize into either cis-DAMN (C) or trans-DAMN (T), and since trans-DAMN (T) has a free energy of $\Delta E= 0.61$ \cite{SanchezEtAl1967} (0.56 \cite{GuptaTandon2012}) kcal mol$^{-1}$ higher than cis-DAMN (C) a Boltzmann factor of $1/(\exp (-\Delta E/RT)+1)$ is included for cis-DAMN and the same with $+\Delta E/RT$ for trans-DAMN.

The rates for hydrolysis and polymerization are similar for concentrations of HCN (H) between approximately 0.01 M and 0.1 M  (equal rates at 0.03 M for pH 7, T=80 $^\circ C$, Fig. 15 of reference \cite{SanchezEtAl1967}). At lower concentrations, hydrolysis dominates while at higher concentrations polymerization dominates \cite{SanchezEtAl1967}.

\item HCN (H) can also thermally polymerize into trans-DAMN (T) which has a free energy  of 0.61 kcal mol$^{-1}$ \cite{SanchezEtAl1967} higher than cis-DAMN (C). We therefore assume that the rate constant for the polymerization into trans-DAMN is the same as that for cis-DAMN multiplied by a temperature dependent Boltzmann factor $1/(\exp (+\Delta E/RT)+1)$.

\item Trans-diaminomaleonitrile, trans-DAMN (T), produced through the thermal reaction \#6, or through the UV photon-induced transformation of cis-DAMN (C) into trans-DAMN (see reaction \#9), is a good catalyst because it has electronic donor parts (–NH$_2$ groups) and acceptor parts (–CN group) linked by a double bond. As such, it can act as a catalyst for the tetramization of 4HCN into cis-DAMN \cite{SanchezEtAl1968}. Cis-DAMN is also a catalyst for the same thermal reactions, but has significantly less activity than trans-DAMN \cite{SanchezEtAl1967} and therefore its catalytic activity is neglected in our analysis. As can be surmised from the discussion of table 7 of reference \cite{SanchezEtAl1967}, including 0.01 M of the tetramer trans-DAMN increases the rate of tetramization by a factor of 12 at 20 $^\circ$C which would correspond to a reduction in the activation energy of 1.45 kcal mol$^{-1}$. This change in the barrier height is therefore included in the rate constant for this catalyzed reaction.

\item  Trans-DAMN also acts as an auto-catalyst for its own thermal production from 4HCN \cite{SanchezEtAl1968} and we assume a similar reduction in barrier height as for its catalysis of the production of cis-DAMN from 4HCN (reaction \#7).

\item  (a) cis-DAMN can transform into trans-DAMN (3), step (2) $\rightarrow (3)$ of figure \ref{fig:AdenineSyn} through a rotation around the double covalent carbon-carbon bond by absorbing a high energy photon (298 nm) to overcome the large energy barrier for rotation, calculated to be 58.03 kcal mol$^{-1}$ \cite{GuptaTandon2012} ($> 4$ eV \cite{BoulangerEtAl2013}). The quantum yield has been measured by Koch and Rodehorst to be $q_9=0.045$ \cite{KochRodehorst1974}.\\
\ \\
(b) Beacuse of the high energy barrier 58.03 kcal mol$^{-1}$ between trans-DAMN and cis-DAMN, thermal energy, even at our high temperatures, is insufficient to significantly reverse the rotation about the double covalent bond. However, trans-DAMN (T) can absorb a photon at 313 nm which would provide it with sufficient energy to isomerize back into cis-DAMN. The quantum efficiency for this reversal was determined to be $q_{9r}=0.020$ by fitting to the experimental data given in Fig. 1 of Koch and Rodehorst \cite{KochRodehorst1974} as explained in the description of figure \ref{fig:ConcenKoch}.

\item The absorption of a photon at 313 nm excites trans-DAMN which then transforms into AIAC through proton transfer from one of the amino groups \cite{BoulangerEtAl2013}.  Although there does not appear to exist a quantum yield for this photochemical reaction in the literature, the quantum yield for trans-DAMN to AICN (T$\rightarrow$ I) has been measured by Koch and Rodehorst \cite{KochRodehorst1974} to be 0.0034. By fitting our model concentration results to the experimental data of Koch and Rodehorst (see discussion of Fig. \ref{fig:ConcenKoch}) it is found that a best value for the quantum yield of trans-DAMN to AIAC (T$\rightarrow$J), $q_{10}$, is 0.006 and therefore the quantum yield for AIAC to AICN (J$\rightarrow$ I) would be $q_{11}=0.0034/0.006=0.5833$.

\item AIAC (J) on absorbing a photon at 275 nm then transforms through photon-induced cyclicization (ring closure) into an azetene intermediate (5 of figure \ref{fig:AdenineSyn}) in an excited state, which then transforms to the N-heterocyclic
carbene (6 of figure \ref{fig:AdenineSyn}) and finally this tautomerizes to give the imidazole AICN (I) \cite{BoulangerEtAl2013}. As noted above, the quantum yield for this process (J$\rightarrow$I) is taken to be 0.5833 to give the overall quantum yield for trans-DAMN to AICN (T$\rightarrow$ I) to be 0.0034 \cite{KochRodehorst1974}. AICN absorbs maximally at wavelength 250 nm.

\item The imidazole, 4-aminoimidazole-5-carbonitrle, AICN (I) created in the previous photochemical reaction (reaction \#11) can hydrolize to 4-aminoimidazole-5-carboxamide, AICA (L). I determined the rate equation for this 1st order reaction from the data of Sanchez et al. \cite{SanchezEtAl1968} at different temperatures (their table 1). From this temperature data, the barrier to hydrolysis can be determined to be 19.93 kcal mol$^{-1}$ and the frequency factor to be $\ln A = 12.974$.

\item The final coupling of a fourth HCN to AICN (I) and its cyclization to form adenine (A) is a very exothermal overall, $\Delta G=-53.7$ kcal mol$^{-1}$, but there are numerous large energy barriers on the path to its completion \cite{RoyEtAl2007}. The first step is the coupling of an HCN molecule to AICN, and this appears to be rate limiting since it has the highest energy barrier, calculated in the gas phase, of 39.7 kcal mol$^{-1}$ \cite{RoyEtAl2007}. However, it is catalyzed by both bulk solvent and specific water molecules which reduce the barrier to 29.6 kcal  mol$^{-1}$, or by ammonium molecules with bulk water solvent which reduce the barrier further to 27.6 kcal  mol$^{-1}$ \cite{RoyEtAl2007}. A number of experimental works \cite{YonemitsuEtAl1974, ZubayMui2001,HillOrgel2002,BarksEtAl2010} have revealed that ammonium formate (Af) could provide a route with an even lower barrier, but the rate is still too slow to allow significant adenine production from AICN and ammonium formate, unless a strong concentration mechanism existed, for example, dehydration \cite{BarksEtAl2010}, or perhaps the build up of concentration inside the vesicle, or the reaction-diffusion self-organizing occurring within the vesicle, as will be considered below.

A solution to this rate problem may exist, however, without requiring high concentrations. As early as 1974 Yonemitsu et al. \cite{YonemitsuEtAl1974} showed that including formamide, the hydrolysis product of HCN (reacction \#1), in aqueous solution, or by itself (neat solution), along with ammonium formate could dramatically speed up the reaction as long as the temperature was above approximately 80 $^\circ$C, leading to a successful industrial patent for the production of adenine from cis-DAMN (C) or trans DAMN (T) and formamide with ammonium formate. From examples 1 and 12 of the experiments of Yonemitsu et al. carried out at 150 and 100 $^\circ$C (using 135 g of formamide, 30 g of ammonium formate, and 2.01 g of DAMN) giving rise to 43.5\% and 30.0\% product of adenine after 5 and 10 hours at those temperatures respectively, it is possible to calculate an activation barrier for the overall reaction of $E_a=6.682$ kcal mol$^{-1}$. Since ammonium formate is a salt, the probable pathway from AICN to adenine would be that proposed by Zubay and Mui \cite{ZubayMui2001} where the ammonium ion NH$_4^+$ attacks the triple NC bond of AICN and the formate ion  HCOO$^-$ attacks the amine NH$_2$ group of AICN (figure 8 of reference \cite{ZubayMui2001}) both catalyzed by the proton transfer process involving formamide (see below), leading to this very low barrier. We therefore assume the reaction to be of second order and determined by the Arrhenius equation of form,
\begin{equation}
k_{13}=exp(- E_a/RT +\ln A),
\end{equation}
where $E_a = 6.682$ kcal  mol$^{-1}$ and the pre-exponential frequency factor A was estimated from the reduced mass dependence of the Langevin model \cite{Herbst2001}, $A=2\pi e\sqrt{\alpha / \mu}$ for a charged ion - neutral molecule system where $e$ is the ion electronic charge, $\alpha$ is the polarizability of the neutral reactant, and $\mu$ is the reduced mass of the reactants \cite{Benallou2019}. Considering all factors being equal except the reduced mass, and then normalizing to the frequency factor of reaction \#12 for the hydrolysis of AICN (I) by the inverse square root of the reduced mass for the reacting species, gives a value of $\ln A = 12.9734$. 

\item There exists a second possible route to adenine from AICN and HCN, without involving ammonium formate but considering the catalytic effect of formamide. Recently, Wang et al. \cite{WangEtAl2013} have studied through ab initio DFT calculations the synthesis of adenine starting from only formamide and propose what they call a ``formamide self-catalytic mechanism''. This mechanism consists of (1) a proton transfer from N to O of formamide to form the imidic acid tautomer, formimidic acid (Fa), obtained in our case through photon-induced proton transfer, reaction \#2; (2) a proton exchange between one imidic tautomer and one amide tautomer, resulting in two formimidic acids; and (3) an interaction between these two imidic acids yielding formimidic acid, a water molecule, and HCN. This formamide self-catalytic mechanism has relevance to the entire adenine synthesis process starting from only formamide since it reduces many of the barriers on route to adenine \cite{WangEtAl2013}. 

Of importance to us here of Wang et al.'s results is the step of the attachment of HCN to the amine group (NH$_2$) of AICN. They show for their particular case of formiminylation of 5-aminoimidazole  (Fig. 13 of reference \cite{WangEtAl2013}) that this reaction can be formamide-catalyzed (as described above) and find the activation energy barrier for this to be 19.9 kcal mol$^{-1}$ (significantly lower than 46.1 kcal mol$^{-1}$ in the noncatalyzed process and 34.0 kcal mol$^{-1}$ in the water-assisted process) and that the subsequent dehydration process to give the amidine (Am) (our case) is calculated to be 14.0 kcal mol$^{-1}$ (34.3 kcal mol$^{-1}$ in the noncatalyzed reaction). 

Therefore, we assume that the attachment of HCN (H) to AICN (I) to form 5-(N'-formamidinyl)-1H-imidazole-4-carbonitrileamidine (Am) to be a formamide catalyzed thermal reaction involving formimidic acid and formamide and we assume the rate of this reaction to be determined by the Arrhenius equation of form
\begin{equation}
k_{14}=exp(- E_a/RT +\ln A)
\end{equation}
where $E_a = 19.9$ kcal  mol$^{-1}$ and the pre-exponential frequency factor A is again estimated from the reduced mass dependence of the Langevin model \cite{Herbst2001}, considering again all factors equal except the reduced mass, and then  normalizing to the reaction \#12 for the hydrolysis of AICN (I) by the inverse square root of the reduced mass for the reacting species, giving a value of $\ln A = 12.613$.

It is important to note that AICN (I) has a conical intersection for a charge transfer from the molecule in the excited state to a neighboring cluster of water molecules \cite{SzablaEtAl2014b}. With AICN left in the charged state, this would significantly increase the rate of attachment, through charge-dipole interaction, to formamide, which has a dipole moment significantly larger than that of water (Table \ref{tab:Abbrev}), effectively changing the reaction from third order to second order, thereby increasing significantly the overall rate of this last attachment of HCN to AICN through this formamide catalyzed reaction.

The possibility of a hot ground state reaction occurring to aid in overcoming the barrier to producing adenine (A) from AICN (I) and HCN (H) could also be considered during daylight periods. These occur within a narrow time window after photon excitation, calculated by Boulanger et al. for a molecule (trans-DAMN) which has a similar conical intersection as AICN, to be about 0.2 ps, which corresponds to the time at which the excess energy on the molecule has been reduced to about 1/3 of its initial value, allowing reactions to proceed with a maximum barrier height of about 30 kcal mol$^{-1}$ \cite{BoulangerEtAl2013}. This possibility, however, will not be included in our model. It would have the overall effect of increasing the rate of the production of adenine.

\item After the attachment of a fifth HCN (H) to AICN (I) to form the amidine (Am), reaction \#14, a subsequent tautomerization is required (calculated to have a high barrier of about 50 kcal mol$^{-1}$) which, once overcome, allows the system to proceed through a subsequent barrier-less cyclicization to form adenine \cite{GlaserEtAl2007}. Such a high barrier to the final cyclicization means that, at the temperatures considered here, it cannot be a thermal reaction. Indeed, the fact that adenine has been found in space and in meteorites where temperatures are expected to be very low, indicated to Glaser et al.  \cite{GlaserEtAl2007} that a photochemical route must be available. They suggested a photon-induced tautomerization of amidine, which absorbs strongly at 250 nm.  Although oscillator strengths for the tautomerization have been calculated by Glaser et al., different {\em ab initio} approaches give significantly different values, so experiment will be required for its reliable determination. Therefore, until such data becomes available, we assume a similar molar extinction coefficient as for AICN and, being conservative, a quantum efficiency of $q_{15}=0.06$ but measure the effect on adenine production for a $\pm 30$\% variation of this parameter value (see table \ref{tab:ParamVariation}). In fact, these parameter variation results indicate that because of the large activation energy required, and the fact that the reactions are of second order, reactions \#14 and \#15 only come into play at very high temperature and can be neglected.

\item The temperature dependent rate equation for the destruction of adenine (A) through hydrolysis to give hypoxathine (Hy) which could then lead to guanine, or through deamination to some amino acids \cite{Frick1952}, was determined in careful experiments by Levy and Miller \cite{LevyMiller1998} (and also by Wang and Hu \cite{WangHu2016}). Zheng and Meng calculated a transition state barrier for hydrolysis of 23.4 kcal  mol$^{-1}$ \cite{ZhengMeng2009}.

\item to 24. These reactions represent the absorption of a photon, in a 20 nm region centered on the wavelength of peak absorption, on the molecule which then decays through internal conversion at a conical intersection to the ground state on sub-picosecond time scales. All molecules listed in this set of photo-reactions are basically photo-stable because of a peaked conical intersection connecting the excited state with the ground state. These reactions, with large quantum efficiencies, represent the bulk of the flow of energy from the incident UVC spectrum to the emitted outgoing ocean surface spectrum in the infrared and therefore contribute most to photon dissipation, or entropy production. 
 
\end{enumerate}

In order to obtain simple kinetic equations for the photochemical reactions listed in table \ref{tab:Reactions}, we assume that the molecules only absorb within a region $\pm 10$ nm of their maximum absorption wavelength $\lambda_{max}$ and that this absorption is at their maximum molar extinction with coefficient $\epsilon$ (table \ref{tab:Abbrev}), and finally that these wavelength regions do not overlap for calculating the shadowing effect of concentrations higher up in the vesicle. We assume that the vesicle is at the ocean surface and the depth coordinate is divided into $i=20$ bins of width $\Delta x=5$ $\mu$m and the time interval for the recursion calculation for the concentrations at a particular depth is 10 ms. The recursion relation for the factor of light intensity $L_\lambda (i,C)$ for a concentration of the molecule $C$, at a depth $x(i)=i\cdot \Delta x$ below the ocean surface will be,
\begin{equation}
   L_\lambda (i,C(i)) = L_\lambda (i-1,C(i-1)) e^{-\Delta x\cdot \alpha_\lambda}\cdot 10^{-\Delta x\cdot \epsilon_\lambda C(i)}
\end{equation}
where $\alpha_\lambda$ is the absorption coefficient of water at wavelength $\lambda$ and $\epsilon_\lambda$ is the molar extinction coefficient of the particular absorbing substance which has concentration $C(i)$ at $x(i)$.

The kinetic equations to give the increment in concentration after each time step $\Delta t \equiv dt$, for use in a discrete recursion relation, at a depth $x$ below the surface are determined from the reactions listed in table \ref{tab:Reactions} to be the following; 
\begin{align}
{dH \over dt} &= D_H{\partial^2 H \over \partial x^2} - k_1H + d\cdot q_3 I_{220}L_{220}(Fa){(1-10^{-\Delta x\epsilon_{220} Fa}) \over \Delta x}  -k_5H^2-k_6H^2-k_7H^2T-k_8H^2T \nonumber\\
&= D_H{\partial^2 H \over \partial x^2} + d\cdot q_3I_{220}L_{220}(Fa) {(1-10^{-\Delta x\epsilon_{220} Fa}) \over \Delta x} -Hk_1 - H^2(k_5+k_6 +T(k_7+k_8)) \label{r1}\\
{dF \over dt} & = D_F{\partial^2 F \over \partial x^2} + k_1H -d\cdot q_2I_{220}L_{220}(F){(1-10^{-\Delta x\epsilon_{220} F}) \over \Delta x} -k_4F -k_{14}IFa\label{r2}\\
{dFa \over dt} & = D_{Fa}{\partial^2 Fa \over \partial x^2} + d\cdot q_2I_{220}L_{220}(F){(1-10^{-\Delta x\epsilon_{220} F}) \over \Delta x} - d\cdot q_3I_{220}L_{220}(Fa) {(1-10^{-\Delta x\epsilon_{220} Fa}) \over \Delta x}\\
{dAf \over dt} & = D_{Af}{\partial^2 Af \over \partial x^2} + k_4F - k_{13}IAf\\
{dC \over dt} & = D_C{\partial^2 C \over \partial x^2} + k_5H^2 +k_7H^2T- d\cdot q_9I_{298}L_{298}(C) {(1-10^{-\Delta x\epsilon_{298} C}) \over \Delta x}\nonumber\\
 &\ \ \ +d\cdot q_{9r}I_{313}L_{313}(T){(1-10^{-\Delta x\epsilon_{313} T}) \over \Delta x}\label{r5}\\
{dT \over dt} &= D_T{\partial^2 T \over \partial x^2} + k_6H^2+k_8H^2T + d\cdot q_9I_{298}L_{298}(C) {(1-10^{-\Delta x\epsilon_{298} C}) \over \Delta x} \nonumber\\
&\ \ \  - d\cdot q_{10}I_{313}L_{313}(T){(1-10^{-\Delta x\epsilon_{313} T}) \over \Delta x} -d\cdot q_{9r}I_{313}L_{313}(T){(1-10^{-\Delta x\epsilon_{313} T}) \over \Delta x}\label{r6}\\
{dJ \over dt} &= D_J{\partial^2 J \over \partial x^2} + d\cdot q_{10}I_{313}L_{313}(T){(1-10^{-\Delta x\epsilon_{313} T})\over \Delta x} - d\cdot q_{11}I_{275}L_{275}(J) {(1-10^{-\Delta x\epsilon_{275} J}) \over \Delta x} \\
{dI \over dt} &= D_I{\partial^2 I \over \partial x^2} + d\cdot q_{11}I_{275}L_{275}(J){(1-10^{-\Delta x\epsilon_{275} J}) \over \Delta x} -k_{12}I - k_{13}IAf - k_{14}IFa \\
{dL \over dt} &= D_L{\partial^2 L \over \partial x^2} + k_{12}I\\ 
{dAm \over dt} &= D_{Am}{\partial^2 Am \over \partial x^2} + k_{14}IFa  - d\cdot q_{15}I_{250}L_{250}(Am){(1-10^{-\Delta x\epsilon_{250} Am}) \over \Delta x}\\
{dA \over dt} &= D_A{\partial^2 A \over \partial x^2} + d\cdot q_{15}I_{250}L_{250}(Am){(1-10^{-\Delta x\epsilon_{250} Am}) \over \Delta x} + k_{13}IAf - k_{16}A\\
{dHy \over dt} &= D_Hy{\partial^2 Hy \over \partial x^2} + k_{16}A \label{r16}
\end{align}
where the differentials are calculated discretely (e.g. $dH/dt\equiv \Delta H/\Delta t$) and all concentration values are calculated at discrete time steps of $\Delta t = 10$ ms and the calculated value of the change (e.g. $\Delta H(j)/\Delta t$) for time step $j$ is summed to the previous value (e.g. $H(j-1)$). The day/night factor $d$ is equal to 1 during the day and 0 at night. $I_{220}, I_{298}, I_{313}, I_{275}$ and $I_{250}$ are the intensities of the photon fluxes at $220, 298, 313, 275$ and $250$ nm respectively (Fig. \ref{fig:Pigments}). $\epsilon_\lambda$ are the coefficients of molar extinction for the relevant molecule at the corresponding photon wavelengths $\lambda$.

\subsection{Vesicle Permeability and Internal Diffusion}

The permeability of the the vesicle wall to the molecule, and the diffusion constant for the molecule within the inner aqueous region of the interior of the vesicle will both decrease with the area of the molecule and with the size of its electric dipole moment (Table \ref{tab:Abbrev}) and increase with temperature. It is interesting to note that almost all of the final and intermediate product molecules have large dipole moments, implying tendency towards entrapment within the vesicle. We assume that the vesicle cannot remain intact at temperatures greater than 95 $^\circ$C but that below this temperature it is completely permeable to H$_2$O, HCN (H) and formimidic acid (Fa) but impermeable to all the other intermediate products due to their large size and large electric dipole moments. Note that ammonium formate would be in its ionic form and therefore also unable to cross the fatty acid membrane. Permeabilities across lipid boundaries are reduced by orders of magnitude if the molecules are polar or are charged \cite{YangHinner2015}.

The diffusion constant D$_Y$ for the molecule $Y$ will depend on the viscosity of the solution inside the vesicle which is dependent on the amount of organic material within the vesicle. Studies of intracellular diffusion of nucleotides indicate three factors influencing diffusion rates, besides temperature, at high solute densities; the viscosity of the environment, collisional interactions dependent on concentration, and binding interactions between molecules \cite{AgarwalEtAl2016}. The diffusion constant of adenine in pure water has been determined to be $D_A=7.2\times 10^{-6}$ cm$^2$ s$^{-1}$ \cite{BowenMartin1964} while the measured diffusion rates in the cytoplasm of different cell types varies between $1.36\times 10^{-6}$ to $7.8\times 10^{-6}$ cm$^2$ s$^{-1}$\cite{AgarwalEtAl2016}. 

Surface films of organics and trace metals, with a high density of lipids and other hydrocarbons produced, for example, by the ultraviolet spectrum of figure \ref{fig:Pigments} on CO$_2$ saturated water \cite{MichaelianRodriguez2019} could have been expected on the ocean surface during the Archean. Diffusion constants in this sea surface microlayer would then be significantly smaller than for the bulk water. Diffusion rates inside the vesicle will depend on the amount of organic material already existing at the air/water interface (this may have varied spatially considerably) captured during the formation of the vesicle, and on the amount of ongoing organic synthesis inside the vesicle. 

All diffusion constants are defined relative to that for adenine through the formula;
\begin{equation}
D_Y = {\mu_AA_A\over \mu_Y A_Y}\cdot D_A,
\label{eq:DiffConst}
\end{equation}
where $A_A$ is the polar surface area and $\mu_A$ the dipole moment of adenine (table \ref{tab:Abbrev}) and investigate two different diffusion constants, the smallest value for adenine in present day cytoplasm and one four orders of magnitude smaller. Using the values given in table \ref{tab:Abbrev} for the molecule dipole moment and the area, we obtain the results given in table \ref{tab:Diff}.

\begin{table}[H]
\caption{Diffusion constants relative to that of adenine for the different intermediate product molecules obtained from equation (\ref{eq:DiffConst}).  Two different multiplicative factors of $D_A =1\times10^{-6}$ and $1\times10^{-10}$ cm$^2$ s$^{-1}$ are used in the calculations.}
\centering
\resizebox{16cm}{!}{
\begin{tabular}{|l|l|l|l|l|l|l|l|l|l|l|l|}
\hline
$D_H$ & $D_F$ & $D_{Fa}$, & $D_{Af}$ & $D_C$ & $D_T$& $D_J$ & $D_I$ & $D_L$ & $D_{Am}$ & $D_A$ & $D_{Hy}$ \\
\hline
7.752 &  2.988 &  11.190 &  6.689 &  0.892 &  4.073 &  4.073 & 1.908 & 1.532 & 1.000 & 1.000 & 2.482\\
\hline
\end{tabular}}
\label{tab:Diff}
\end{table}
Cyclical boundary conditions are assumed for diffusion, except for HCN (H) and formimidic acid (Fa) which can permeate the vesicle wall and therefore at the wall they are given their fixed value specified in the initial conditions of the environment outside the vesicle (see below). The second order derivatives for calculating the diffusion were obtained using the second order finite difference method with double precision variables.

\subsection{Initial Conditions}

Miyakawa, Cleaves and Miller \cite{MiyakawaEtAla2002} estimated the steady state bulk ocean concentration of HCN at the origin of life assuming production through electric discharge on atmospheric methane to produce radicals which attack N$_2$, leading to an input rate to the oceans of 100 nmole cm$^{-2}$ y$^{-1}$, and loss of HCN due to hydrolysis and destruction at submarine vents with a 10 million year recycling time of all ocean water for an ocean of 3 Km average depth. For an ocean of pH 6.5 and temperature of 80 $^\circ$C, they obtained a value of [HCN]$=1.0\times 10^{-10}$ M \cite{MiyakawaEtAla2002}. 

However, as mentioned above, HCN can also be produced through the solar Lyman alpha line (121.6 nm) photo-lysing N$_2$ in the upper atmosphere giving atomic nitrogen which then combines with CH and CH$_2$ to give HCN, or through 145 nm photolysis of CH$_4$ leading to a CH$^*$ radical which attacks N$_2$ to give HCN  \cite{TrainerEtAl2012}. Including this UV production would increase the input of HCN concentration to the oceans by a factor of at least 6 \cite{Zahnle1986, ChybaSagan1992, StriblingMiller1986}. Furthermore, the first $\sim 100$ $\mu$m of the ocean surface is now known to be a unique region, known as the hydrodynamic boundary layer, in which surface tension leads to enriched organics with densities up to $10^4$ times that of organic material in the water column slightly below \cite{Hardy1982}. Trace metal enhancement in this microlayer can be one to three orders of magnitude greater than in the bulk \cite{Hardy1982,ZhangEtAl2004}. Langmuir circulation, Eddy currents, and the scavenging action of bubbles tends to concentrate organic materials into this surface film. If disturbed or mixed, the film rapidly reestablishes its integrity. This high density of organic material trapped through hydrophobic and ionic interactions at the ocean surface leads to significantly lower rates of diffusion at the surface microlayer as compared to the ocean bulk \cite{Hardy1982}. Little diffusion and turbulence therefore imply little mixing. The ocean microlayer is therefore a very stable layer which, of course, would not be recycled through ocean vents. Finally, although HCN is very soluble in bulk water, recent molecular dynamic simulations have shown that it concentrates to about an order of magnitude larger at the air-water interface due to lateral HCN dipole-dipole interactions, and that it evaporates at lower rates than does water \cite{FabianEtAl2014}. 

Therefore, rather than assuming the low bulk concentrations of Miyakawa et al., we instead consider two higher initial surface concentrations for HCN (H) ($6\times 10^{-5}$ and $6\times 10^{-4}$ M) and formimidic acid (Fa) ($1\times 10^{-5}$ and $1\times 10^{-4}$ M), the latter resulting from a photochemical tautomerization of formamide, the hydrolysis product of HCN (reactions \#1 and \#2 of table \ref{tab:Reactions}). We also allow for the perturbation of the system by considering the probable existence of small and sparse patches of much higher concentrations, up to $0.1$ M of both these molecules, justified by the above mentioned characteristics of the ocean microlayer and the dipole-dipole interaction between HCN molecules. The initial concentrations of all other reactants and products inside the vesicle (assumed impermeable to these) are taken to be $1.0\times 10^{-10}$ M.

\section{Results}
\label{sec:Results}
\subsection{Validation of Model}

The rates of tetramization of 4HCN (H) to cis-DAMN (C) and trans-DAMN (T) are given by the terms $k_5H^2$ and $k_6H^2$ of Eqs. (\ref{r5}) and ({\ref{r6}), reactions \#5 and \#6, respectively. The rate of hydrolysis of HCN (H) into formamide (F) is given by the term $k_1H$ of Eq. (\ref{r2}), reaction \#1.  The ratio of these rates $H^2(k_5+k_6)/Hk_1$ for pH 7.0 at different concentrations of HCN and as a function of temperature is plotted in figure \ref{fig:TetraHydr} along with experimental values derived from the data of Sanchez et al. \cite{SanchezEtAl1967} for the point at which tetramization and hydrolysis rates are equal (Ratio$=1$).

\begin{figure}[H]
\centering
\includegraphics[width=8 cm]{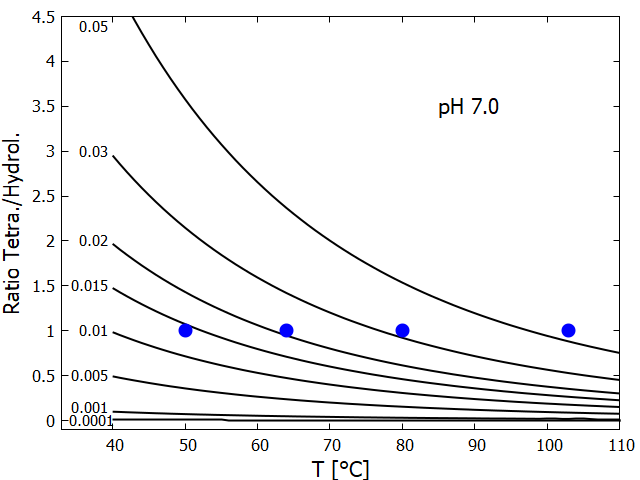}
\caption{The ratio of the rates of tetramization to hydrolysis as a function of temperature as determined by our model for aqueous solutions of HCN at different concentrations [M] (given at the beginning of the corresponding trace) at pH 7.0. The experimental data points in blue for [HCN] = 0.05, 0.03, 0.02, 0.015 M at Ratio=1 were obtained by linearly extrapolating to pH 7.0 from the closest two data points of Fig. 15 of Sanchez et al. \cite{SanchezEtAl1967}.}
\label{fig:TetraHydr}
\end{figure}

From figure \ref{fig:TetraHydr} it can be seen that ratio of the rates of HCN tetramization to hydrolysis increases with HCN concentration and with lower temperature (Fig. \ref{fig:TetraHydr}). For this reason, eutectic concentration at freezing temperatures was deemed to be the most probable route from HCN to the nucleobases, giving rise to the ``cold origin of life'' scenarios. However, not withstanding the fact that this contradicts the available geochemical evidence of high temperatures during the Archean, it will be shown here that high temperatures could also have led to significant concentrations of the nucleobases for the following reasons, 1) the ocean surface microlayer is a region of orders of magnitude higher organic density than the bulk, 2) through UVC photochemistry on HCN inside a fatty acid vesicle, a build up of product molecules would occur for those unable to permeate the vesicle wall, 3) as will be shown below, hydrolysis of HCN leads to formamide (F), and a subsequent hydrolysis to ammonium formate (Af), the former of which is an important catalyst, and the latter a necessary component, for the final attachment of a 5th HCN molecule to AICN (I) to give adenine (reaction \#13) which occurs with great efficacy above temperatures of 80 $^\circ$C \cite{YonemitsuEtAl1974}, and 4) besides reaction \#13, there is an alternative route to adenine, reaction  \#14 which has a high activation energy so would occur only at very high temperatures ($>95$ $^\circ$C).

Experiments have been performed by Koch and Rodehorst \cite{KochRodehorst1974}  concerning the photo-transmutation of cis-DAMN (C) into trans-DAMN (T) and then into AICN (I) (Fig. 1 of reference \cite{KochRodehorst1974}) which are the important photochemical steps in our model. This occurs through three photochemical reactions $\gamma_{298} + C\rightarrow T$ , $\gamma_{313} + T\rightarrow J$, $\gamma_{275} + J\rightarrow I$, the intermediate being AIAC (J) (see figure \ref{fig:AdenineSyn}). Our model can be compared to these experimental results given that the light source used by Koch and Rodehorst was stipulated as being a Rayonet RPR3000 A lamp which peaks in intensity at 305 nm with  $\sim 10$\% smaller and similar output at both 313 and 298 nm, and about 10\% of the latter at 275 nm (see Fig. 13 of reference \cite{Shvydkiv2012}). These ratios of Rayonet RPR3000 A lamp light intensity at 313:298:275 nm of 1.0:1.0:0.1 were used in our model and all initial concentrations set to zero except that of cis-DAMN (C), which was set to 0.00145 M (Fig. 1 of reference \cite{KochRodehorst1974}). The day/night light cycling was disabled and the two quantum efficiencies, unavailable in the literature, for $\gamma_{313} + T\rightarrow J$ and $\gamma_{313} + T\rightarrow C$, were adjusted to $q_{10}=0.006$ and $q_{9r}=0.020$ to give a best fit of the model to the experimental data. Note that determining $q_{10}$ in this manner then determines $q_{11}$ since $q_{10}\times q_{11} =0.0034$ \cite{KochRodehorst1974}. The overall intensity of the light on sample was adjusted to give the correct time scale. The results are plotted in figure \ref{fig:ConcenKoch} and agree well with experiment.

\begin{figure}[H]
\centering
\includegraphics[width=10 cm]{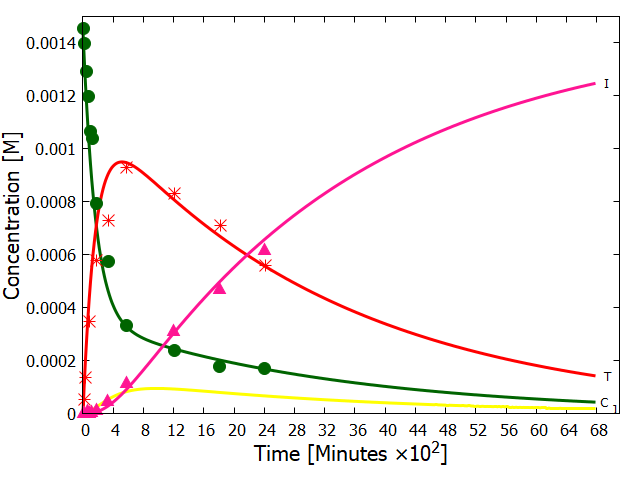}
\caption{The concentrations of cis-DAMN (C, dark-green), trans-DAMN (T, red), AIAC (J, yellow), and AICN (I, pink) obtained as a function of time from our model and compared with the experimental data points of Koch and Rodehorst (Fig. 1 of reference \cite{KochRodehorst1974}) starting with a concentration of cis-DAMN of 0.00145 M. The overall light intensity and the quantum efficiencies $q_{9r}$ and $q_{10}$ were adjusted to give the best fit.}
\label{fig:ConcenKoch}
\end{figure}

At the photostationary state under the Rayonet lamp, Koch and Rodehorst find that the remaining DAMN is distributed between its two isomers trans (T) and cis (C) with proportions of 80\% and 20\% respectively \cite{KochRodehorst1974}. Our model at close to the stationary state, at 6800 minutes (Fig. \ref{fig:ConcenKoch}), gives these proportions as 77\% and 23\% respectively.
 
Using instead the UV light intensities of the Archean surface UV solar spectrum (Fig. \ref{fig:Pigments}) gives the concentration profiles shown in figure \ref{fig:ConcenKoch2}. The difference between figures \ref{fig:ConcenKoch} and \ref{fig:ConcenKoch2} are due to the difference in light spectra, principally due to the light intensity at 298 nm (responsible for the C $\rightarrow$T isomerization) in the solar spectrum arriving at the Archean Earth surface being an order of magnitude smaller than that of the Rayonet lamp used in the experiments.

\begin{figure}[H]
\centering
\includegraphics[width=10 cm]{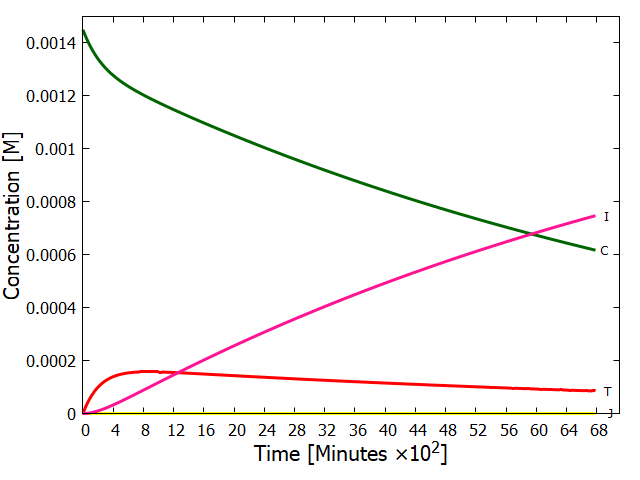}
\caption{The concentrations of cis-DAMN (C, dark-green), trans-DAMN (T, red), AIAC (J, yellow), and AICN (I, pink) obtained as a function of time from our model using the light spectrum of the Archean surface (Fig. \ref{fig:Pigments}) starting with a concentration of cis-DAMN of 0.00145 M.}
\label{fig:ConcenKoch2}
\end{figure}

The catalytic effect of trans-DAMN on the tetramization of HCN (reaction \#7) was incorporated into the model by reducing the energy of the activation barrier such as to give the same amplification factor of 12 due to the catalytic effect of the inclusion of 0.01 M trans-DAMN in the HCN solution observed in the experiments of Sanchez et al. \cite{SanchezEtAl1967} at a temperature 20 $^\circ$C (see discussion of reaction \#7 after table \ref{tab:Reactions}).

All other parameters employed in the model, such as activation barrier energies, pre-exponential frequency factors, and quantum efficiencies (except $q_{15}$), were taken directly from experiment, or by fitting to experimental rate versus temperature data, or taken from accurate first principles calculations as described in detail after table \ref{tab:Reactions}. However, to determine the variability of our model results with respect to possible inaccuracies in the parameters, in table \ref{tab:ParamVariation} the critical parameters of the model (those quantum efficiencies not determined directly by experiment, or chemical reactions with high activation energies) are varied by $\pm 30$\%, and the effect on the final adenine concentration is noted after 30 Archean days at 80 $^\circ$C.

\begin{table}[H]
\caption{The concentration of adenine [M] produced after 30 Archean days at 80 $^\circ$C determined by the model as a function of a $\pm 30$\% variation of the most sensitive parameters of the model with respect to their nominal values listed in the table (see also table \ref{tab:Reactions}). The initial concentrations were [H]$_0= 6.0\times10^{-5}$ M, [F]$_0=1.0\times10^{-5}$ M, [Fa]$_0=1.0\times10^{-5}$ M with all other concentrations [Y]$_0=1.0\times10^{-10}$ M and the diffusion constant was $D_A =1\times10^{-6}$  cm$^2$ s$^{-1}$. One perturbation of the system of [H] and [Fa] to 0.1 M for 2 minutes occurs at 10.4 Archean days (see figure \ref{fig:Concen80e-5}).}
\centering
\resizebox{16cm}{!}{
\begin{tabular}{|l|ll|l|l|l|l|}
\hline
\# & reaction & parameter & nominal value &  -30\% & nominal & +30\% \\  
\hline
9b & $\gamma_{313}$+T$\rightarrow$ C & $q_{9r}$ & 0.020 &   8.222e-6 & 7.292e-6 & 6.528e-6\\
12 & I ${\stackrel{k_{12}}{\rightharpoonup}}$ L hydrolysis of AICN & $E_{a12}$ & 19.93 kcal mol$^{-1}$ &  1.093e-6 & 7.292e-6 & 7.311e-6\\
13 & I:F + Af ${\stackrel{k_{13}}{\rightharpoonup}}$ A + F & $E_{a13}$ & 6.68  kcal mol$^{-1}$ &  7.311e-6 & 7.292e-6 & 6.636e-6\\
14 & I:F + Fa ${\stackrel{k_{14}}{\rightharpoonup}}$ Am + Fa +H$_2$O & $E_{a14}$ & 19.90 kcal mol$^{-1}$ &  7.292e-6  & 7.292e-6 & 7.292e-6\\
15 & $\gamma_{250}$+Am$\rightarrow$ A & $q_{15}$ &0.060  &  7.292e-6   & 7.292e-6 & 7.292e-6 \\
16 & A$\rightarrow$ Hy hydrolysis of adenine & expn. of $k_{16}$ & -5902 & 1.586e-9 &7.292e-6 &7.306e-6\\
\hline
\end{tabular}}
\label{tab:ParamVariation}
\end{table}

From table \ref{tab:ParamVariation} it can be seen that at 80 $^\circ$C the parameter variability with greatest affect on the concentration of adenine are the first order hydrolysis reactions, \#12 for hydrolysis of AICN and \#16 for hydrolysis of adenine itself. Reducing the activation barrier for adenine destruction through hydrolysis determined by Levy and Miller \cite{LevyMiller1998} (half-life of adenine of 8.0 years at 80 $^\circ$C at neutral pH) by 30\% leads to only a single order magnitude increase in adenine over its initial value of $1\times 10^{-10}$ M after 30 days as compared to the almost 5 orders of magnitude increase which occurs when using the nominal value. It is noted that the hydrolysis of adenine leads to guanine, or through deamination to some amino acids \cite{Frick1952}. Changing the parameters for reactions \#14 and \#15 does not affect adenine production because this route to adenine production through Am only comes into play at temperatures above $\sim 95$ $^\circ$C because of the high activation energy and the fact that reaction \#14 is of second-order. Most of the adenine production at 80 $^\circ$C occurs through reaction \#13.

\subsection{Evolution of the Concentration Profile}

In figures \ref{fig:Concen90e-5} through \ref{fig:Concen80e-5D-10} I present the concentrations as a function of time in Archean days (16 hours) of the relevant molecules in the photochemical synthesis of adenine inside the vesicle obtained by solving simultaneously the differential kinetic equations, (\ref{r1}) through (\ref{r16}), for the initial conditions and diffusion constants listed in the figure captions.

The concentration profiles of the molecules evolve over time because of accumulation of photo-products within the vesicle and because of a deliberate external perturbation, effectuated at 10.4 Archean days, of the non-linear system which leads it to a new stationary state which processes the environmental precursor molecule HCN into adenine at a greater rate. This leads to greater dissipative efficacy of the system, i.e. to a concentration profile of the molecules which dissipates more efficiently the incident UVC spectrum.

\begin{figure}[H]
\begin{center}
\includegraphics[width=19 cm]{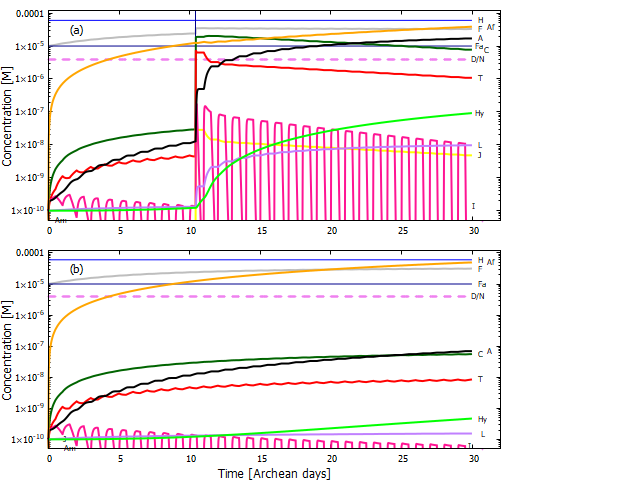} 
\end{center}
\caption{(a) Concentrations as a function of time in Archean days (16 hours) of the precursor and product molecules; H - HCN, F - formamide Fa - formimidic acid, Af - ammonium formate , C - cis DAMN, A - adenine, I - AICN, T - trans DAMN, J - AIAC, L - AICA, Am - Amidine, Hy - hypoxanthine, dissipatively structured on route to the synthesis of adenine (black trace). The initial conditions are; temperature T=90 $^\circ$C, initial concentrations [H]$_0= 6.0\times10^{-5}$ M, [F]$_0=1.0\times10^{-5}$ M, [Fa]$_0=1.0\times10^{-5}$ M and all other initial concentrations [Y]$_0=1.0\times10^{-10}$ M. The diffusion constant exponential factor was $1.0\times10^{-6}$ (e.g. $D_A=1.0\times10^{-6}$ cm$^2$ s$^{-1}$). There is one perturbation of the system corresponding to the vesicle floating into a region of HCN (H) and formimidic acid (Fa) of concentration 0.1 M for two minutes at 10.4 Archean days (vertical line at the top of the graph). A new stationary state at higher adenine concentration is reached after the perturbation. The violet horizontal dashed line, D/N, identifies alternate periods of daylight (violet) and night (blank). After 30 Archean days, the concentration of adenine within the vesicle (black trace) has grown by more than five orders of magnitude, from $1.0\times10^{-10}$ to $1.7\times10^{-5}$ M. (b) Same as (a) but without perturbation  giving a two orders of magnitude smaller final concentration of adenine compared to the case with perturbation (a). 
}
\label{fig:Concen90e-5}
\end{figure}

\begin{figure}[H]
\begin{center}
\includegraphics[width=19 cm]{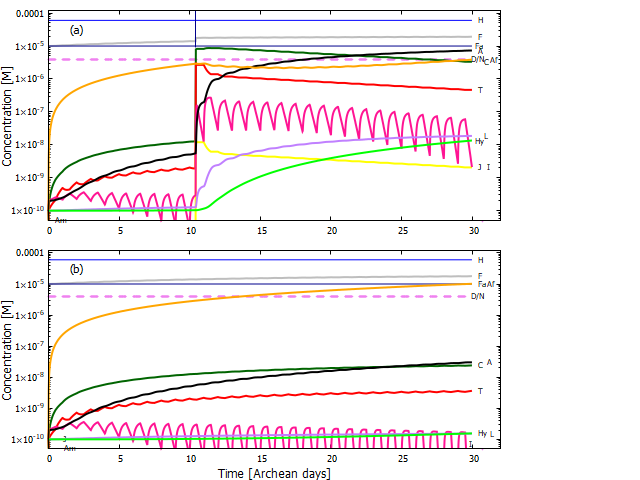}
\end{center}
\caption{(a)The same as for Fig. \ref{fig:Concen90e-5} except for a temperature of 80 $^\circ$C. The adenine concentration reaches $7.3\times 10^{-6}$ M. (b) The same without perturbation. The adenine concentration reaches $3.0\times 10^{-8}$ M.}
\label{fig:Concen80e-5}
\end{figure}

The dynamics observed in the concentrations profiles displayed in figures  \ref{fig:Concen90e-5} and \ref{fig:Concen80e-5} is a function of both the external perturbations affecting the system and of its inherent non-linearity. First, given the fixed concentrations of HCN (H) and formimidic acid (Fa) in the environment, to which the vesicle is permeable, photochemical reactions occur during daylight hours (denoted by the violet colored sections of the horizontal dashed line labeled as D/N). This gives rise to the observable diurnal oscillations in the concentrations of trans-DAMN (T) and AICN (I) since these are direct products of photochemical reactions.

At 10.4 Archean days, the vesicle is perturbed by assuming it passes through a region of high density of HCN (H) and formimidic acid (Fa)  (0.1 M) for a 2 minute period (vertical blue line at top of (a) graphs). This sudden impulse in HCN and Fa concentration gives rise to rapid increases in all concentrations within the vesicle, in particular for formamide (F), the hydrolysis product of H, which is an important catalyst for reaction \#13 which produces adenine (A) from AICN (I) (see table \ref{tab:ParamVariation}) and this reaction route is the most important for adenine production as it has a low activation barrier energy. Ammonium formate (Af) is used up in this reaction so it's concentration decreases after the perturbation. More importantly, however, immediately after the perturbation there is a greater production of trans-DAMN (T) in the vesicle and since T acts as a catalyst for the polymerization of HCN (H) (reactions \# 7 and \#8), this will produce a greater metabolism of H into DAMN within the vesicle and therefore leads to a stronger diffusion of H into the vesicle from the outside environment as long as T remains higher than before the perturbation. In other words, the reason that a short impulse of HCN and formimidic acid gives rise to an important increase in the rate of production of adenine is that the vesicle's semi-permeable wall, together with the set of equations describing the photochemical and chemical reactions, Eqs. (\ref{r1}-\ref{r16}), form a non-linear system which therefore has more than one stationary state solution at any given time. 

The perturbation causes the system to leave the attraction basin of one solution determined by its initial conditions and evolve towards a different, and more probable, stationary state of much higher rate of production of adenine (the rate is given by the slope of the black trace of figure \ref{fig:Concen80e-5}, more obvious on the linear graph of figure \ref{fig:Concen80e-5Linear}). The second stationary state is more probable than the initial because its concentration profile is more dissipative, i.e. with more molecules having conical intersections dissipating the absorbed photon energy rapidly into heat. In the new state, there is therefore less probability of photochemical reactions changing the concentration profile. However, the thermodynamic driving force for evolution to the new stationary state after the perturbation is that the system obtains a greater photon dissipative efficacy. For systems where local equilibrium is valid (section \ref{sec:Thermodynamics}), this is the same as saying that the entropy production of the system increases (see figure \ref{fig:EntProd80e-5}). This is an example of dissipative structuring, in this case, of the concentration profile, which, I suggest, is the physics and chemistry behind biological evolution, seen at its earliest stages.

\begin{figure}[H]
\begin{center}
\includegraphics[width=15 cm]{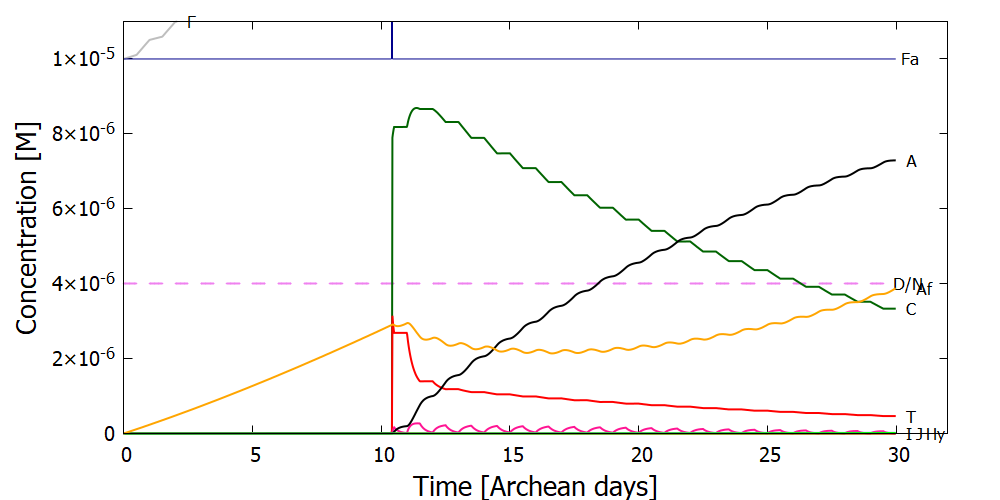}
\end{center}
\caption{The same as for Fig. \ref{fig:Concen80e-5}(a) except plotted on a linear scale. There is a large increase in the rate of production of adenine (slope of black line) after the transient perturbation at 10.4 Archean days of 2 minutes duration. After the perturbation there is a greater metabolism of HCN (H) from the environment due to the non-linearity of the system.}
\label{fig:Concen80e-5Linear}
\end{figure}

Figure \ref{fig:Concen80e-4} shows the results obtained by increasing the concentrations of HCN (H) and formimidic acid (Fa) of the environment by an order of magnitude to $6\times 10^{-4}$ and $1\times 10^{-4}$ M respectively. Comparing the adenine production in figure \ref{fig:Concen80e-5}(a), obtained with a single perturbation of H and Fa to 0.1 M for two minutes with a background environmental concentration of H of $6\times 10^{-5}$ M, with figure \ref{fig:Concen80e-4}(b) without perturbation but a background concentration of H of 10 times higher at $6\times 10^{-4}$ M, emphasizes again the fact that it is not the cumulative exposure of the vesicle to environmental HCN (H) concentration that most affects the rate of production of adenine, but rather the non-linearity which allows a perturbation to evolve the system into a new production regime (new stationary state).

\begin{figure}[H]
\begin{center}
\includegraphics[width=19 cm]{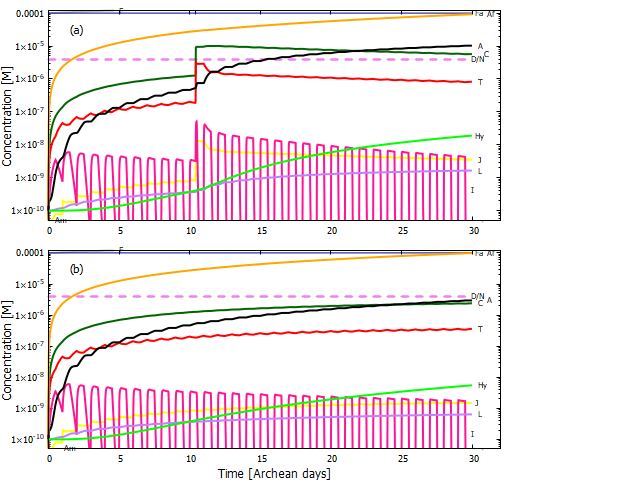} 
\end{center}
\caption{(a) The same as for Fig. \ref{fig:Concen80e-5} (a), 80 $^\circ$C, except with concentrations of HCN (H) at $6\times 10^{-4}$ and formamide (F) and formimidic acid (Fa) at $1\times 10^{-4}$ M. After 30 days, the concentration of adenine (black trace) reaches a value of $1.0\times 10^{-5}$ M. (b) The same as (a) but without perturbation. The adenine concentration reaches $2.9\times 10^{-6}$ M.}
\label{fig:Concen80e-4}
\end{figure}

Figure \ref{fig:Concen80e-5D-10} shows the results with a diffusion exponential four orders of magnitude smaller, at $1.0\times10^{-10}$ cm$^2$ s$^{-1}$. The small diffusion constant allows the coupling of the reactions with diffusion leading to spatial symmetry breaking of the concentration profiles (see also figures \ref{fig:ConProf80e-5D-10-1} and \ref{fig:ConProf80e-5D-10-2}). 

\begin{figure}[H]
\begin{center}
\includegraphics[width=19 cm]{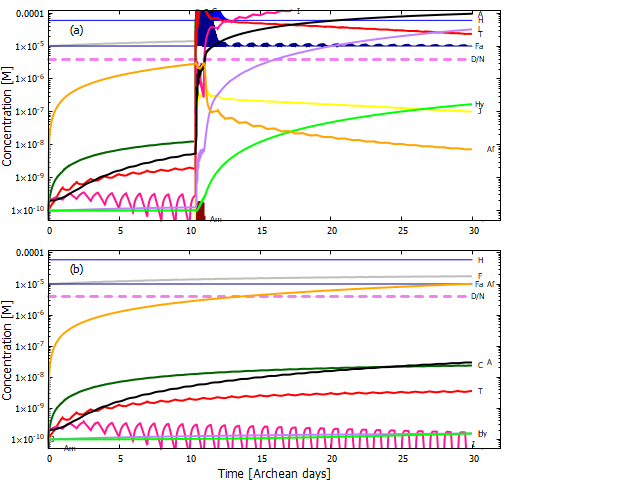} 
\end{center}
\caption{The same as Fig. \ref{fig:Concen80e-5}, 80 $^\circ$C, except with the diffusion exponential four orders of magnitude smaller, $1.0\times10^{-10}$ cm$^2$ s$^{-1}$ (e.g. $D_A=1.0\times10^{-10}$ cm$^2$ s$^{-1}$). Eleven bins in depth $x$ below the ocean surface are plotted until reaching the bottom of the 100 $\mu$m (0.01 cm) vesicle. The top of the vesicle is at a depth of 0.00025 cm below the ocean surface. The small diffusion constant allows spatial symmetry breaking of the concentration profiles. This  results in thicker lines since the 11 different depth bins are plotted in this figure.}
\label{fig:Concen80e-5D-10}
\end{figure}

Figures \ref{fig:ConProf80e-5D-10-1} and \ref{fig:ConProf80e-5D-10-2} plot the  product concentration profiles as a function of depth below the ocean surface for the initial conditions of figure \ref{fig:Concen80e-5D-10} at the time of 10.7 Archean days (5 hours after the perturbation). The coupling of reaction to diffusion leads to a non-homogeneous distribution of products within the vesicle, with a general increase in concentration towards the center of the vesicle. The tendency days after the perturbation is towards homogeneity. Without perturbation, the concentration profiles remain homogeneous over the vesicle.
\begin{figure}[H]
\begin{center}
\includegraphics[width=12 cm]{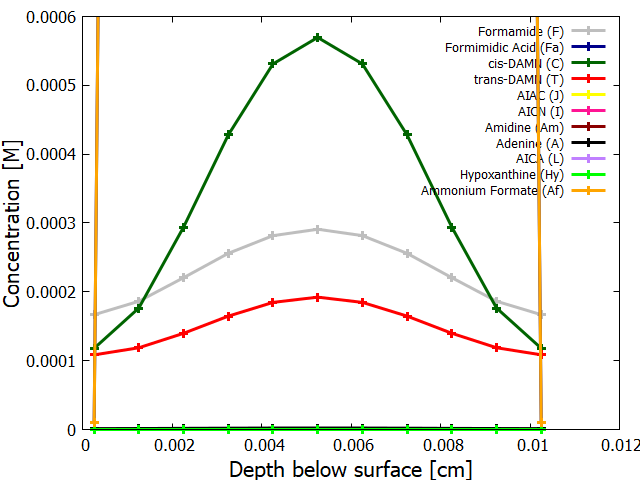}
\end{center}
\caption{The concentration profile of the products as a function of depth below the ocean surface (the top of the vesicle is at a depth of 0.00025 cm below the surface) for the initial conditions of figure \ref{fig:Concen80e-5D-10} and taken at the time of 10.7 Archean days (5 hours after the perturbation). Twelve bins in depth $x$ below the ocean surface are plotted until reaching the bottom of the 100 $\mu$m (0.01 cm) diameter vesicle.}
\label{fig:ConProf80e-5D-10-1}
\end{figure}

\begin{figure}[H]
\begin{center}
\includegraphics[width=12 cm]{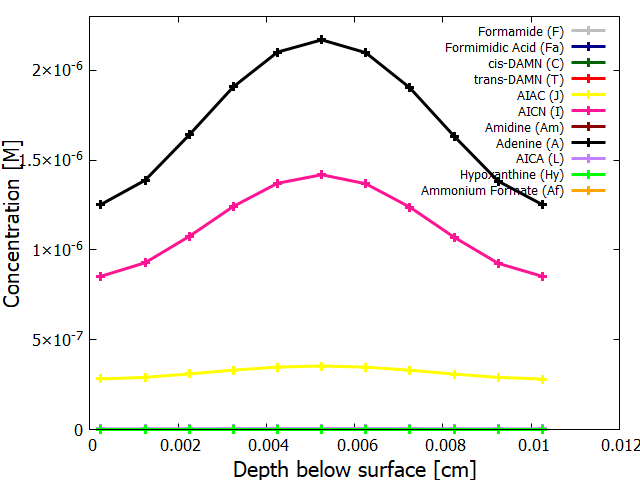}
\end{center}
\caption{The same as figure \ref{fig:ConProf80e-5D-10-1} except with an expanded y-scale to emphasize the products of lesser concentration taken at the time of 10.7 Archean days (5 hours after the perturbation). }
\label{fig:ConProf80e-5D-10-2}
\end{figure}

The temperature dependence of the amount of product molecules produced after 30 Archean days is given in figure \ref{fig:TempDep}. Ammonium formate (Af) is produced by the hydrolysis of first HCN (H) to formamide (F) (reacrion \#1) and then hydrolysis of formamide to Af (reaction \#4). Both of these reactions have high activation energies and this results in Af only being produced in quantities at temperatures greater than 80 $^\circ$C. Most of the adenine (A) production occurs through reaction \#13 which consumes Af and therefore high temperatures are important to the production of adenine for this set of reactions.
\begin{figure}[H]
\begin{center}
\includegraphics[width=16 cm]{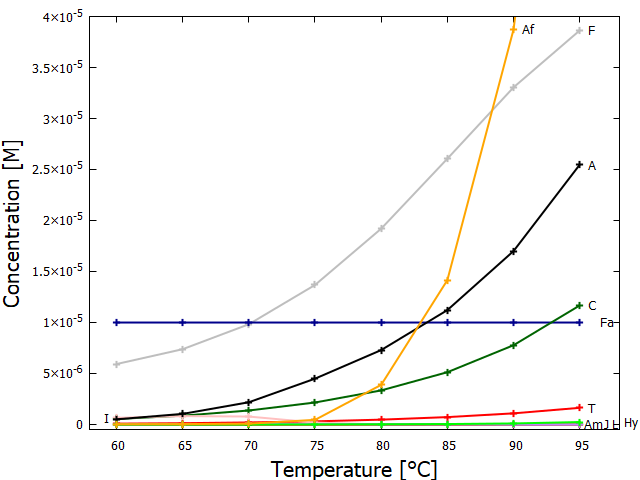}
\end{center}
\caption{The temperature dependence of the concentrations of the product molecules, with the initial conditions, [H]$_0=6\times 10^{-5}$,  [F]$_0=1\times 10^{-5}$,  [Fa]$_0=1\times 10^{-5}$ M, and all other molecules  [Y]$_0=1\times 10^{-10}$ and the diffusion constant $D_A=1.0\times10^{-6}$, with one perturbation at 10.4 Archean days.}
\label{fig:TempDep}
\end{figure}

It is instructive to compare our overall non-equilibrium results obtained with the model of UVC production of adenine from HCN within a lipid vesicle with the quasi-equilibrium experiments of Ferris et al. \cite{FerrisEtAl1978}. Starting with a high 0.1 M concentration of HCN in water (pH 9.2), and allowing this solution to polymerize in the dark at room temperature for 7 months, and then subjecting these polymers to hydrolysis at 110 $^\circ$C for 24 hours, Ferris et al. obtain an adenine yield of 1 mg l$^{-1}$  (equivalent to a concentration of $7.4\times 10^{-6}$ M  - the molar mass of adenine being 135.13 g mol$^{-1}$). Our model gives a similar adenine concentration of $7.3\times 10^{-6}$ M within 30 days (Fig. \ref{fig:Concen80e-5}), starting from a much lower and more realistic initial concentration of HCN of only $6.0\times 10^{-5}$ M (with only one perturbation of 0.1 M for two minutes) and a more realistic neutral pH of 7.0 at 80 $^\circ$C and under a UVC flux integrated from 210 – 280 nm of about 4 W m$^{-2}$ during daylight hours (Fig. \ref{fig:Pigments}). At 90 $^\circ$C adenine concentration increases to $1.7\times 10^{-5}$ M (Fig. \ref{fig:Concen90e-5}).

In figure \ref{fig:EntProd80e-5}(a) I plot the entropy production as a function of time in Archean days due to the photon dissipation by the corresponding molecular concentration profile as represented by reactions 17 to 24 of table \ref{tab:Reactions}. In general, the entropy production is an increasing function of time. These photo-reactions represent the terms $d_JP/dt$ of equation (\ref{eq:EntProdChg}), and even though the terms $d_XP/dt$ which represent the variation of the entropy production due to rearrangement of the chemical affinities (the free forces $X$), are negative definite (corresponding to the structuring of the molecules) consistent with the Glansdorf-Prigogine universal evolutionary criterion, the total entropy production $dP/dt =d_JP/dt + d_XP/dt$ increases. This is due to the the fact that $d_JP/dt$ represents the entropy production due to the chemical/photochemical reactions plus the dissipation of the photons which are flowing through the system and being converted from short wavelength UV into long wavelength infrared (dissipated) light, and this photon flow captured by the system increases over the evolution of the concentration profile of the intermediate products within the vesicle, particularly after the 2 minute perturbation of the system at 10.4 days by allowing it to float through a patch of high concentration of HCN and formimidic acid (Fa) of 0.1 M. In figure \ref{fig:EntProd80e-5}(b) I plot the same entropy production but for the case in which there is no perturbation of the system (the environmental concentrations of HCN and formimidic acid (Fa) are kept constant at $6\times 10^{-5}$  and $1\times 10^{-5}$ M respectively for both cases). In this latter case, the entropy production remains almost 3 orders of magnitude lower.
 
\begin{figure}[H]
\begin{center}
\includegraphics[width=16 cm]{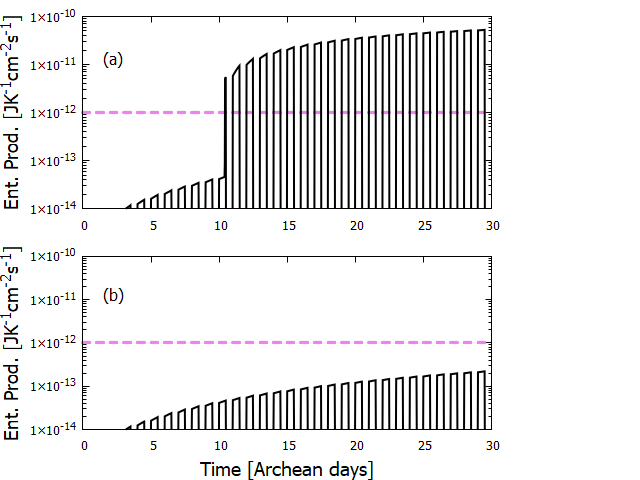}
\end{center}
\caption{(a) The entropy production as a function of time during the UVC photochemical dissipative structuring process leading to adenine within a vesicle floating at a sea surface at a temperature of 80 $^\circ$C. Only the entropy production due to photon dissipation is included. This entropy production increases monotonically as photochemical reactions convert HCN into the photon dissipative product molecules, including adenine. During the day, entropy production is due to the dissipation of the UVC light into heat by the existing product concentration profile. At night, entropy production goes to zero (although thermal chemical reactions still occur during the night, this entropy production is small and not included). At 10.4 Archean days, the system is perturbed for two minutes and the entropy production increases discretely by almost 3 orders of magnitude and remains high. (b) The same but for no perturbation of the system.}
\label{fig:EntProd80e-5}
\end{figure}

\section{Discussion}

Not only do high temperatures increase the build up of product concentration (Fig. \ref{fig:TempDep}) but they would also foment phosphorylation with phosphate salts and formamide, favoring the formation of acyclonucleosides and the phosphorylation and trans-phosphorylation of nucleosides which only occurs efficiently at temperatures above $70$ $^\circ$C \cite{Schoffstall1976,CostanzoEtAl2007}.

Besides entrapment of product molecules inside the vesicle, another concentration mechanism for this system arises through the coupling between reaction and diffusion in the non-linear regime which leads to the breaking of spatial symmetry (e.g. the Belousov-Zhabotinsky reaction \cite{Prigogine1967}). For low diffusion rates, the homogeneous stationary state is no longer stable with respect to a space dependent perturbation and intermediate products become preferentially concentrated at the center. This we find only significant for very low diffusion rates of $ D_A\sim 1\times10^{-10}$ cm$^2$ s$^{-1}$ (Figs. \ref{fig:Concen80e-5D-10}, \ref{fig:ConProf80e-5D-10-1} and \ref{fig:ConProf80e-5D-10-2}).

The other important purine of RNA and DNA, guanine, can be produced from AICA (L) (the hydrolysis product of AICN, reaction \#12 of table \ref{tab:Reactions}), through a thermal reaction with either cyanogen (CN)$_2$ or cyanate (OCN$^-$). Cyanogen can be generated from HCN (H) either photochemically \cite{AireyDainton1966} or thermally \cite{FoxHarada1961}; cyanate is obtained from cyanogen through hydrolysis \cite{SanchezEtAl1967}. The production of guanine from AICA would increase the photon dissipation of the system, as can be seen by comparing the molar extinction coefficients and wavelengths of maximum absorption of these two molecules, and therefore the concentration of guanine would increase, or, in other words, be dissipatively selected for by the same non-linear, non-equilibrium, thermodynamic mechanism of perturbations leading to new states with an increase in the rate of purine production and dissipation, as explained in the previous section and in section \ref{sec:Thermodynamics}.  

Regarding the pyrimidines cytosine, uracil, and thymine, Ferris, Sanchez and Orgel \cite{FerrisEtAl1968} showed that on heating to 100 $^\circ$C a 5:1 ratio of cyanate with cyanoacetylene, cytosine was formed in yields of 19\%. In this reaction, cytosine is formed mainly in a sequence involving the stable intermediate cyanovinylurea. Cyanogen or cyanoformamide can replace cyanate in this synthesis. Cytosine hydrolizes quite readily to uracil, and when uracil is reacted with formic acid in dilute aqueous solutions at 100–140 $^\circ$C, thymine is formed \cite{ChoughuleyEtAl1977}. All of the purines and pyrimidines can therefore be obtained within the same non-equilibrium non-linear vesicle model by assuming only HCN and some acetylene (C$_2$H$_2$) dissolved in a water solvent at high temperature under UVC light.

Inorganic catalysts have not been included but can increase the rate of purine production. For example Cu$^{+2}$ ions have a large effect in increasing the rate constant for the conversion of HCN (H) to cis-DAMN (C) \cite{SanchezEtAl1967} (reaction \# 5). Cu$^{+2}$ ions also reduce the energy difference (but not the barrier crossing height) between the isomers  formimidic and formamidic acid of formamide \cite{LunaEtAl1998}. Metal ions would have been in high abundance at the ocean surface microlayer \cite{Hardy1982,GrammatikaZimmerman2001}.

\section{Summary and Conclusions}

Understanding the process involved in the origin of life requires the delineation of a coherent physical-chemical framework for the various continuous and sustained dissipative processes involved; synthesis, proliferation, and evolution towards molecular complexes of ever greater dissipative efficacy. Early life may have been a particular form of non-equilibrium structuring; {\em microscopic} dissipative structuring of carbon based molecules under Archean UVC light. The synthesized products, the fundamental molecules, were ``self-organized'' pigments which absorbed strongly in the UVC region and were endowed with peaked conical intersections allowing the efficient dissipation of this absorbed light into heat. These dissipative structures attained stability once endowed with a conical intersection to internal conversion as a result of the reduced quantum efficiency for dexcitation through further photochemical reaction pathways. It was, however, not a fortuitous coincidence, nor a requirement for photo-stability, that the fundamental molecules of life have these photochemical characteristics (Fig. \ref{fig:Pigments}) but rather that these characteristics are, in fact, the ``design'' goals of dissipative structuring. Such molecules with peaked conical intersections and presenting broad absorption would then form a basis set of molecules for the subsequent construction of all early dissipative structures and processes of life.  

The initial dissipative structuring at the origin of life must necessarily have occurred in the long wavelength UVC region where there was enough energy to directly break and reform carbon double covalent bonds while not enough energy to disassociate these molecules through successive ionization. Photochemical reactions in this wavelength region provide a much richer suite of microscopic mechanisms for dissipative structuring than do thermal chemical reactions. These mechanisms include tautomerizations, dissasociations, radicalizations, isomerizations, charge transfers, additions, and substitutions. Unlike macroscopic dissipative structures such as hurricanes or convection cells, at normal temperatures these microscopic dissipative structures remain intact even after the removal of the imposed light potential driving their synthesis due to strong covalent bonding between atoms. The corresponding macroscopic dissipative structures are the concentration profiles of these molecules. 

As an example, I presented a simple kinetic model of UVC photochemical reactions, based on published experimental and {\em ab initio} data, for the synthesis of adenine from HCN in water solvent within a lipid vesicle permeable to HCN, H$_2$O and formimidic acid (the photon-tautomerized hydrolysis product of HCN), but impermeable to the reaction products, floating at the surface of a hot ocean and under physical conditions consistent with those offered by the geochemical fossil evidence of the early Archean.

The results presented here indicate that given UVC light continuously incident over a dilute aqueous solution of HCN at high temperature, significant dissipative structuring of adenine will occur, and if this occurs within a lipid vesicle enclosure, significant concentrations of adenine will build up within a short time period. There is no need to begin with large initial concentrations of HCN by invoking low temperature eutectic concentration and there is no need for alkaline conditions in order to favor HCN polymerization over hydrolysis since successive hydrolysis lead to formamide and ammonium formate which are catalysts for the important final step of the attachment of the last HCN to AICN (reaction \# 13). Destruction of adenine through hydrolysis does not compete significantly with its production through this proposed route, and, in fact, provides a route to the synthesis of guanine.

Perturbations caused by the vesicle floating into patches of higher concentration of HCN and formimidic acid that could have existed at isolated regions of the Archean ocean surface microlayer could have provoked the non-linear autocatalytic system into new states of higher adenine productivity. This leads to a discontinuous increase in ``metabolism'' of precursor HCN molecules from the environment transformed into UVC pigment molecules. Evolution is also towards concentration profiles of product molecules with an absorption maximum closer to the peak intensity of the incident UVC spectrum and with peaked conical intersections to internal conversion, both increasing the overall efficacy of dissipation of the incident UVC solar spectrum. The Glansdorff-Prigogine criterion mandating decreasing contributions to the entropy production due to the reorganization of the free forces (the chemical affinities over the temperature) determines stability and guides evolution. Concentration of adenine within our vesicle increased by 5 orders of magnitude, from $10^{-10}$ to $10^{-5}$ M, over the very short period of only 30 Archean days and the {\em total} entropy production, including the most important contribution due to the variation of the flows of energy (photons) through the system, increased by more than 5 orders of magnitude over the same period (figure \ref{fig:EntProd80e-5}).

For very low diffusion rates, there can be significant coupling of reactions with diffusion, leading to non-homogeneous distributions of some of the intermediate products, with greater concentration of these at the center of the vesicle. Such spatial symmetry breaking could facilitate further structuring such as polymerization of the nucleobases into nucleic acid.

Dissipative structuring under light as the fundamental creative force in biology appears to have been ongoing, from the initial dissipation at the UVC wavelengths of the Archean by the fundamental molecules of life, to the dissipation of wavelengths up to the red-edge (700 nm) by the organic pigments of today \cite{Michaelian2009, Michaelian2011, MichaelianSimeonov2015}. Beyond the red-edge, starting at about 1200 nm, water in the ocean surface microlayer absorbs strongly and dissipates photons into heat efficiently. There is, therefore, still a wavelength region between 700 and 1200 nm which remains to be conquered by future evolution of pigments. The simultaneous coupling of biotic with abiotic irreversible processes, such as the water cycle and ocean and air currents, culminating in an efficient global dissipating system known as the biosphere, increases further the efficacy of solar photon dissipation into the far infrared much beyond 1200 nm \cite{Michaelian2009,Michaelian2012}. 

Empirical evidence for selection in nature towards states of increased dissipation exists on vastly different size and time scales. For example, the increase in photon absorption and dissipation efficacy of a plant leaf over its life-cycle \cite{Gates1980}, the proliferation of photon absorbing pigments over the entire Earth surface, the correlation between ecosystem succession and increased dissipation \cite{UlanowiczHannon1987,SchneiderKay1994}, and the general increase of biosphere efficacy in photon dissipation over evolutionary history, including, for example, the plant-induced increases in the water cycle \cite{Kleidon2008, Michaelian2012a} and animal dispersal of nutrients required for pigment synthesis \cite{Michaelian2016}. There is also evidence for this at the microscopic scale, for example in the increased rates of energy dissipation per unit biomass of the living cell over its evolutionary history \cite{Zotin1984}. Here I have suggested how evolutionary increases in dissipation occur even at the nanoscale, i.e. the sequential increase in photon dissipation at each step along the dissipative synthesis of the nucleobases from precursor molecules under a UVC photon potential. 

Any planet around any star giving off light in the long wavelength UVC region, but with protection against shorter wavelength light which could destroy molecules through successive ionization, should therefore have its own concentration profile of dissipatively structured carbon based fundamental molecules (UVC pigments) who's characteristics would depend on the exact nature of the local UV environment and the precursor and solvent molecules available. Examples may include the sulfur containing UV pigments found in the clouds of Venus \cite{LimayeEtAl2018}, the UV absorbing thiophenes \cite{HeinzSchulze2020} and the red chlorophyll-like pigments \cite{Pershin2002} found on the surface of Mars, the UVC and UVB absorbing poly-aromatic hydrocarbons (PAHs) found in the atmosphere and on the surface of Titan \cite{LopezPuertasEtAl2013}, on the surface of asteroids, and in interstellar space \cite{MichaelianSimeonov2015}. The observation that thiophenes and PAHs found on mars, on asteroids, and in space are of generally large size can be understood from within this non-equilibrium thermodynamic perspective since, without the possibility of vibrational dissipation through hydrogen bonding to solvent molecules, these molecules would have ``grown'' to large sizes through dissipative selection in order to support many low frequency vibrational modes which would increase dissipation by pushing the absorbed photon energy towards the infrared. 

Dissipative structuring, dissipative proliferation, and dissipative selection are the necessary and sufficient ingredients for an explanation in physical-chemical terms of the synthesis, proliferation, and evolution of organic molecules on planets, comets, asteroids, and interstellar space \cite{MichaelianSimeonov2015}, and, in particular, for contributing to an understanding of the origin and evolution of life on Earth.

%%%%%%%%%%%%%%%%%%%%%%%%%%%%%%%%%%%%%%%%%%
\vspace{6pt} 

%%%%%%%%%%%%%%%%%%%%%%%%%%%%%%%%%%%%%%%%%%
%

\noindent {\bf \Large Acknowledgments\\}
The author is grateful to Carlos Bunge, Iván Santamaría-Holek, and anonymous referees for their revision of, and suggestions on, the manuscript. This research was funded by DGAPA-UNAM grant number IN104920.

\appendix
\section{Relation to Evolution through Statistical Mechanical Fluctuation Theorems Employing Linear Stability Analysis}
\label{sec:FlucTheo}
Since some recent works have considered the evolution of driven dissipative (non-equilibrium) systems from a more restricted statistical mechanical framework employing fluctuation theorems and linear stability analysis, here I establish the relationship between this statistical linear theory and the more complete thermodynamic analysis employing CIT theory as given in section \ref{sec:Thermodynamics}. In fact, Glansdorff and Prigogine performed exactly this comparison for chemical reactions in their 1971 book ``Thermodynamic Theory of Structure, Stability and Fluctuations'' \cite{GlansdorffPrigogine1971}. As emphasized by Glansdorff and Prigogine, statistical fluctuation theorems with linear stability analysis is only a simplified caricature of CIT formalism, corresponding to queries only in the neighborhood of a stationary state (small fluctuations). As we have seen in section \ref{sec:Thermodynamics}, for non-linear systems, the new stable states which can be reached after a fluctuation depend on the type of fluctuation (for example, the paths in generalized phase space on the excited potential energy surface of molecules proceeding through a conical intersection leading to a particular photochemical reaction). Not all fluctuations have the same outcome, and therefore the direction of evolution of the system can depend strongly on the particular fluctuation. Therefore, both stochastic and deterministic components must be included in a proper analysis of instabilities around a stationary state for a real non-linear system, and it is precisely this which gives a {\em history} to such systems. Therefore, contrary to what has been claimed regarding these systems in the recent literature, a statistical analysis employing fluctuation theorems and linear stability theory is not sufficient to describe the evolution of a non-linear dissipative system. There are, however, a few general results for {\em average} variables and small fluctuations about the stationary state and the connection of these results to CIT theory is made clear in this appendix.
 
The the relation between the probability of a particular fluctuation occurring and entropy production was first considered by Einstein \cite{Einstein1905} who showed that for a Markovian and ergodic system and under a Gaussian distribution of fluctuations, the probability $P$ of a fluctuation of any size at the equilibrium state was related to the change in entropy $\Delta S$ afforded to the system by
\begin{equation}
P \propto \exp[\Delta S/k_B],
\label{eq:probfluc}
\end{equation}
where $k_B$ is the Boltzmann constant. Since at equilibrium entropy $S$  is maximum, $\Delta S$ due to a fluctuation must be negative, and thus the probabilities for fluctuations which lead the system away from equilibrium towards smaller entropy become exponentially smaller with the size of the decrease in entropy. Expanding the entropy around the equilibrium state gives,
\begin{equation}
S =  S_{eq} + (\delta S)_{eq}+{1 \over 2} (\delta^2 S)_{eq} + \dots
\end{equation}
For an isolated system at equilibrium $S$ is a maximum, so $(\delta S)_{eq} = 0$, and assuming {\em small  fluctuations} $\Delta S$, gives $\Delta S = S - S_{eq} \approx 1/2 (\delta^2 S)_{eq}<0$, (ignoring higher order terms in the expansion) and equation (\ref{eq:probfluc}) can be rewritten as
\begin{equation}
P \propto \exp[(\delta^2 S)_{eq}/2k_B].
\label{eq:probfluc2}
\end{equation}
The quantity $(\delta^2 S)_{eq}$ is known as the {\em excess entropy} (due to a small fluctuation around equilibrium). Since $(\delta S)_{eq}=0$ and $(\delta^2 S)_{eq}<0$, $(\delta S)_{eq}$ is a Lyapunov function which establishes the local stability of the equilibrium state, which (as well as global stability) is a well known property. 

This equilibrium Fluctuation Theorem was realized to also apply to non-equilibrium situations as long as the time scales associated with the fluctuations acting on the system are much shorter than the time scales associated with changes in the external boundary conditions. This was developed in detail by Onsager in 1931 \cite{Onsager1931a,Onsager1931b}, and a few decades later by Callen and Welton \cite{CallenWelton1951}, Onsager and Machlup \cite{OnsagerMachlup1953}, Kubo \cite{Kubo1966}, Prigogine and Nicolis \cite{PrigogineNicolis1971} and more recently by Evans et al. \cite{EvansEtAl1993} and Evans and Searles \cite{EvansSearles2002}. In particular, Prigogine and Nicolis \cite{PrigogineNicolis1971} made the theorem quantitative for non-equilibrium stationary states by extending Einstein's result to give \cite{GlansdorffPrigogine1971},
\begin{equation}
P \propto \exp[(\delta^2S)_0/2k_B]                                      
\end{equation}
where the excess entropy $(\delta^2S)_0$  is now calculated  around  a stationary non-equilibrium state. 

The work of Onsager, Prigogine, and Nicolis was generalized by Evans et al. \cite{EvansEtAl1993} and later by Gaspard and Andrieux \cite{Gaspard2013,Gaspard2004,AndrieuxGaspard2007,AndrieuxEtAl2009} and given the name the ``Stationary State Current-Fluctuation Theorem''. According to this theorem, the probability $P$ of observing a set of flows ${\bf J}_\alpha$ in the long time limit $\ t \rightarrow \infty$ (at the stationary state) with respect to that of observing their time reversed flows $-{\bf J}_\alpha$ is given by \cite{EvansEtAl1993}, 
\begin{equation}
{P({\bf J}_\alpha) \over P(-{\bf J}_\alpha)} \approx \exp\left[{{{\bf A}_\alpha \cdot {\bf J}_\alpha \over k_BT}}\cdot Vt\right] = \exp\left[{d_iS/dt \over k_B}\cdot t\right], \ for\ \ t \rightarrow \infty,
\label{eq:CurrentFluc}
\end{equation}
where ${\bf A}_\alpha$ are the set of affinities (e.g. dependent on the molecular concentrations), ${\bf J}_\alpha$ are the flows (e.g. the reaction rates), and $V$ is the volume. Equation (\ref{eq:CurrentFluc}) is valid for the equilibrium as well as the nonequilibrium stationary states, independently of the values of the affinities, for Markovian or semi-Markovian stochastic processes, but only if the large-deviation properties of the process are well defined in the long-time limit. The last term of equation (\ref{eq:CurrentFluc}) is derived from equation (\ref{eq:entprod}) section \ref{sec:Thermodynamics}, under the assumption of local equilibrium, and is just the exponential of the entropy production attributed to the set of flows divided by the Boltzmann constant $k_B$ times the observational time $t$. Therefore, given the possibility of two (or more) sets of flows ${\bf J}_\alpha$, ${\bf J}_\beta, ...$ corresponding to two (or more) sets of free affinities ${\bf A}_\alpha$, ${\bf A}_\beta, ...$ leading to two (or more) different values of the entropy production, current fluctuations which lead the system towards the states of greater dissipation are generally favored on average. As with the Fluctuation Theorem, the Current-Fluctuation Theorem is not limited in its validity to macroscopic systems (the thermodynamic limit) nor is it limited to systems in local equilibrium. Local thermodynamic equilibrium is required only to validate the concept of entropy density in non-equilibrium situations and thus to associate dissipation with entropy production (the last term of Eq. (\ref{eq:CurrentFluc})).

The problem for our system operating in the non-linear regime where multiple stationary states are possible, however, is that, dependent on the size and type of the external perturbations, the large-deviation (large fluctuation) properties of the process may not be well defined in the long-time limit.  As mentioned above, to determine the evolution of our system, both stochastic and deterministic aspects have to be considered. 

Therefore, in the general case of non-linear systems, for whatever sized fluctuations about the stationary state, the probability of evolution, constrained by the Glansdorff-Prigogine criterion, equation (\ref{eq:GPC}) section \ref{sec:Thermodynamics}, from one stationary state to another has to be evaluated through a local stability analysis which ultimately concerns the size of the fluctuations, the size of the barriers in $d_XP$, and the size of the catchment basins of neighboring stationary states in a generalized phase space. For example, concentration profiles of the intermediate or final molecules in our reaction scheme which have the most conical intersections leading to internal conversion (direct and rapid photon dissipation into heat), rather than leading to other photochemical reactions, will be those with the largest catchment basin in this generalized phase space for large fluctuations and will thus become the most probable in the long time limit. This more general stochastic+deterministic rule for the evolution among stationary states of non-linear dissipative systems, I have termed {\em thermodynamic selection} \cite{Michaelian2009,Michaelian2011,Michaelian2016} or more precisely as {\em dissipative selection} in order to emphasize its nature as a trend towards selection of  stationary states with ever greater dissipative efficacy.

%\bibliographystyle{abbrvnat}
%\bibliographystyle{unsrtnat}

%\bibliography{Collection2017-5}

\begin{thebibliography}{156}
\providecommand{\natexlab}[1]{#1}
\providecommand{\url}[1]{\texttt{#1}}
\expandafter\ifx\csname urlstyle\endcsname\relax
  \providecommand{\doi}[1]{doi: #1}\else
  \providecommand{\doi}{doi: \begingroup \urlstyle{rm}\Url}\fi

\bibitem[Anders(1989)]{Anders1989}
E.~Anders.
\newblock Pre-biotic organic matter from comets and asteroids.
\newblock \emph{Nature}, 342:\penalty0 255–257, 1989.
\newblock URL \url{https://doi.org/10.1038/342255a0}.

\bibitem[Chyba et~al.(1990)Chyba, Thomas, Brookshaw, and Sagan]{ChybaEtAl1990}
C.~F. Chyba, P.J. Thomas, L.~Brookshaw, and C.~Sagan.
\newblock Cometary delivery of organic molecules to the early earth.
\newblock \emph{Science}, 249:\penalty0 366--373, 1990.

\bibitem[Chyba and Sagan(1992)]{ChybaSagan1992}
C.~F. Chyba and C.~Sagan.
\newblock Endogenous production, exogenous delivery and impact-shock synthesis
  of organic molecules: an inventory for the origins of life.
\newblock \emph{Nature}, 355:\penalty0 125–132, 1992.
\newblock URL \url{https://doi.org/10.1038/355125a0}.

\bibitem[Brack(1998)]{Brack1998}
A.~Brack, editor.
\newblock \emph{The Molecular Origins of Life: Assembling Pieces of the
  Puzzle}.
\newblock Cambridge University Press, 1998.
\newblock \doi{10.1017/CBO9780511626180}.

\bibitem[Sutherland(2017)]{Sutherland2017}
J.~Sutherland.
\newblock Opinion: Studies on the origin of life — the end of the beginning.
\newblock \emph{Nat Rev Chem}, 1:\penalty0 0012, 2017.
\newblock \doi{https://doi.org/10.1038/s41570-016-0012}.

\bibitem[Miller et~al.(1976)Miller, Urey, and Oró]{MillerEtAl1973}
S.~L. Miller, H.~C. Urey, and J.~Oró.
\newblock Origin of organic compounds on the primitive earth and in meteorites.
\newblock \emph{J. Mol. Evol.}, 9:\penalty0 59--72, 1976.

\bibitem[Oró et~al.(1990)Oró, S.L., and Lazcano]{OroEtAl1990}
J.~Oró, Miller S.L., and A.~Lazcano.
\newblock The origin and early evolution of life on earth.
\newblock \emph{Ann. Rev. Earth Planet. Sci.}, 18:\penalty0 317--356, 1990.

\bibitem[Cleaves and Miller(2007)]{CleavesMiller2007}
HJ~Cleaves and SL~Miller.
\newblock Organic chemistry on the primitive earth and beyond.
\newblock \emph{Systems biology}, 1, 2007.

\bibitem[Schr\"odinger(1944)]{Schrodinger1944}
Erwin Schr\"odinger.
\newblock \emph{What is Life? The Physical Aspect of the Living Cell}.
\newblock Cambridge University Press, 1944.

\bibitem[Moberg(2020)]{Moberg2019}
Christina Moberg.
\newblock Schrödinger's what is life?—the 75th anniversary of a book that
  inspired biology.
\newblock \emph{Angewandte Chemie International Edition}, 59\penalty0
  (7):\penalty0 2550--2553, 2020.
\newblock \doi{https://doi.org/10.1002/anie.201911112}.
\newblock URL
  \url{https://onlinelibrary.wiley.com/doi/abs/10.1002/anie.201911112}.

\bibitem[Glansdorff(1977)]{Glansdorff1977}
P.~Glansdorff.
\newblock \emph{Living Systems as Energy Converters}, chapter Energetic
  evolution of complex network of reactions.
\newblock North-Holland Publishing Company, 1977.

\bibitem[Michaelian(2009)]{Michaelian2009}
K.~Michaelian.
\newblock Thermodynamic origin of life.
\newblock \emph{ArXiv}, 2009.
\newblock \doi{10.5194/esd-2-37-2011}.
\newblock URL \url{http://arxiv.org/abs/ 0907.0042}.

\bibitem[Michaelian(2011)]{Michaelian2011}
K.~Michaelian.
\newblock {Thermodynamic dissipation theory for the origin of life}.
\newblock \emph{Earth Syst. Dynam.}, 224:\penalty0 37--51, 2011.

\bibitem[Michaelian and Simeonov(2015)]{MichaelianSimeonov2015}
K.~Michaelian and A.~Simeonov.
\newblock { Fundamental molecules of life are pigments which arose and
  co-evolved as a response to the thermodynamic imperative of dissipating the
  prevailing solar spectrum}.
\newblock \emph{Biogeosciences}, 12:\penalty0 4913--4937, 2015.

\bibitem[Michaelian(2016)]{Michaelian2016}
Karo Michaelian.
\newblock \emph{{Thermodynamic Dissipation Theory of the Origina and Evolution
  of Life: Salient characteristics of RNA and DNA and other fundamental
  molecules suggest an origin of life driven by UV-C light}}.
\newblock Self-published. Printed by CreateSpace. Mexico City.
  ISBN:9781541317482., 2016.

\bibitem[Michaelian(2017)]{Michaelian2017}
K.~Michaelian.
\newblock Microscopic dissipative structuring and proliferation at the origin
  of life.
\newblock \emph{Heliyon}, 3:\penalty0 e00424, 2017.
\newblock \doi{10.1016/j.heliyon.2017.e00424}.

\bibitem[Michaelian(2018)]{Michaelian2018}
Karo Michaelian.
\newblock Homochirality through photon-induced denaturing of rna/dna at the
  origin of life.
\newblock \emph{Life}, 8\penalty0 (2), 2018.
\newblock ISSN 2075-1729.
\newblock \doi{10.3390/life8020021}.
\newblock URL \url{http://www.mdpi.com/2075-1729/8/2/21}.

\bibitem[Michaelian and Santillan(2019)]{MichaelianSantillan2019}
K.~Michaelian and N.~Santillan.
\newblock Uvc photon-induced denaturing of dna: A possible dissipative route to
  archean enzyme-less replication.
\newblock \emph{Heliyon}, 5:\penalty0 e01902, 2019.
\newblock URL \url{https://www.heliyon.com/article/e01902}.

\bibitem[Michaelian and Rodriguez(2019)]{MichaelianRodriguez2019}
K.~Michaelian and O.~Rodriguez.
\newblock Prebiotic fatty acid vesicles through photochemical dissipative
  structuring.
\newblock \emph{Revista Cubana de Química}, 31\penalty0 (3):\penalty0
  354--370, 2019.

\bibitem[Mejía~Morales and Michaelian(2020)]{MejiaMichaelian2020}
Julián Mejía~Morales and Karo Michaelian.
\newblock Photon dissipation as the origin of information encoding in rna and
  dna.
\newblock \emph{Entropy}, 22\penalty0 (9), 2020.
\newblock ISSN 1099-4300.
\newblock \doi{10.3390/e22090940}.
\newblock URL \url{https://www.mdpi.com/1099-4300/22/9/940}.

\bibitem[Prigogine et~al.(1972{\natexlab{a}})Prigogine, Nicolis, and
  Babloyantz]{PrigogineEtAl1972}
I.~Prigogine, G.~Nicolis, and A.~Babloyantz.
\newblock Thermodynamics of evolution.
\newblock \emph{Physics Today}, 25:\penalty0 23--28, 1972{\natexlab{a}}.
\newblock URL \url{https://doi.org/10.1063/1.3071090}.

\bibitem[Babloyantz and Hiernaux(1975)]{BabloyantzHiernaux1975}
A.~Babloyantz and J.~Hiernaux.
\newblock Models for cell differentiation and generation of polarity in
  diffusion-governed morphogenetic fields.
\newblock \emph{Bulletin of Mathematical Biology}, 37:\penalty0 637 -- 657,
  1975.
\newblock ISSN 0092-8240.
\newblock \doi{https://doi.org/10.1016/S0092-8240(75)80051-6}.
\newblock URL
  \url{http://www.sciencedirect.com/science/article/pii/S0092824075800516}.

\bibitem[Michaelian(2005)]{Michaelian2005}
K.~Michaelian.
\newblock Thermodynamic stability of ecosystems.
\newblock \emph{Journal of Theoretical Biology}, 237\penalty0 (3):\penalty0 323
  -- 335, 2005.
\newblock ISSN 0022-5193.
\newblock \doi{https://doi.org/10.1016/j.jtbi.2005.04.019}.
\newblock URL
  \url{http://www.sciencedirect.com/science/article/pii/S0022519305001839}.

\bibitem[Kleidon et~al.(2010)Kleidon, Malhi, and Cox]{KleidonEtAl2010}
Axel Kleidon, Yadvinder Malhi, and Peter~M. Cox.
\newblock Maximum entropy production in environmental and ecological systems.
\newblock \emph{Philosophical Transactions of the Royal Society B: Biological
  Sciences}, 365\penalty0 (1545):\penalty0 1297--1302, 2010.
\newblock \doi{10.1098/rstb.2010.0018}.
\newblock URL
  \url{https://royalsocietypublishing.org/doi/abs/10.1098/rstb.2010.0018}.

\bibitem[Prigogine and Nicolis(1971)]{PrigogineNicolis1971}
I.~Prigogine and G.~Nicolis.
\newblock Biological order, structure and instabilities.
\newblock \emph{Quarterly Reviews of Biophysic}, 4:\penalty0 107--144, 1971.

\bibitem[Prigogine et~al.(1972{\natexlab{b}})Prigogine, Nicolis, and
  Babloyantz]{Prigogine2EtAl1972}
I.~Prigogine, G.~Nicolis, and A.~Babloyantz.
\newblock Thermodynamics of evolution.
\newblock \emph{Physics Today}, 25:\penalty0 38--44, 1972{\natexlab{b}}.
\newblock \doi{10.1063/1.3071140}.
\newblock URL \url{https://doi.org/10.1063/1.3071140}.

\bibitem[Kleidon(2009)]{Kleidon2009}
Axel Kleidon.
\newblock {Maximum entropy production and general trends in biospheric
  evolution}.
\newblock \emph{Paleontological Journal}, \penalty0 (Biosphere Origin and
  Evolution), 2009.

\bibitem[Michaelian(2012{\natexlab{a}})]{Michaelian2012}
K.~Michaelian.
\newblock Biological catalysis of the hydrological cycle: lifes thermodynamic
  function.
\newblock \emph{Hydrol. Earth Syst. Sci.}, 16:\penalty0 2629--2645,
  2012{\natexlab{a}}.
\newblock \doi{10.5194/hess-16-2629-2012}.
\newblock URL \url{www.hydrol-earth-syst-sci.net/16/2629/2012/}.

\bibitem[Michaelian(2012{\natexlab{b}})]{Michaelian2012a}
K.~Michaelian.
\newblock \emph{The Biosphere}, chapter The biosphere: A thermodynamic
  imperative.
\newblock INTECH, 2012{\natexlab{b}}.
\newblock ISBN 979-953-307-504-3.

\bibitem[Michaelian and Simeonov(2017)]{MichaelianSimeonov2017}
K.~Michaelian and A.~Simeonov.
\newblock Thermodynamic explanation of the cosmic ubiquity of organic pigments.
\newblock \emph{Astrobiol. Outreach}, 5:\penalty0 156, 2017.

\bibitem[Prigogine(1967)]{Prigogine1967}
I.~Prigogine.
\newblock \emph{{Introduction to Thermodynamics Of Irreversible Processes}}.
\newblock John Wiley \& Sons, third edition, 1967.

\bibitem[Boltzmann(1974)]{Boltzmann1886}
L.~Boltzmann.
\newblock \emph{Ludwig Boltzmann: Theoretical physics and philosophical
  problems: Selected writings}.
\newblock 1974.

\bibitem[Berkner and Marshall(1964)]{BerknerMarshall1964}
L.V. Berkner and L.C. Marshall.
\newblock \emph{Origin and Evolution of the Oceans and Atmosphere}, pages
  102--126.
\newblock J. Wiley and Sons, 1964.

\bibitem[Sagan(1973)]{Sagan1973}
C.~Sagan.
\newblock {Ultraviolet Selection Pressure on the Earliest Organisms}.
\newblock \emph{J. Theor. Biol.}, 39:\penalty0 195--200, 1973.

\bibitem[Cnossen et~al.(2007)Cnossen, Sanz-Forcada, Favata, Zegers, and
  Arnold]{CnossenEtAl2007}
I.~Cnossen, J.~Sanz-Forcada, O.~Favata, F.and~Witasse, T.~Zegers, and N.~F.
  Arnold.
\newblock The habitat of early life: Solar x-ray and uv radiation at earth’s
  surface 4–3.5 billion years ago.
\newblock \emph{J. Geophys. Res.}, 112:\penalty0 E02008, 2007.
\newblock \doi{10.1029/2006JE002784}.

\bibitem[Schuurman and Stolow(2018)]{SchuurmanStolow2018}
M.~S. Schuurman and A.~Stolow.
\newblock Dynamics at conical intersections.
\newblock \emph{Annu. Rev. Phys. Chem.}, 69:\penalty0 427--450, 2018.

\bibitem[Ferris and Orgel(1966)]{FerrisOrgel1966}
J.~P. Ferris and L.~E. Orgel.
\newblock An unusual photochemical rearrangement in the synthesis of adenine
  from hydrogen cyanide.
\newblock \emph{J. Am. Chem. Soc.}, 88:\penalty0 1074--1074, 1966.

\bibitem[Sagan and Khare(1971)]{SaganKhare1971}
Carl Sagan and Bishun~N. Khare.
\newblock Long-wavelength ultraviolet photoproduction of amino acids on the
  primitive earth.
\newblock \emph{Science}, 173\penalty0 (3995):\penalty0 417--420, 1971.
\newblock ISSN 0036-8075.
\newblock \doi{10.1126/science.173.3995.417}.
\newblock URL \url{https://science.sciencemag.org/content/173/3995/417}.

\bibitem[Ruiz-Bermejo et~al.(2013)Ruiz-Bermejo, Zorzano, and
  Osuna-Esteban]{Ruiz-BermejoEtAl2013}
M.~Ruiz-Bermejo, M.~P. Zorzano, and S.~Osuna-Esteban.
\newblock Simple organics and biomonomers identified in hcn polymers: An
  overview.
\newblock \emph{Life}, 3:\penalty0 421--448, 2013.
\newblock \doi{10.3390/life3030421}.

\bibitem[Cleaves and Miller(1998)]{CleavesMiller1998}
H.~James Cleaves and Stanley~L. Miller.
\newblock Oceanic protection of prebiotic organic compounds from uv radiation.
\newblock \emph{Proceedings of the National Academy of Sciences}, 95\penalty0
  (13):\penalty0 7260--7263, 1998.
\newblock ISSN 0027-8424.
\newblock \doi{10.1073/pnas.95.13.7260}.
\newblock URL \url{https://www.pnas.org/content/95/13/7260}.

\bibitem[Cockell(2000)]{Cockell2000}
Charles~S. Cockell.
\newblock The ultraviolet history of the terrestrial planets — implications
  for biological evolution.
\newblock \emph{Planetary and Space Science}, 48\penalty0 (2):\penalty0 203 --
  214, 2000.
\newblock ISSN 0032-0633.
\newblock \doi{https://doi.org/10.1016/S0032-0633(99)00087-2}.
\newblock URL
  \url{http://www.sciencedirect.com/science/article/pii/S0032063399000872}.

\bibitem[Mulkidjanian et~al.(2003)Mulkidjanian, Cherepanov, and
  Galperin]{MulkidjanianEtAl2003}
Armen~Y. Mulkidjanian, Dmitry~A. Cherepanov, and Michael~Y. Galperin.
\newblock Survival of the fittest before the beginning of life: selection of
  the first oligonucleotide-like polymers by uv light.
\newblock \emph{BMC Evolutionary Biology}, 3\penalty0 (1):\penalty0 12, May
  2003.
\newblock ISSN 1471-2148.
\newblock \doi{10.1186/1471-2148-3-12}.
\newblock URL \url{https://doi.org/10.1186/1471-2148-3-12}.

\bibitem[Oparin(1924)]{Oparin1924}
A.~I. Oparin.
\newblock \emph{Proiskhozhdenie zhizny (The origin of life)}.
\newblock Weidenfeld and Nicholson, 1924.

\bibitem[S.(1929)]{Haldane1929}
Haldane J.~B. S.
\newblock Origin of life.
\newblock \emph{Rationalist Annual}, 148:\penalty0 3–10, 1929.

\bibitem[Urey(1952)]{Urey1952}
Harold~C. Urey.
\newblock On the early chemical history of the earth and the origin of life.
\newblock \emph{PNAS}, 38\penalty0 (4):\penalty0 351--363, 1952.
\newblock ISSN 0027-8424.
\newblock \doi{10.1073/pnas.38.4.351}.
\newblock URL \url{https://www.pnas.org/content/38/4/351}.

\bibitem[Sagan(1957)]{Sagan1957}
C.~Sagan.
\newblock Radiation and the origin of the gene.
\newblock \emph{Evolution}, 11:\penalty0 40--55, 1957.

\bibitem[Baly et~al.(1927)Baly, Stephen, and Hood]{BalyEtAl1927}
Edward Charles~Cyril Baly, W.~E. Stephen, and N.~R. Hood.
\newblock The photosynthesis of naturally occurring compounds.\&\#x2014;ii. the
  photosynthesis of carbohydrates from carbonic acid by means of visible light.
\newblock \emph{Proceedings of the Royal Society of London. Series A,
  Containing Papers of a Mathematical and Physical Character}, 116\penalty0
  (773):\penalty0 212--219, 1927.
\newblock \doi{10.1098/rspa.1927.0132}.
\newblock URL
  \url{https://royalsocietypublishing.org/doi/abs/10.1098/rspa.1927.0132}.

\bibitem[Miller(1953)]{Miller1953}
Stanley~L. Miller.
\newblock A production of amino acids under possible primitive earth
  conditions.
\newblock \emph{Science}, 117\penalty0 (3046):\penalty0 528--529, 1953.
\newblock ISSN 0036-8075.
\newblock \doi{10.1126/science.117.3046.528}.
\newblock URL \url{https://science.sciencemag.org/content/117/3046/528}.

\bibitem[Oró and Kimball(1962)]{OroKimball1962}
J.~Oró and A.P. Kimball.
\newblock Synthesis of purines under possible primitive earth conditions: Ii.
  purine intermediates from hydrogen cyanide.
\newblock \emph{Archives of Biochemistry and Biophysics}, 96\penalty0
  (2):\penalty0 293 -- 313, 1962.
\newblock ISSN 0003-9861.
\newblock \doi{https://doi.org/10.1016/0003-9861(62)90412-5}.
\newblock URL
  \url{http://www.sciencedirect.com/science/article/pii/0003986162904125}.

\bibitem[Ponnamperuma et~al.(1963{\natexlab{a}})Ponnamperuma, Sagan, and
  Mariner]{PonnamperumaEtAl1963}
C.~Ponnamperuma, C.~Sagan, and R.~Mariner.
\newblock Synthesis of adenosine triphosphate under possible primitive earth
  conditions.
\newblock \emph{Nature}, 199:\penalty0 222--226, 1963{\natexlab{a}}.

\bibitem[Ponnamperuma and Mariner(1963)]{PonnamperumaEtAl1963b}
Ponnamperuma and R.~Mariner.
\newblock Formation of ribose and deoxyribose by ultraviolet irradiation of
  formaldehyde in water.
\newblock \emph{Rad. Res.}, 19:\penalty0 183, 1963.

\bibitem[Ponnamperuma et~al.(1963{\natexlab{b}})Ponnamperuma, Mariner, and
  Sagan]{PonnamperumaEtAl1963c}
Ponnamperuma, R.~Mariner, and C.~Sagan.
\newblock Formation of adenosine by ultraviolet irradiation of a solution of
  adenine and ribose.
\newblock \emph{Nature}, 198:\penalty0 1199--1200, 1963{\natexlab{b}}.

\bibitem[Mej{\'i}a and Michaelian(2018)]{MejiaMichaelian2018}
J.~Mej{\'i}a and K.~Michaelian.
\newblock Information encoding in nucleic acids through a
  dissipation-replication relation.
\newblock \emph{ArXiv}, 2018.
\newblock \doi{10.3390/e22090940}.
\newblock URL \url{https://arxiv.org/abs/1804.05939}.

\bibitem[Yarus et~al.(2009)Yarus, Widmann, and Knight]{YarusEtAl2009}
Michael Yarus, Joseph Widmann, and Rob Knight.
\newblock {RNA-Amino Acid Binding: A Stereochemecal Era for the Genetic Code}.
\newblock \emph{J Mol Evol}, 69:\penalty0 406--429, 2009.
\newblock \doi{10.1007/s00239-009-9270-1}.

\bibitem[Vitas and Dobovi{\v{s}}ek(2018)]{VitasDobovisek2018}
Marko Vitas and Andrej Dobovi{\v{s}}ek.
\newblock In the beginning was a mutualism - on the origin of translation.
\newblock \emph{Origins of Life and Evolution of Biospheres}, Apr 2018.
\newblock ISSN 1573-0875.
\newblock \doi{10.1007/s11084-018-9557-6}.
\newblock URL \url{https://doi.org/10.1007/s11084-018-9557-6}.

\bibitem[Rayleigh(1873)]{Rayleigh1873}
John~W. Rayleigh.
\newblock Some general theorems relating to vibrations.
\newblock \emph{Proc. Math. Soc. London}, 4:\penalty0 357--368, 1873.

\bibitem[Glansdorff and Prigogine(1971)]{GlansdorffPrigogine1971}
P.~Glansdorff and I.~Prigogine.
\newblock \emph{Thermodynamic Theory of Structure, Stability and Fluctuations}.
\newblock Wiley - Interscience., 1971.

\bibitem[Orr-Ewing(2014)]{Orr-Ewing}
A.~Orr-Ewing.
\newblock Reaction dynamics –relaxation pathways.
\newblock \emph{Lecture Notes}, pages 1--36, 2014.
\newblock URL \url{http://iramis.cea.fr/meetings/MTS2/publies/Orr-Ewing MOLIM
  lecture 2 for publication.pdf}.

\bibitem[Roberts et~al.(2014)Roberts, Marroux, Grubb, Ashfold, and
  Orr-Ewing]{RobertsEtAl2014}
Gareth~M. Roberts, Hugo J.~B. Marroux, Michael~P. Grubb, Michael N.~R. Ashfold,
  and Andrew~J. Orr-Ewing.
\newblock On the participation of photoinduced n–h bond fission in aqueous
  adenine at 266 and 220 nm: A combined ultrafast transient electronic and
  vibrational absorption spectroscopy study.
\newblock \emph{The Journal of Physical Chemistry A}, 118\penalty0
  (47):\penalty0 11211--11225, 2014.
\newblock \doi{10.1021/jp508501w}.
\newblock URL \url{https://doi.org/10.1021/jp508501w}.
\newblock PMID: 25296392.

\bibitem[Kleinermanns et~al.(2013)Kleinermanns, Nachtigallová, and
  de~Vries]{KleinermannsEtAl2013}
Karl Kleinermanns, Dana Nachtigallová, and Mattanjah~S. de~Vries.
\newblock Excited state dynamics of dna bases.
\newblock \emph{International Reviews in Physical Chemistry}, 32\penalty0
  (2):\penalty0 308--342, 2013.
\newblock \doi{10.1080/0144235X.2012.760884}.
\newblock URL \url{https://doi.org/10.1080/0144235X.2012.760884}.

\bibitem[Barbatti et~al.(2010)Barbatti, Aquino, Szymczak, Nachtigallová,
  Hobza, and Lischka]{BarbattiEtAl2010}
M.~Barbatti, A.J. Aquino, J.J. Szymczak, D.~Nachtigallová, P.~Hobza, and
  H.~Lischka.
\newblock Relaxation mechanisms of uv-photoexcited dna and rna nucleobases.
\newblock \emph{Proc Natl Acad Sci U S A}, 107\penalty0 (50):\penalty0
  21453--21458, 2010.
\newblock \doi{doi:10.1073/pnas.1014982107}.

\bibitem[Polli et~al.(2010)Polli, Altoè, Weingart, Spillane, Manzoni, Brida,
  Tomasello, Orlandi, Kukura, Mathies, Garavelli, and Cerullo]{PolliEtAl2010}
Dario Polli, Piero Altoè, Oliver Weingart, Katelyn~M. Spillane, Cristian
  Manzoni, Daniele Brida, Gaia Tomasello, Giorgio Orlandi, Philipp Kukura,
  Richard~A. Mathies, Marco Garavelli, and Giulio Cerullo.
\newblock Conical intersection dynamics of the primary photoisomerization event
  in vision.
\newblock \emph{Nature}, 467:\penalty0 440--443, 2010.
\newblock \doi{10.1038/nature09346}.
\newblock URL \url{https://doi.org/10.1038/nature09346}.

\bibitem[Serrano-Perez et~al.(2013)Serrano-Perez, de~Vleeschouwer, de~Proft,
  Mendive-Tapia, Bearpark, and Robb]{Serrano-PerezEtAl2013}
J.~J. Serrano-Perez, F.~de~Vleeschouwer, F.~de~Proft, D.~Mendive-Tapia, M.~J.
  Bearpark, and M.~A. Robb.
\newblock How the conical intersection seam controls chemical selectivity in
  the photocycloaddition of ethylene and benzene.
\newblock \emph{J. Org. Chem.}, 78:\penalty0 1874--1886, 2013.

\bibitem[Trainer et~al.(2012)Trainer, Jimenez, Yung, Toon, and
  Tolbert]{TrainerEtAl2012}
M.~G. Trainer, J.~L. Jimenez, Y.~L. Yung, O.~B. Toon, and M.~A. Tolbert.
\newblock Nitrogen incorporation in ch$_4$-n$_2$ photochemical aerosol produced
  by far uv irradiation.
\newblock \emph{NASA archives}, 2012.

\bibitem[Ritson and Sutherland(2012)]{RitsonSutherland2012}
D.~Ritson and J.~Sutherland.
\newblock Prebiotic synthesis of simple sugars by photoredox systems chemistry.
\newblock \emph{Nature Chem.}, 4:\penalty0 895–899, 2012.
\newblock URL \url{https://doi.org/10.1038/nchem.1467}.

\bibitem[Das et~al.(2019)Das, Ghule, and Vanka]{DasEtAl2019}
Tamal Das, Siddharth Ghule, and Kumar Vanka.
\newblock Insights into the origin of life: Did it begin from hcn and h2o?
\newblock \emph{ACS Central Science}, 5\penalty0 (9):\penalty0 1532--1540,
  2019.
\newblock \doi{10.1021/acscentsci.9b00520}.
\newblock URL \url{https://doi.org/10.1021/acscentsci.9b00520}.

\bibitem[Pflüger(1875)]{Pfluger1875}
E.~Pflüger.
\newblock Beitragë zur lehre von der respiration. i. ueber die physiologische
  verbrennung in den lebendigen organismen.
\newblock \emph{Arch. Ges. Physiol.}, 10:\penalty0 641–644, 1875.

\bibitem[Minard and Matthews(2004)]{MinardMatthews2004}
R.~D. Minard and C.~N. Matthews.
\newblock Hcn world: Establishing proteininucleic acid life via hydrogen
  cyanide polymers.
\newblock \emph{Abstr. Pap. Am. Chem. Soc.}, 228:\penalty0 U963--U963, 2004.

\bibitem[Matthews(2004)]{Matthews2004}
C.~N. Matthews.
\newblock \emph{Series: Cellular Origin and Life in Extreme Habitats and
  Astrobiology}, volume~6, chapter The HCN World, pages 121--135.
\newblock Kluwer, Dordrecht, 2004.

\bibitem[Neveu et~al.(2013)Neveu, Kim, and Benner]{NeveuEtAl2013}
Marc Neveu, Hyo-Joong Kim, and Steven~A. Benner.
\newblock The “strong” rna world hypothesis: Fifty years old.
\newblock \emph{Astrobiology}, 13\penalty0 (4):\penalty0 391--403, 2013.
\newblock \doi{10.1089/ast.2012.0868}.
\newblock URL \url{https://doi.org/10.1089/ast.2012.0868}.
\newblock PMID: 23551238.

\bibitem[Boulanger et~al.(2013)Boulanger, Anoop, Nachtigallova, Thiel, and
  Barbatti]{BoulangerEtAl2013}
E.~Boulanger, A.~Anoop, D.~Nachtigallova, W.~Thiel, and M.~Barbatti.
\newblock Photochemical steps in the prebiotic synthesis of purine precursors
  from hcn.
\newblock \emph{Angew. Chem. Int.}, 52:\penalty0 8000--8003, 2013.

\bibitem[Or\'o(1960)]{Oro1960}
J.~Or\'o.
\newblock \emph{Biochem. Biophys. Res. Commun.}, 2:\penalty0 407–412, 1960.

\bibitem[Sanchez et~al.(1967)Sanchez, Ferris, and Orgel]{SanchezEtAl1967}
R.~A. Sanchez, J.~P. Ferris, and L.~E. Orgel.
\newblock Studies in prebiodc synthesis ii: Synthesis of purine precursors and
  amino acids from aqueous hydrogen cyanide.
\newblock \emph{J. Mol. Biol.}, 80:\penalty0 223--253, 1967.

\bibitem[Sanchez et~al.(1968)Sanchez, Ferris, and Orgel]{SanchezEtAl1968}
R.~A. Sanchez, J.~P. Ferris, and L.~E. Orgel.
\newblock Studies in prebiodc synthesis iv: Conversion of
  4-aminoimidazole-5-carbonitrile derivatives to purines.
\newblock \emph{J. Mol. Biol.}, 38:\penalty0 121--128, 1968.

\bibitem[Roy et~al.(2007)Roy, Najafian, and von Rague~Schleyer]{RoyEtAl2007}
D.~Roy, K.~Najafian, and P.~von Rague~Schleyer.
\newblock Chemical evolution: The mechanism of the formation of adenine under
  prebiotic conditions.
\newblock \emph{PNAS}, 104\penalty0 (44):\penalty0 17272–17277, October 2007.

\bibitem[Stribling and Miller(1986)]{StriblingMiller1986}
R.~Stribling and S.~L. Miller.
\newblock Energy yields for hydrogen cyanide and formaldehyde syntheses: The
  hcn and amino acid concentrations in the primitive ocean.
\newblock \emph{Origins Life}, 17:\penalty0 261–273, 1986.

\bibitem[Sanchez et~al.(1966)Sanchez, Ferris, and Orgel]{SanchezEtAl1966}
R.~Sanchez, J.~Ferris, and L.~E. Orgel.
\newblock Conditions for purine synthesis: Did prebiotic synthesis occur at low
  temperatures?
\newblock \emph{Science}, 153\penalty0 (3731):\penalty0 72--73, 1966.
\newblock ISSN 0036-8075.
\newblock \doi{10.1126/science.153.3731.72}.
\newblock URL \url{https://science.sciencemag.org/content/153/3731/72}.

\bibitem[Miller and Lazcano(1995)]{MillerLazcano1995}
S.~L. Miller and A.~Lazcano.
\newblock The origin of life – did it occur at high temperatures?
\newblock \emph{Mol. Evol.}, 41:\penalty0 689–692, 1995.

\bibitem[Bada and Lazcano(2002)]{BadaLazcano2002}
Jeffrey~L. Bada and Antonio Lazcano.
\newblock Some like it hot, but not the first biomolecules.
\newblock \emph{Science}, 296:\penalty0 1982--1983, 2002.

\bibitem[Miyakawa et~al.(2002{\natexlab{a}})Miyakawa, Cleaves, and
  Miller]{MiyakawaEtAlb2002}
S.~Miyakawa, H.~J. Cleaves, and S.~L. Miller.
\newblock The cold origin of life: B. implications based on pyrimidines and
  purines produced from frozen ammonium cyanide solutions.
\newblock \emph{Origins of Life and Evolution of the Biosphere}, 32:\penalty0
  209--218, 2002{\natexlab{a}}.

\bibitem[Hardy(1982)]{Hardy1982}
J.~T. Hardy.
\newblock The sea-surface microlayer (1982) biology, chemistry and
  anthropogenic enrichment.
\newblock \emph{Prog. Oceanogr.}, 11:\penalty0 307--328, 1982.

\bibitem[Grammatika and Zimmerman(2001)]{GrammatikaZimmerman2001}
M.~Grammatika and W.~B. Zimmerman.
\newblock Microhydrodynamics of flotation processes in the sea-surface layer.
\newblock \emph{Dynam. Atmos. Oceans}, 34:\penalty0 327--348, 2001.

\bibitem[Fábián et~al.(2014)Fábián, Szőri, and Jedlovszky]{FabianEtAl2014}
Balázs Fábián, Milán Szőri, and Pál Jedlovszky.
\newblock Floating patches of hcn at the surface of their aqueous solutions –
  can they make “hcn world” plausible?
\newblock \emph{The Journal of Physical Chemistry C}, 118\penalty0
  (37):\penalty0 21469--21482, 2014.
\newblock \doi{10.1021/jp505978p}.
\newblock URL \url{https://doi.org/10.1021/jp505978p}.

\bibitem[Oró(1995)]{Oro1995}
J.~Oró.
\newblock Chemical synthesis of lipids and the origin of life.
\newblock \emph{J Biol Phys}, 20:\penalty0 135--147, 1995.
\newblock \doi{https://doi.org/10.1007/BF00700430}.

\bibitem[Walde et~al.(1994)Walde, Wick, Fresta, Mangone, and
  Luisi]{WaldeEtAl1994}
Peter Walde, Roger Wick, Massimo Fresta, Annarosa Mangone, and Pier~Luigi
  Luisi.
\newblock Autopoietic self-reproduction of fatty acid vesicles.
\newblock \emph{Journal of the American Chemical Society}, 116\penalty0
  (26):\penalty0 11649--11654, 1994.
\newblock \doi{10.1021/ja00105a004}.
\newblock URL \url{https://doi.org/10.1021/ja00105a004}.

\bibitem[Deamer(2017)]{Deamer2017}
D.W. Deamer.
\newblock The role of lipid membranes in life’s origin.
\newblock \emph{Life}, 7:\penalty0 5, 2017.
\newblock URL \url{https://doi.org/10.3390/life7010005}.

\bibitem[Fan et~al.(2014)Fan, Fang, and Ma]{FanEtAl2014}
Ye~Fan, Yun Fang, and Lin Ma.
\newblock The self-crosslinked ufasome of conjugated linoleic acid:
  Investigation of morphology, bilayer membrane and stability.
\newblock \emph{Colloids and Surfaces B: Biointerfaces}, 123:\penalty0 8 -- 14,
  2014.
\newblock ISSN 0927-7765.
\newblock \doi{https://doi.org/10.1016/j.colsurfb.2014.08.028}.
\newblock URL
  \url{http://www.sciencedirect.com/science/article/pii/S0927776514004512}.

\bibitem[Han and Calvin(1969)]{HanCalvin1969}
Jerry Han and Melvin Calvin.
\newblock Occurrence of fatty acids and aliphatic hydrocarbons in a 3.4
  billion-year-old sediment.
\newblock \emph{Nature}, 224\penalty0 (5219):\penalty0 576--577, 1969.

\bibitem[Van~Hoeven et~al.(1969)Van~Hoeven, Maxwell, and
  Calvin]{VanHoevenEtAl1969}
William Van~Hoeven, JR~Maxwell, and Melven Calvin.
\newblock Fatty acids and hydrocarbons as evidence of life processes in ancient
  sediments and crude oils.
\newblock \emph{Geochimica et Cosmochimica Acta}, 33\penalty0 (7):\penalty0
  877--881, 1969.

\bibitem[Karhu and Epstein(1986)]{KarhuEpstein1986}
J.~Karhu and S.~Epstein.
\newblock The implication of the oxygen isotope records in coexisting cherts
  and phosphates.
\newblock \emph{Geochim. Cosmochim. Acta}, 50:\penalty0 1745–1756, 1986.

\bibitem[Knauth(1992)]{Knauth1992}
L.~P. Knauth.
\newblock \emph{Lecture Notes in Earth Sciences \#43}, chapter Isotopic
  Signatures and Sedimentary Records, pages 123--152.
\newblock Springer-Verlag, Berlin, 1992.

\bibitem[Knauth and Lowe(2003)]{KnauthLowe2003}
L.~P. Knauth and D.~R. Lowe.
\newblock High archean climatic temperature inferred from oxygen isotope
  geochemistry of cherts in the 3.5 ga swaziland group, south africa.
\newblock \emph{Geol. Soc. Am. Bull.}, 115:\penalty0 566--580, 2003.

\bibitem[Schoffstall(1976)]{Schoffstall1976}
A.~M. Schoffstall.
\newblock Prebiotic phosphorylation of nucleosides in formamide.
\newblock \emph{Origins Life Evol Biosphere}, 7:\penalty0 399–412, 1976.
\newblock URL \url{https://doi.org/10.1007/BF00927935}.

\bibitem[Costanzo et~al.(2007)Costanzo, Saladino, Crestini, Ciciriello, and
  Di~Mauro]{CostanzoEtAl2007}
G.~Costanzo, R.~Saladino, C.~Crestini, F.~Ciciriello, and E.~Di~Mauro.
\newblock Nucleoside phosphorylation by phosphate minerals.
\newblock \emph{J Biol Chem.}, 282\penalty0 (23):\penalty0 16729--16735, 2007.
\newblock URL \url{doi:10.1074/jbc.M611346200}.

\bibitem[{Turing}(1952)]{Turing1952}
A.~M. {Turing}.
\newblock {The Chemical Basis of Morphogenesis}.
\newblock \emph{Philosophical Transactions of the Royal Society of London
  Series B}, 237\penalty0 (641):\penalty0 37--72, August 1952.
\newblock \doi{10.1098/rstb.1952.0012}.

\bibitem[Petersen et~al.(2008)Petersen, Dahl, Jensen, Poulsen, Thøgersen, and
  Keiding]{PetersenEtAl2008}
Christian Petersen, Niels~Henning Dahl, Svend~Knak Jensen, Jens~Aage Poulsen,
  Jan Thøgersen, and Søren~Rud Keiding.
\newblock Femtosecond photolysis of aqueous formamide.
\newblock \emph{The Journal of Physical Chemistry A}, 112\penalty0
  (15):\penalty0 3339--3344, 2008.
\newblock \doi{10.1021/jp7110764}.
\newblock URL \url{https://doi.org/10.1021/jp7110764}.
\newblock PMID: 18321081.

\bibitem[Basch et~al.(1968)Basch, Robin, and Kuebler]{BaschEtAl1968}
Harold Basch, M.~B. Robin, and N.~A. Kuebler.
\newblock Electronic spectra of isoelectronic amides, acids, and acyl
  fluorides.
\newblock \emph{The Journal of Chemical Physics}, 49\penalty0 (11):\penalty0
  5007--5018, 1968.
\newblock \doi{10.1063/1.1669992}.
\newblock URL \url{https://doi.org/10.1063/1.1669992}.

\bibitem[Lelj and Adamo(1995)]{LeljAdamo1995}
F.~Lelj and C.~Adamo.
\newblock Solvent effects on isomerization equilibria: An energetic analysis in
  the framework of density functional theory.
\newblock \emph{Theoretica chimica acta}, 91:\penalty0 199–214, 1995.
\newblock \doi{10.1007/BF01114987}.
\newblock URL \url{https://doi.org/10.1007/BF01114987}.

\bibitem[Koch and Rodehorst(1974)]{KochRodehorst1974}
T.H. Koch and R.M. Rodehorst.
\newblock Quantitative investigation of the photochemical conversion of
  diaminomaleonitrile to diaminofumaronitrile and 4-amino-5-cyanoimidazole.
\newblock \emph{J. Am. Chem. Soc.}, 96:\penalty0 6707–6710, 1974.

\bibitem[Gupta and Tandon(2012)]{GuptaTandon2012}
V.P. Gupta and Poonam Tandon.
\newblock Conformational and vibrational studies of isomeric hydrogen cyanide
  tetramers by quantum chemical methods.
\newblock \emph{Spectrochimica Acta Part A: Molecular and Biomolecular
  Spectroscopy}, 89:\penalty0 55 -- 66, 2012.
\newblock ISSN 1386-1425.
\newblock \doi{https://doi.org/10.1016/j.saa.2011.12.030}.
\newblock URL
  \url{http://www.sciencedirect.com/science/article/pii/S1386142511011048}.

\bibitem[Ferris et~al.(1978)Ferris, Joshi, Edelson, and
  Lawless]{FerrisEtAl1978}
JP~Ferris, PC~Joshi, EH~Edelson, and JG~Lawless.
\newblock Hcn: a plausible source of purines, pyrimidines and amino acids on
  the primitive earth.
\newblock \emph{Journal of molecular evolution}, 11\penalty0 (4):\penalty0
  293--311, 1978.

\bibitem[Glaser et~al.(2007)Glaser, Hodgen, Farrelly, and
  McKee]{GlaserEtAl2007}
R.~Glaser, B.~Hodgen, D.~Farrelly, and E.~McKee.
\newblock Adenine synthesis in interstellar space: mechanisms of prebiotic
  pyrimidine-ring formation of monocyclic hcn-pentamers.
\newblock \emph{Astrobiology}, 7\penalty0 (3):\penalty0 455‐470, 2007.
\newblock \doi{doi:10.1089/ast.2006.0112}.

\bibitem[Cavaluzzi and Borer(2004)]{CavaluzziBorer2004}
M.~J. Cavaluzzi and P.~N. Borer.
\newblock Revised uv extinction coefficients for nucleoside-5'- monophosphates
  and unpaired dna and rna.
\newblock \emph{Nucleic Acids Research}, 32\penalty0 (1):\penalty0 e13, 2004.
\newblock \doi{10.1093/nar/gnh015}.

\bibitem[Franz and Gianturco(2014)]{FranzGianturco2014}
J.~Franz and F.A. Gianturco.
\newblock Low-energy positron scattering from dna nucleobases: the effects from
  permanent dipoles.
\newblock \emph{Eur. Phys. J. D}, 68\penalty0 (279), 2014.
\newblock URL \url{https://doi.org/10.1140/epjd/e2014-50072-0}.

\bibitem[Stimson and Reuter(1943)]{StimsonReuter1943}
Miriam~Michael Stimson and Mary~Agnita Reuter.
\newblock Ultraviolet absorption spectra of nitrogenous heterocycles. vii. the
  effect of hydroxy substitutions on the ultraviolet absorption of the series:
  Hypoxanthine, xanthine and uric acid1.
\newblock \emph{Journal of the American Chemical Society}, 65\penalty0
  (2):\penalty0 153--155, 1943.
\newblock \doi{10.1021/ja01242a006}.
\newblock URL \url{https://doi.org/10.1021/ja01242a006}.

\bibitem[Miyakawa et~al.(2002{\natexlab{b}})Miyakawa, Cleaves, and
  Miller]{MiyakawaEtAla2002}
S.~Miyakawa, H.~J. Cleaves, and S.~L. Miller.
\newblock The cold origin of life: A. implications based on the hydrolytic
  stabilities of hydrogen cyanide and formamide.
\newblock \emph{Origins of Life and Evolution of the Biosphere}, 32:\penalty0
  195--208, 2002{\natexlab{b}}.
\newblock URL \url{https://doi.org/10.1023/A:1016514305984}.

\bibitem[Kua and Thrush(2016)]{KuaThrush2016}
Jeremy Kua and Kyra~L. Thrush.
\newblock Hcn, formamidic acid, and formamide in aqueous solution: A
  free-energy map.
\newblock \emph{The Journal of Physical Chemistry B}, 120\penalty0
  (33):\penalty0 8175--8185, 2016.
\newblock \doi{10.1021/acs.jpcb.6b01690}.
\newblock URL \url{https://doi.org/10.1021/acs.jpcb.6b01690}.
\newblock PMID: 27016454.

\bibitem[Maier and Endres(2000)]{MaierEndres2000}
Günther Maier and Jörg Endres.
\newblock Isomerization of matrix-isolated formamide: Ir-spectroscopic
  detection of formimidic acid.
\newblock \emph{European Journal of Organic Chemistry}, 2000\penalty0
  (6):\penalty0 1061--1063, 2000.
\newblock
  \doi{10.1002/(SICI)1099-0690(200003)2000:6<1061::AID-EJOC1061>3.0.CO;2-5}.
\newblock URL
  \url{https://chemistry-europe.onlinelibrary.wiley.com/doi/abs/10.1002/}.

\bibitem[Duvernay et~al.(2005)Duvernay, Trivella, Borget, Coussan, Aycard, and
  Chiavassa]{DuvernayEtAl2005}
F.~Duvernay, A.~Trivella, F.~Borget, S.~Coussan, J.~P. Aycard, and
  T.~Chiavassa.
\newblock Matrix isolation fourier transform infrared study of
  photodecomposition of formimidic acid.
\newblock \emph{J. Phys. Chem. A}, 109:\penalty0 11155– 11162, 2005.

\bibitem[Barks et~al.(2010)Barks, Buckley, Grieves, Di~Mauro, Hud, and
  Orlando]{BarksEtAl2010}
H.~L. Barks, R.~Buckley, G.~A. Grieves, E.~Di~Mauro, N.~V. Hud, and T.~M.
  Orlando.
\newblock Guanine, adenine, and hypoxanthine production in uv-irradiated
  formamide solutions: Relaxation of the requirements for prebiotic purine
  nucleobase formation.
\newblock \emph{ChemBioChem}, 11\penalty0 (9):\penalty0 1240--1243, 2010.
\newblock \doi{10.1002/cbic.201000074}.
\newblock URL
  \url{https://chemistry-europe.onlinelibrary.wiley.com/doi/abs/10.1002/cbic.201000074}.

\bibitem[Gingell et~al.(1997)Gingell, Mason, Zhao, Walker, and
  Siggel]{GingellEtAl1997}
J.M. Gingell, N.J. Mason, H.~Zhao, I.C. Walker, and M.R.F. Siggel.
\newblock Vuv optical-absorption and electron-energy-loss spectroscopy of
  formamide.
\newblock \emph{Chemical Physics}, 220\penalty0 (1):\penalty0 191 -- 205, 1997.
\newblock ISSN 0301-0104.
\newblock \doi{https://doi.org/10.1016/S0301-0104(97)00137-7}.
\newblock URL
  \url{http://www.sciencedirect.com/science/article/pii/S0301010497001377}.

\bibitem[Yonemitsu et~al.(1974)Yonemitsu, Isshiki, and
  Kijima]{YonemitsuEtAl1974}
E.~Yonemitsu, T.~Isshiki, and Y.~Kijima.
\newblock Process for preparing adenine, 1974.
\newblock URL \url{https://patents.google.com/patent/US4059582A/en}.
\newblock US Patent 4,059,582.

\bibitem[Zubay and Mui(2001)]{ZubayMui2001}
G.~Zubay and T.~Mui.
\newblock Prebiotic synthesis of nucleotides.
\newblock \emph{Orig Life Evol Biosph}, 31:\penalty0 87–102, 2001.
\newblock URL \url{https://doi.org/10.1023/A:1006722423070}.

\bibitem[Wang et~al.(2013)Wang, Gu, Nguyen, Springsteen, and
  Leszczynski]{WangEtAl2013}
Jing Wang, Jiande Gu, Minh~Tho Nguyen, Greg Springsteen, and Jerzy Leszczynski.
\newblock From formamide to purine: A self-catalyzed reaction pathway provides
  a feasible mechanism for the entire process.
\newblock \emph{The Journal of Physical Chemistry B}, 117\penalty0
  (32):\penalty0 9333--9342, 2013.
\newblock \doi{10.1021/jp404540x}.
\newblock URL \url{https://doi.org/10.1021/jp404540x}.
\newblock PMID: 23902343.

\bibitem[Levy and Miller(1998)]{LevyMiller1998}
M.~Levy and S.~L. Miller.
\newblock The stability of the rna bases: Implications for the origin of life.
\newblock \emph{Proc. Natl. Acad. Sci. USA}, 95:\penalty0 7933–7938, 1998.

\bibitem[Wang and Hu(2016)]{WangHu2016}
Shiliang Wang and Anguang Hu.
\newblock Comparative study of spontaneous deamination of adenine and cytosine
  in unbuffered aqueous solution at room temperature.
\newblock \emph{Chemical Physics Letters}, 653:\penalty0 207 -- 211, 2016.
\newblock ISSN 0009-2614.
\newblock \doi{https://doi.org/10.1016/j.cplett.2016.05.001}.
\newblock URL
  \url{http://www.sciencedirect.com/science/article/pii/S0009261416302810}.

\bibitem[Wang et~al.(1991)Wang, Nichols, Feyereisen, Gutowski, Boatz, Haymet,
  and Simons]{WangEtAl1991}
Xiao~Chuan Wang, Jeff Nichols, Martin Feyereisen, Maciej Gutowski, Jerry Boatz,
  A.~D.~J. Haymet, and Jack Simons.
\newblock Ab initio quantum chemistry study of formamide-formamidic acid
  tautomerization.
\newblock \emph{The Journal of Physical Chemistry}, 95\penalty0 (25):\penalty0
  10419--10424, 1991.
\newblock \doi{10.1021/j100178a032}.
\newblock URL \url{https://doi.org/10.1021/j100178a032}.

\bibitem[Cataldo et~al.(2009)Cataldo, Patanè, and Compagnini]{CataldoEtAl2009}
Franco Cataldo, Giacomo Patanè, and Giuseppe Compagnini.
\newblock Synthesis of hcn polymer from thermal decomposition of formamide.
\newblock \emph{Journal of Macromolecular Science, Part A}, 46\penalty0
  (11):\penalty0 1039--1048, 2009.
\newblock \doi{10.1080/10601320903245342}.
\newblock URL \url{https://doi.org/10.1080/10601320903245342}.

\bibitem[Hill and Orgel(2002)]{HillOrgel2002}
A.~Hill and L.E. Orgel.
\newblock Synthesis of adenine from hcn tetramer and ammonium formate.
\newblock \emph{Orig Life Evol Biosph}, 32:\penalty0 99–102, 2002.
\newblock URL \url{https://doi.org/10.1023/A:1016070723772}.

\bibitem[Herbst(2001)]{Herbst2001}
Eric Herbst.
\newblock The chemistry of interstellar space.
\newblock \emph{Chem. Soc. Rev.}, 30:\penalty0 168--176, 2001.
\newblock \doi{10.1039/A909040A}.
\newblock URL \url{http://dx.doi.org/10.1039/A909040A}.

\bibitem[Benallou(2019)]{Benallou2019}
Abdelilah Benallou.
\newblock A new mechanistic insight of dna base adenine formation from pentamer
  hcn in the gas phase of interstellar clouds.
\newblock \emph{Journal of Taibah University for Science}, 13\penalty0
  (1):\penalty0 105--111, 2019.
\newblock \doi{10.1080/16583655.2018.1543163}.
\newblock URL \url{https://doi.org/10.1080/16583655.2018.1543163}.

\bibitem[Szabla et~al.(2014)Szabla, Šponer, Šponer, Sobolewski, and
  Góra]{SzablaEtAl2014b}
Rafał Szabla, Judit~E. Šponer, Jiří Šponer, Andrzej~L. Sobolewski, and
  Robert~W. Góra.
\newblock Solvent effects on the photochemistry of
  4-aminoimidazole-5-carbonitrile{,} a prebiotically plausible precursor of
  purines.
\newblock \emph{Phys. Chem. Chem. Phys.}, 16:\penalty0 17617--17626, 2014.
\newblock \doi{10.1039/C4CP02074J}.
\newblock URL \url{http://dx.doi.org/10.1039/C4CP02074J}.

\bibitem[Frick(1952)]{Frick1952}
G.~Frick.
\newblock Formation of amino-acids in hydrolysis of adenine.
\newblock \emph{Nature}, 169:\penalty0 758–759, 1952.
\newblock URL \url{https://doi.org/10.1038/169758a0}.

\bibitem[Zheng and Meng(2009)]{ZhengMeng2009}
H.~Zheng and F.~Meng.
\newblock Theoretical study of water-assisted hydrolytic deamination mechanism
  of adenine.
\newblock \emph{Struct Chem}, 20:\penalty0 943–949, 2009.
\newblock URL \url{https://doi.org/10.1007/s11224-009-9495-z}.

\bibitem[Yang and Hinner(2015)]{YangHinner2015}
N.~J. Yang and M.~J. Hinner.
\newblock Getting across the cell membrane: An overview for small molecules,
  peptides, and proteins.
\newblock \emph{Methods Mol Biol.}, 1266:\penalty0 29–53, 2015.
\newblock \doi{10.1007/978-1-4939-2272-7_3}.

\bibitem[Agarwal et~al.(2008)Agarwal, Clancy, and Harvey]{AgarwalEtAl2016}
S.~Agarwal, C.~Clancy, and R.~Harvey.
\newblock Mechanisms restricting diffusion of intracellular camp.
\newblock \emph{Sci. Rep.}, 6:\penalty0 19577, 2008.
\newblock \doi{10.1038/srep19577}.

\bibitem[Bowen and Martin(1964)]{BowenMartin1964}
William~J. Bowen and Harold~L. Martin.
\newblock The diffusion of adenosine triphosphate through aqueous solutions.
\newblock \emph{Archives of Biochemistry and Biophysics}, 107\penalty0
  (1):\penalty0 30 -- 36, 1964.
\newblock ISSN 0003-9861.
\newblock \doi{https://doi.org/10.1016/0003-9861(64)90265-6}.
\newblock URL
  \url{http://www.sciencedirect.com/science/article/pii/0003986164902656}.

\bibitem[Zahnle(1986)]{Zahnle1986}
Kevin~J. Zahnle.
\newblock Photochemistry of methane and the formation of hydrocyanic acid (hcn)
  in the earth's early atmosphere.
\newblock \emph{Journal of Geophysical Research: Atmospheres}, 91\penalty0
  (D2):\penalty0 2819--2834, 1986.
\newblock \doi{10.1029/JD091iD02p02819}.
\newblock URL
  \url{https://agupubs.onlinelibrary.wiley.com/doi/abs/10.1029/JD091iD02p02819}.

\bibitem[Zhang et~al.(2004)Zhang, Liu, Liu, Yu, and Wang]{ZhangEtAl2004}
Z.~Zhang, C.~Liu, L.~Liu, L.~Yu, and Z.~Wang.
\newblock Study on dissolved trace metals in sea surface microlayer in daya
  bay.
\newblock \emph{Chinese Journal of Oceanology and Limnology}, 22\penalty0
  (1):\penalty0 54 -- 63, 2004.

\bibitem[{Oksana Shvydkiv}(2012)]{Shvydkiv2012}
{Oksana Shvydkiv}.
\newblock \emph{Microphotochemistry – a New Resources-Efficient Synthesis
  Tool Approach}.
\newblock PhD thesis, Chemical Sciences Dublin City University, 2012.
\newblock URL \url{https://core.ac.uk/download/pdf/147603119.pdf}.
\newblock The School of Chemical Sciences Dublin City University.

\bibitem[Airey and Dainton(1966)]{AireyDainton1966}
P.~L. Airey and Frederick~Sydney Dainton.
\newblock The photochemistry of aqueous solutions of fe(ii) ii. processes in
  acidified solutions of potassium ferrocyanide at 25\&\#xb0;c.
\newblock \emph{Proceedings of the Royal Society of London. Series A.
  Mathematical and Physical Sciences}, 291\penalty0 (1427):\penalty0 478--486,
  1966.
\newblock \doi{10.1098/rspa.1966.0109}.
\newblock URL
  \url{https://royalsocietypublishing.org/doi/abs/10.1098/rspa.1966.0109}.

\bibitem[Fox and Harada(1961)]{FoxHarada1961}
Sidney~W. Fox and Kaoru Harada.
\newblock Synthesis of uracil under conditions of a thermal model of
  prebiological chemistry.
\newblock \emph{Science}, 133\penalty0 (3468):\penalty0 1923--1924, 1961.
\newblock ISSN 0036-8075.
\newblock \doi{10.1126/science.133.3468.1923}.
\newblock URL \url{https://science.sciencemag.org/content/133/3468/1923}.

\bibitem[Ferris et~al.(1968)Ferris, Sanchez, and Orgel]{FerrisEtAl1968}
J.~P. Ferris, R.~A. Sanchez, and L.~E. Orgel.
\newblock Studies in prebiotic synthesis iii. synthesis of pyrimidines from
  cyanoacetylene and cyanate.
\newblock \emph{J. Mol. Biol.}, 33:\penalty0 693--704, 1968.

\bibitem[Choughuley et~al.(1977)Choughuley, Subbaraman, Kazi, and
  Chadha]{ChoughuleyEtAl1977}
A.S.U. Choughuley, A.S. Subbaraman, Z.A. Kazi, and M.S. Chadha.
\newblock A possible prebiotic synthesis of thymine: Uracil-formaldehyde-formic
  acid reaction.
\newblock \emph{Biosystems}, 9\penalty0 (2):\penalty0 73 -- 80, 1977.
\newblock ISSN 0303-2647.
\newblock \doi{https://doi.org/10.1016/0303-2647(77)90014-4}.
\newblock URL
  \url{http://www.sciencedirect.com/science/article/pii/0303264777900144}.
\newblock Fifth international conference on the origin of life.

\bibitem[Luna et~al.(1998)Luna, Morizur, Tortajada, Alcamí, Mó, and
  Yáñez]{LunaEtAl1998}
A.~Luna, J.-P. Morizur, J.~Tortajada, M.~Alcamí, O.~Mó, and M.~Yáñez.
\newblock Role of cu+ association on the formamide $\rightarrow$ formamidic
  acid $\rightarrow$ (aminohydroxy)carbene isomerizations in the gas phase.
\newblock \emph{The Journal of Physical Chemistry A}, 102\penalty0
  (24):\penalty0 4652--4659, 1998.
\newblock \doi{10.1021/jp980629c}.
\newblock URL \url{https://doi.org/10.1021/jp980629c}.

\bibitem[Gates(1980)]{Gates1980}
D.~M. Gates.
\newblock \emph{Biophysical Ecology}.
\newblock Springer-Verlag, 1980.
\newblock ISBN 0-387-90414-X.

\bibitem[Ulanowicz and Hannon(1987)]{UlanowiczHannon1987}
R.E. Ulanowicz and B.M. Hannon.
\newblock Life and the production of entropy.
\newblock \emph{Proc R Soc Lond B}, 232:\penalty0 181--192, 1987.

\bibitem[Schneider and Kay(1994)]{SchneiderKay1994}
E.~D. Schneider and J.~J. Kay.
\newblock Complexity and thermodynamics: towards a new ecology.
\newblock \emph{Futures}, 24:\penalty0 626--647, 1994.

\bibitem[Kleidon(2008)]{Kleidon2008}
Axel Kleidon.
\newblock Entropy production by evapotranspiration and its geographic
  variation.
\newblock \emph{Soil \& Water Res.}, 3:\penalty0 S89--S94, 2008.

\bibitem[Zotin(1984)]{Zotin1984}
A.~I. Zotin.
\newblock \emph{Bioenergetic trends of evolutionary progress of organisms},
  pages 451--458.
\newblock De Gruyter, 1984.

\bibitem[Limaye et~al.(2018)Limaye, Mogul, Smith, Ansari, Słowik, and
  Vaishampayan]{LimayeEtAl2018}
Sanjay~S. Limaye, Rakesh Mogul, David~J. Smith, Arif~H. Ansari, Grzegorz~P.
  Słowik, and Parag Vaishampayan.
\newblock Venus' spectral signatures and the potential for life in the clouds.
\newblock \emph{Astrobiology}, 18\penalty0 (9):\penalty0 1181--1198, 2018.
\newblock \doi{10.1089/ast.2017.1783}.
\newblock URL \url{https://doi.org/10.1089/ast.2017.1783}.
\newblock PMID: 29600875.

\bibitem[Heinz and Schulze-Makuch(2020)]{HeinzSchulze2020}
Jacob Heinz and Dirk Schulze-Makuch.
\newblock Thiophenes on mars: Biotic or abiotic origin?
\newblock \emph{Astrobiology}, 20\penalty0 (4):\penalty0 552--561, 2020.
\newblock \doi{10.1089/ast.2019.2139}.
\newblock URL \url{https://doi.org/10.1089/ast.2019.2139}.
\newblock PMID: 32091933.

\bibitem[Pershin(2002)]{Pershin2002}
S.~Pershin.
\newblock Correlation of “chlorophyll” and water index on mars surface.
\newblock In \emph{Microsymposium 36, MS079}, 2002.

\bibitem[L{\'o}pez-Puertas et~al.(2013)L{\'o}pez-Puertas, Dinelli, Adriani,
  Funke, Garc{\'i}a-Comas, Moriconi, D’Aversa, Boersma, and
  Allamandola]{LopezPuertasEtAl2013}
Manuel L{\'o}pez-Puertas, Bianca~Maria Dinelli, Alberto Adriani, Bernd Funke,
  Maya Garc{\'i}a-Comas, M~L Moriconi, Emiliano D’Aversa, Cornelis Boersma,
  and Louis~J. Allamandola.
\newblock Large abundances of polycyclic aromatic hydrocarbons in titan's upper
  atmosphere.
\newblock \emph{The Astrophysical Journal}, 770:\penalty0 132, 2013.

\bibitem[Einstein(1905)]{Einstein1905}
A.~Einstein.
\newblock On the movement of small particles suspended in stationary liquids
  required by the molecular-kinetic theory of heat.
\newblock \emph{Ann. Phys.}, \penalty0 (17):\penalty0 549, 1905.

\bibitem[Onsager(1931{\natexlab{a}})]{Onsager1931a}
L.~Onsager.
\newblock {Reciprocal Relations in Irreversible Processes, I}.
\newblock \emph{Phys. Rev.}, 37:\penalty0 405--426, 1931{\natexlab{a}}.

\bibitem[Onsager(1931{\natexlab{b}})]{Onsager1931b}
L.~Onsager.
\newblock {Reciprocal Relations in Irreversible Processes, II}.
\newblock \emph{Phys. Rev.}, 38:\penalty0 2265, 1931{\natexlab{b}}.

\bibitem[Callen and Wellton(1951)]{CallenWelton1951}
H.~B. Callen and T.~A. Wellton.
\newblock Irreversibility and generalized noise.
\newblock \emph{Physical Review}, 83:\penalty0 34--40, 1951.

\bibitem[Onsager and Machlup(1953)]{OnsagerMachlup1953}
L.~Onsager and S.~Machlup.
\newblock Fluctuations and irreversible processes.
\newblock \emph{Phys. Rev.}, 91:\penalty0 1505--1512, 1953.

\bibitem[Kubo(1966)]{Kubo1966}
R.~Kubo.
\newblock The fluctuation-dissipation theorem.
\newblock \emph{Rep. Prog. Phys.}, 29:\penalty0 255--284, 1966.

\bibitem[Evans et~al.(1993)Evans, Cohen, and Morriss]{EvansEtAl1993}
Denis~J. Evans, E.~G.~D. Cohen, and G.~P. Morriss.
\newblock Probability of second law violations in shearing steady states.
\newblock \emph{Phys. Rev. Lett.}, 71:\penalty0 2401--2404, Oct 1993.
\newblock \doi{10.1103/PhysRevLett.71.2401}.
\newblock URL \url{https://link.aps.org/doi/10.1103/PhysRevLett.71.2401}.

\bibitem[Evans et~al.(2002)Evans, Cohen, and Morriss]{EvansSearles2002}
D.~J. Evans, E.~G.~D. Cohen, and G.~P. Morriss.
\newblock The fluctuation theorem.
\newblock \emph{Advances in Physics}, 51\penalty0 (7):\penalty0 1529--1585,
  2002.

\bibitem[Gaspard(2013)]{Gaspard2013}
P.~Gaspard.
\newblock \emph{Engineering of Chemical Complexity}, chapter Self-Organization
  at the Nanoscale Scale inFar-From-Equilibrium Surface Reactions and
  Copolymerizations, pages 51--77.
\newblock World Scientific, 2013.
\newblock ISBN 978-9814390453.

\bibitem[Gaspard(2004)]{Gaspard2004}
Pierre Gaspard.
\newblock Fluctuation theorem for nonequilibrium reactions.
\newblock \emph{The Journal of Chemical Physics}, 120\penalty0 (19):\penalty0
  8898--8905, 2004.
\newblock \doi{10.1063/1.1688758}.
\newblock URL \url{https://doi.org/10.1063/1.1688758}.

\bibitem[Andrieux and Gaspard(2007)]{AndrieuxGaspard2007}
D.~Andrieux and P.~Gaspard.
\newblock Fluctuation theorem for currents and schnakenberg network theory.
\newblock \emph{J. Stat. Phys.}, 127:\penalty0 107--131, 2007.

\bibitem[Andrieux et~al.(2009)Andrieux, Gaspard, Monnai, and
  Tasaki]{AndrieuxEtAl2009}
D~Andrieux, P~Gaspard, T~Monnai, and S~Tasaki.
\newblock The fluctuation theorem for currents in open quantum systems.
\newblock \emph{New Journal of Physics}, 11\penalty0 (4):\penalty0 043014, apr
  2009.
\newblock \doi{10.1088/1367-2630/11/4/043014}.
\newblock URL \url{https://doi.org/10.1088%2F1367-2630%2F11%2F4%2F043014}.

\end{thebibliography}

\end{document}